\documentclass[epj,iicol,numbook]{svjour}

\usepackage{float}
\usepackage{epsfig}
\usepackage{dsfont}
\usepackage{amsmath}
\usepackage{braket}
\usepackage{amsfonts}
\usepackage{amssymb}
\usepackage{graphicx}  
\usepackage[dvipsnames]{xcolor}
\usepackage{todonotes}
\usepackage{subcaption}
\usepackage{bbm}
\usepackage{tikz}
\usepackage{multirow}
\usepackage{lmodern}
\usepackage{braket}
\usepackage{wrapfig}
\usepackage{comment}
\usepackage{xcolor}
\usepackage{mathtools}
\usepackage{cuted}
\RequirePackage[colorlinks,citecolor=blue,urlcolor=magenta,linkcolor=blue]{hyperref}
\usepackage[square,numbers,sort&compress]{natbib}

\begin{document}
	
	\title{Analyzing quantum gravity spillover in the semiclassical regime}
	
	\author{Harkirat Singh Sahota\thanks{harkirat221@gmail.com, ph17078@iisermohali.ac.in} \and Kinjalk Lochan\thanks{kinjalk@iisermohali.ac.in}}
	
	\institute{Department of Physical Sciences, Indian Institute of Science Education \& Research (IISER) Mohali, Sector 81 SAS Nagar, Manauli PO 140306 Punjab India.}
	
	\date{Received: date / Revised version: date}
	\abstract{
		One of the standard approaches of incorporating the quantum gravity (QG) effects into the semiclassical analysis is to adopt the notion of a quantum-corrected spacetime arising from the QG model. This procedure assumes that the expectation value of the metric variable effectively captures the relevant QG subtleties in the semiclassical regime. We investigate the viability of this effective geometry approach for the case of dust dominated and a dark energy dominated universe. We write the phase space expressions for the geometric observables and construct corresponding Hermitian operators. A general class of operator ordering of these observables is considered, and their expectation values are calculated for a unitarily evolving wave packet. In the case of dust dominated universe,  the expectation value of the Hubble parameter matches the ``semiclassical” expression, the expression computed from the scale factor expectation value. In the case of the Ricci scalar, the relative difference between the semiclassical expression and quantum expectation is maximum at singularity and decays for late time. For a cosmological constant driven universe, the difference between the semiclassical expressions and the expectation value is most pronounced far away from the bounce point, hinting at the persistent quantum effect at the late time. The parameter related to the shape of the distribution appears as a control parameter in these models. In the limit of a sharply peaked distribution, the expectation value of the observables matches with their semiclassical counterpart, and the usage of effective geometry approach is justified.
	\PACS{{PACS-key}{discribing text of that key} \and {PACS-key}{discribing text of that key} }}
	
	\titlerunning{Analyzing quantum gravity spillover in the semiclassical regime} \authorrunning{Sahota \& Lochan}
	\maketitle
	\section{Introduction}\label{Sec1}
	
	The notion of observables in canonical quantum gravity has been at the forefront of the issues that plague the theory; (for recent reviews; see, e.g., \cite{Tambornino:2011vg,anderson_2012} and references therein). Classically, at the level of canonical description of the singular systems, i.e., the systems with gauge degrees of freedom, the whole phase space is no longer the physical space \cite{henneaux_quantization_1992,Gitman:1990qh,Date:2010xr,prokhorov_hamiltonian_2011}. The gauge redundancy is encoded in a set of functions of phase space variables which are constrained to vanish on-shell. These functions are called the constraints of the theory, and they define the physical space, called the constraint surface, in the phase space. The constraints generate the gauge transformations on the phase space, and not all functions of phase space variables correspond to the physical (Dirac) observables. The phase space functions that are invariant under the gauge transformations are considered the Dirac observables. This implies that these functions need to have weakly vanishing Poisson bracket with the generators of these gauge transformations \cite{Gitman:1990qh,henneaux_quantization_1992,Date:2010xr,prokhorov_hamiltonian_2011}.

	General relativity is a famous example of the singular systems, which has diffeomorphism and time reparametrization constraints appearing at the canonical level \cite{Dirac:1958sc,arnowitt_dynamical_1959,PhysRev.114.924}, that generate the gauge transformations in the phase space, i.e., diffeomorphism and time reparametrization transformations \cite{Bergmann:1972ud,Lee:1990nz,Pons:1996av}. The observables in general relativity, therefore, must be invariant under diffeomorphisms and time reparametrizations. However, the discussion surrounding the systems with time reparametrization symmetry is tricky as the Hamiltonian of such systems is itself a constraint. This means that the Hamiltonian generates the gauge transformations, implying the dynamics in a generally covariant system is just the unfolding of the gauge transformation. It seems that the physical observables in such a system are frozen in time or, in other words, are constants of motion; this is commonly referred to as the problem of time in QG \cite{kuchar_time_2011,Isham:1992ms}. 
	
	A possible resolution to this conundrum comes from the understanding that the canonical Hamiltonian generates the evolution in the coordinate time, which due to general covariance, is indeed redundant. What we observe is the evolution of the dynamical fields with respect to the other fields. This idea is best implemented in the context of relational quantum dynamics, see e.g., \cite{Tambornino:2011vg,Chataignier:2021ncn}. The approach, in a nutshell, is an amalgamation of the different manifestations of relational notions available, called the `{\it trinity of relational quantum dynamics}' \cite{Hoehn:2019fsy,Hoehn:2020epv}. The system can be described equally well by a clock-neutral picture of Dirac quantization (first quantize then constrain), the relational Schr\"{o}dinger picture of the Page-Wootters formalism \cite{PhysRevD.27.2885,Wootters1984}, and the relational Heisenberg picture resulting from quantum symmetry reduction \cite{universe5050116,Hohn:2018toe}. A detailed exposition of these approaches can be found in \cite{Hoehn:2019fsy}.
	
	The motivation to have the discussion about observables is twofold - how the QG effects are incorporated in the observations, e.g., in effective geometry approach \cite{Ashtekar:2009mb,Agullo:2012sh,Agullo:2012fc,Agullo:2013ai} and how singularity resolution is addressed in the canonical approach to quantum gravity \cite{dewitt_quantum_1967,kiefer_quantum_2012,kiefer_singularity_2019-1}. The operational approach to incorporate the effects of quantum gravity is to introduce the notion of quantum-corrected spacetime coming from the quantization of the background geometry. It can be achieved in various ways, e.g., in the dressed metric approach, where the evolution of quantum fields on quantum geometry defined by a physical state $\Psi_o$ is mathematically equivalent to their evolution on an effective classical background geometry ``dressed" with quantum corrections \cite{Agullo:2012sh}. The quantum state $\Psi_o$ is conjectured to be sharply peaked on the classical trajectory. The dressed metric is defined in such a way that it captures the moments of the field variables appearing in the Schr\"odinger equation for the perturbation variables. Therefore, the main argument is that the dynamics of the quantum fields is only sensitive to the expectation values that are captured by the dressed metric.
	
	Quantum corrections to the background geometry can also be encoded via a quantum-corrected Friedmann equation coming from, e.g., polymerized Hamiltonian constraint \cite{Ashtekar:2006uz,Ashtekar:2006wn} or de Broglie-Bohm quantization \cite{Peter:2005hm,Peter:2006id,Peter:2006hx,Peter:2008qz} or via semiclassical description of affine quantization scheme \cite{Bergeron:2017ddo,Malkiewicz:2020fvy,Martin:2021dbz,Martin:2022ptk}. We will colloquially term these approaches as effective geometry approach where the singular background geometry is replaced by a quantum-corrected singularity-free spacetime.
	A pertinent question to ask here is whether it is justified to use the semiclassical expressions for the Hubble parameter and Ricci scalar (in the non-minimal setting, for example) instead of the expectation value of these observables coming from the QG model. Most of these approaches to quantum gravity rely on the expectation of one dynamical variable, e.g., scale factor or volume, to completely capture the quantum effects. But being a canonical theory, it is worthwhile to explore whether there exists any inconsistency in the expectations of complex product operators such as the Hubble parameter and the Ricci scalar are made up from the conjugate variables. It is equivalent to checking if the expectation of the position operator captures all the quantum characteristics for a free particle. In this work, we will write phase space expressions for these observables. Since these observables are made up of the scale factor and its conjugate momentum, knowing the expectation value of the scale factor alone may not be sufficient. In principle, one should scrutinize any scheme that involves quantum-corrected spacetime through the effective metric and check if any significant departure is observed between the semiclassical expressions (quantities computed from the expectation value of the metric, i.e., effective metric) of the observables of interest and their expectation value.

	Furthermore, DeWitt's criteria is widely used as a marker for the non-existence of singularity in a quantum gravity model. It states that ``\textit{A singularity is said to be avoided if} $\Psi\rightarrow 0$ \textit{in the vicinity of the classical singularity}" \cite{dewitt_quantum_1967}. DeWitt's criteria has been applied for various systems to check for singularity resolution, e.g., \cite{PhysRevD.11.768,PhysRevD.28.2402,hajicek_singularity_2001,Dabrowski:2006dd,bergeron_singularity_2015,liu_singularity_2014,kamenshchik_quantum_2013,sami_avoidance_2006,kiefer_quantum_2015,Kamenshchik:2016rtr,Alonso-Serrano:2018zpi,Bouhmadi-Lopez:2019zvz,kiefer_singularity_2019,kiefer_singularity_2019-1,Jalalzadeh:2022bgz}. There also exists criteria of singularity avoidance based on the vanishing of Klein-Gordon like current associated with Wheeler-DeWitt equation in the vicinity of the singularity and the spreading of the wave packet near singularity \cite{Dabrowski:2006dd,Kiefer:2019bxk}. Studying the spectrum or the expectation value of configuration variables of the model e.g., scale factor or volume operator, is another way to infer the singularity structure in a quantum model \cite{Bojowald:2001xe,Alvarenga:2001nm,Bojowald:2001vw,Bojowald:2002gz,Gryb:2018whn,Gielen:2020abd,Gielen:2021igw,Gielen:2022tzi,Malkiewicz:2019azw}. In general relativity, however, a singular configuration is characterized by the divergence of the curvature invariants and metric variable may indicate the presence of coordinate singularity only. One would therefore expect that the prediction of singularity resolution in these models is robust if the Hermitian operators associated with the curvature invariants have finite expectation values at the singular configurations. Since such operators contain both canonically conjugate operators, the expectation value of only one operator is not guaranteed to capture all the quantum gravity effects. An extensive account of the various singularity resolution criteria used in the context of quantum cosmology can be found in \cite{Thebault:2022dmv}.
	
    However, within the framework of Dirac observables it is not possible to obtain any local, dynamical observable corresponding to any geometric quantity of interest, e.g., scale factor, the Hubble parameter, or any local curvature invariants (which appear at the semiclassical level), since the former demands a complete spacetime independence, by construction. Therefore, one needs to resort to the relational approach via a parameterized system set-up \cite{Chataignier:2021ncn,Hoehn:2019fsy} where one observes the growth of any of the variable of interest viz-a-viz another degree of freedom, while the complete system satisfies the geometric constraints. Still, there remains a question of what phase space functions can be classified as valid observables.  In this work, this question is addressed by following Kucha\v{r}'s proposal for the observables in general relativity \cite{kuchar_canonical_1993,barbour2008constraints,Bojowald_2011}. The main idea is to distinguish between conventional gauge systems and parameterized systems. In the case of parameterized systems, Kucha\v{r} proposed to relax the weakly vanishing of the Poisson bracket of observables with the Hamiltonian constraint in order to capture the dynamics of geometry. This enables one to obtain the growth of spatial diffeomorphism invariant observables w.r.t. to another component. In the context of the FLRW model with fluid as matter, the spatial diffeomorphism constriants are already taken care of via symmetry reduction at the action level itself and one can consider all functions of the scale factor and its conjugate momentum as observables at the classical level. Nevertheless, at the quantum level one needs to worry about which variable should be treated as time and what phase space combination will remain a quantum observable.

	The problem of time in this work is dealt with by working with Brown-Kucha\v{r} dust \cite{brown_dust_1995} and cosmological constant \cite{Henneaux:1989zc,Unruh:1988in,Kuchar:1991xd} as matter, and the fluid degree of freedom provides the notion of a clock in the system. These models are phenomenologically constructed, such that the momentum conjugate to the fluid variable appears linearly in the Hamiltonian constraint. Classically, the fluid variable is linearly related to the comoving time in the appropriate gauge. The Wheeler-DeWitt equation takes the form of Schr\"{o}dinger equation with dust variable as the Schr\"{o}dinger time \cite{Amemiya_2009}. In the quantum picture, unitarity is demanded with respect to the fluid clock. Dynamics in the quantum model tell us about the behavior of gravitational degrees of freedom with respect to the matter degree of freedom, thereby implementing the relational notion of observables. 
	
	After quantization, one would ideally like to obtain the self-adjoint extension of all the relevant observables appearing in theory and study their spectral properties. In the context of infinite dimensional Hilbert space, the self-adjoint operator is a Hermitian operator with the added condition that the domain of the adjoint of the operator is equal to the domain of the operator \cite{Gieres_2000,Bonneau_2001,Gitman2012}. The blueprint that we will follow in this work is to write the self-adjoint extension for the Hamiltonian operator to ensure unitary evolution in the model and construct wave packets from the eigenfunctions of the Hamiltonian operator. We will primarily concern ourselves with the Hermitian extension of the other operators, whose behavior we are interested in. The Hermiticity condition suffices in this case because it ensures that the expectation values are real. 
	
	In this work, we aim to address three questions: (I) Writing the phase space functions corresponding to the observables that depict singularity in the classical picture, and studying their behavior in the quantum model, thereby checking the robustness of DeWitt's criteria. (II) The status of the operator ordering ambiguity in this quantum model. (III) The domain of validity of the effective geometry approach, where the QG signatures are investigated by replacing the scale factor expectation in the classical expressions. These questions are address in the context of a minisuperspace model of gravity in which the system has a finite number of degrees of freedom. Although there are conceptual issues regarding such symmetry reduction before quantization \cite{PhysRevD.40.3982}, yet this toy model is a perfect playground that is relatively easier to handle analytically and also captures the essence of the subtleties associated with the quantization of gravity. 
	
	We will address these issues in the case of a flat-FLRW (Friedmann-Lema\^{i}tre-Robertson-Walker) model with the Brown -Kucha\v{r} dust \cite{brown_dust_1995} and the cosmological constant \cite{Henneaux:1989zc} as the clock. Following Kucha\v{r}'s prescription of observables in the QG models, we will write the quantum observables that correspond to the Hubble parameter, Ricci scalar, and other curvature invariants. We will start with the discussion on observables in generally covariant systems in Sec. \ref{Sec2}. The classical and quantum description of the FLRW model with Brown-Kucha\v{r} dust is given in Sec. \ref{Sec3}, and we will discuss singularity resolution in this model. In Sec. \ref{Sec4}, we will write Hermitian extensions of the observables of importance and address the viability of the effective geometry approach. We will extend this analysis for a perfect fluid model in Sec. \ref{Sec5} and \ref{Sec6} and investigate whether the generic features observed in the earlier case can be seen in this case as well. Finally, we will summarize the results in Sec. \ref{Conclusion} with some remarks.
	
	
	\section{Observables and gauge invariance in generally covariant systems}\label{Sec2}
	Physical observables in a gauge theory are supposed to be invariant under the gauge transformations \cite{Gitman:1990qh,henneaux_quantization_1992}. The canonical analysis of general relativity establishes that the redundancy associated with the choice of coordinates does, in fact, have a direct correspondence with the gauge transformations in the geometrodynamical phase space \cite{Bergmann:1972ud,Lee:1990nz,Pons:1996av}. The total Hamiltonian of general relativity is a constraint that generates the time reparameterization and diffeomorphisms. Thus, the direct implementation of Dirac's ideas about gauge systems leads to the counterintuitive notion of frozen dynamics \cite{Isham:1992ms,kuchar_time_2011}. 
	
	The notion of time and observables in generally covariant systems is discussed in the context of relational dynamics. Due to time reparameterization invariance, the dynamics with respect to the coordinate time is indeed redundant, and a physically meaningful change is observed relationally. Instead of observing change with respect to an absolute external time, the dynamics in a generally covariant system is observed through an internal clock. In a gauge theory, all degrees of freedom are not dynamical, and one can, in principle, choose any internal degree of freedom as a clock. Since one can choose the clock variable at their whim, there exist multiple choices of clocks in the model \cite{Isham:1992ms,kuchar_time_2011}. The inequivalence of clock choice at the quantum level, as conjectured by Gotay and Demaret \cite{PhysRevD.28.2402}, has been investigated in recent works \cite{Gielen:2020abd,Gielen:2021igw,Gielen:2022tzi,Malkiewicz:2019azw}, where it is shown that the quantum model with different clock choices leads to different quantum dynamics. Choosing the fluid clock, a slow clock (which encounters a singularity in finite time), according to the Gotay-Demaret terminology, leads to the singularity resolution. While the model with the scale factor clock, which is a fast clock (which reaches the singular point asymptotically), does not resolve the singularity \cite{Gielen:2022tzi}.
	
	The mainstream implementation of relational dynamics is achieved through Rovelli's proposal of partial and complete observables \cite{Rovelli_1990,Rovelli_1991}. A partial observable is a `physical quantity to which we can associate a measuring procedure leading to a number,' with the assumption that one can associate a measuring procedure to an arbitrary phase space function. A partial observable is a phase space function, and it does not have to commute (weakly) with the constraints of the theory. A complete observable is a `quantity whose value can be predicted by the theory'. Therefore, a complete observable has to commute with the constraints and is, in fact, the Dirac observable. Relational ideas are incorporated by considering two partial observables, an internal clock $T$ and a phase space function $f$, calculating the value of $f$ at a {\it time} at which $T$ takes the value $\tau$. The value of $f$ at the ``clock time" $\tau$ is a constant of motion for the flows generated by the Hamiltonian constraint and therefore gives a one-parameter family of complete (Dirac) observables.
	
	This vague statement is cast succinctly in mathematical language in \cite{Dittrich:2004cb,Dittrich:2005kc}. The flow $\alpha^t_C(x)$ of a phase space point $x$ generated by constraint $C$ with a Hamiltonian vector field $\chi_C(f)=\{C,f\}$ satisfies
	\begin{align}
		\frac{d}{dt}\alpha^t_C(x)=\chi_C(\alpha^t_C(x)),\quad\text{and}\quad\alpha^t_C(f)(x)=f(\alpha^t_C(x)).
	\end{align}
	With these definitions, for two partial observables $f$ and $T$, one can associate a family of complete observables labeled by a parameter $\tau$,  $F_{[f,T]}(\tau,x)$ defined as
	\begin{align}
		F_{[f,T]}(\tau,x)=\alpha^t_C(f)(x)\big|_{\alpha^t_C(T)(x)=\tau}.\label{RelObs}
	\end{align}
	Let $f$, $T$ be two phase space functions and $x\in\mathcal{M}$ be a phase space point, fulfilling the condition $\alpha^t_C(f)(x)=\alpha^s_C(f)(x)\;\;\forall\; s,t\in \mathbb{R}$ for which $\alpha^t_C(T)(x)=\alpha^s_C(T)(x)$, then $F_{[f,T]}(\tau,x)$ is invariant under the flow generated by $C$. The discussion on related implementations of relational dynamics and their interplay can be found in \cite{Hoehn:2019fsy,Hoehn:2020epv}.
	
	A rather straightforward implementation of this idea is achieved in the context of reduced phase space quantization, where classically, the gauge is fixed, and then the quantization is carried out \cite{BARVINSKY1993237,Barvinsky:2013aya}. For the case analysis in this work, the matter degree of freedom is used as the clock in the model. The Hamiltonian constraint generates redundant dynamics with respect to the comoving time. With the appropriate gauge choice, the fluid variable is linearly related to the coordinate time, and one can write the relations between gravitational variables and dust variables, which are gauge invariant.
	
	In the case of fluid models under consideration, the momentum conjugate to the dust variable $T$ is equal to the negative of the gravitational part of the Hamiltonian constraint $P_T=-H_g(a,p_a)$, and that generates the dynamics with respect to the dust variable. The self-adjointness of the operator corresponding to the gravitational Hamiltonian will ensure the unitary evolution in the Schr\"{o}dinger time, i.e., the fluid clock. The expectation values of the various gravitational observables are obtained as a function of the fluid variable, and these relations are invariant under time reparameterization transformations. Therefore, the relational notion of dynamics is implemented by construction in the quantum model.
	
	Still, one has to address the following question: Which phase space functions are to be considered observables in this model? A Dirac observable is a function of phase space variables that has a vanishing Poisson bracket with Hamiltonian constraint and diffeomorphism constraint, which is a highly nonlocal quantity \cite{Torre:1993fq}. The gauge-invariant notion of relational observables introduced above is inappropriate for addressing the questions that we are interested in, as the construction of the Dirac observables corresponding to the Hubble parameter or the Ricci scalar is an untamable task. To this end, we will follow Kucha\v{r}'s philosophy on observables in general relativity.
    
     Kucha\v{r} questions the notion of Dirac observables in generally covariant systems \cite{kuchar_canonical_1993}, where the Hamiltonian itself is a constraint. Contrasting the conventional gauge system and the generally covariant system, Kucha\v{r} argued for a different notion of the physical observable for systems with time reparameterization invariance. The main argument can be summarized as the physically observable quantities need not commute with the Hamiltonian constraint. The phase space functions that commute with all constraints are termed perennials and are not of interest in our analysis. We will follow Kucha\v{r}'s proposal and consider any function of phase space variables as observable while keeping in mind that the relational picture is implemented by construction in the quantum model under consideration.
	\section{FLRW Model with Brown Kucha\v{r} Dust}\label{Sec3}
	In this section, we will discuss the FLRW model coupled to the Brown-Kucha\v{r} dust. We will start with the canonical description of the model and write the observables as a function of phase space variables in the subsection \ref{CFBK}. The classical model has two disjoint solutions, which represent a universe expanding from the Big Bang singularity and a universe collapsing to the Big Crunch singularity. The matter source is the pressureless dust, parameterized via the Brown-Kucha\v{r} formalism \cite{brown_dust_1995,maeda_unitary_2015}. In subsection \ref{QFBK}, we will quantize this model following \cite{kiefer_singularity_2019} and discuss the singularity resolution in this quantum model. 
	
	\subsection{Classical Model}\label{CFBK}
	
	Line element and Ricci scalar for a homogeneous and isotropic FLRW spacetime with constant spatial curvature $k$ are given as
	\begin{align}
		ds^2=&-\mathcal{N}^2(\tau)d\tau^2+a^2(\tau)\left[\frac{dr^2}{1-kr^2}+r^2d\Omega^2\right],\\
		\mathcal{R}=&\frac{6}{\mathcal{N}^2}\left[-\frac{\dot{\mathcal{N}}\dot{a}}{\mathcal{N}a}+\frac{\ddot{a}}{a}+\left(\frac{\dot{a}}{a}\right)^2\right]+\frac{6k}{a^2},\label{RiCl}
	\end{align}
	where $a(\tau)$ is the scale factor and $\mathcal{N}$ is the lapse function. We start with the Einstein-Hilbert action with the Gibbons-Hawking-York (GHY) term,
	\begin{align}
		\mathcal{S}=&\frac{1}{2\kappa}\int d^4x\sqrt{-g}\mathcal{R}-\frac{1}{2\kappa}\int_{\partial\mathcal{M}}d^3x\sqrt{h}\mathcal{K}.
	\end{align}
	where $h$ is the determinant of the induced metric and $\mathcal{K}$ is the extrinsic curvature. The GHY term is included to make the variational problem well-defined, canceling the boundary terms that are coming from the terms involving double derivatives in the EH action. The action for the FLRW model takes the form
	\begin{align}
		\mathcal{S}=&\frac{3V_0}{\kappa}\int d\tau\left[-\frac{a\dot{a}^2}{\mathcal{N}}+k\mathcal{N}a\right].
	\end{align}
	Here we have integrated over the fiducial cell of volume $V_0$ for convenience, even though the integral is over the whole spacetime. After performing Legendre's transformation, we get the Hamiltonian of the system. 
	\begin{align}
		\mathcal{H}=\mathcal{N}\left[-\frac{\kappa}{6V_0}\frac{p_a^2}{2a}-ka\right].\label{HFLRW}
	\end{align}
	The Hamiltonian Constraint is in the square bracket, which generates the time reparametrization invariance transformation. We will use the Brown-Kucha\v{r} dust as the matter source \cite{brown_dust_1995,maeda_unitary_2015}, which is parameterized by a set of non-canonical scalar fields $\rho,\;T,\;W_a$ and $S^a$ with $a=1,2$ and $3$ via the action
	\begin{equation}
		S_{D}=-\frac{1}{2}\int_{\mathcal{M}}d^4x\sqrt{-g}\rho(g^{\mu\nu}U_\mu U_\nu+1).
	\end{equation}
	Here $U_\mu=-\partial_\mu T+W_a\partial_\mu S^a$ is the 4-vector parameterized via the aforementioned scalar fields. The equation of motion of field $\rho$ ensures the timelike nature of the 4-vector $U_\mu$ and the stress-energy tensor corresponding to this matter action is $T_{\mu\nu}=\rho U_\mu U_\nu$. Thus, the 4-vector $U_\mu$ is interpreted as the 4-velocity of the fluid, and $\rho$ is the energy density of the fluid. The ADM decomposition of the matter action yields
	\begin{align}
		S_{D}=\int d\tau d^3x\left(P_T\partial_0T+P_a\partial_0S^a-NH^D-N^iH_i^D\right),
	\end{align}
	where $P_T$ and $P_a$ are momentum conjugate to the fields $T$ and $S^a$, while $H^D$ and $H_i^D$ are the Hamiltonian and diffeomorphism constraints for the Brown-Kucha\v{r} dust, given by
	\begin{align}
		H^D=&\sqrt{P_T^2+h^{ij}H_i^DH_j^D},\\
		H_i^D&=P_T\nabla_iT+P_a\nabla_iS^a.
	\end{align}
	The fields $\rho$ and $W_a$ are non-dynamical and are related to the phase space variables via
	\begin{align}
		W_a=&-P_T^{-1}P_a,\\
		\rho=&\frac{P_T^2}{\sqrt{h(h^{ij}H_i^DH_j^D+P_T^{2})}}.
	\end{align}
	For the case of a symmetry-reduced model such as the FLRW model, we have $S^a\equiv0$, $T=T(\tau)$ which implies $W_a=0$, $\displaystyle\rho=\rho(\tau)=P_T/\sqrt{h}$, $H_i^D=0$ and $H^D=P_T$. In this case, the Hamiltonian constraint for the flat-FLRW model with Brown-Kucha\v{r} dust is given by,
	\begin{align}
		\mathcal{H}=\mathcal{N}\left(-\frac{\kappa}{6V_0}\frac{p_a^2}{2a}+V_0P_T\right)=\mathcal{N}\mathsf{H}.\label{HamBK}
	\end{align}
	In the further analysis, we will choose $\displaystyle\kappa/6V_0=1$ and rescale the dust variable as $\displaystyle V_0P_T\rightarrow P_T$. The momentum conjugate to the dust proper time appears linearly in the Hamiltonian constraint, and the quantization of this model will yield a Schr\"{o}dinger-like equation with the dust variable appearing as Schr\"{o}dinger time. In this case, the momentum conjugate to the dust variable is a perennial (complete observable) and is identified with the energy of the dust, $V_0\rho a^3$, which is indeed a constant of motion. On the other hand, the scale factor and the momentum conjugate to the scale factor are not perennials. The equations of motion for this model are
	\begin{align}
		\begin{aligned}
			\dot{T}&=\{T,\mathcal{H}\}=\mathcal{N},\\
			\dot{P_T}&=\{P_T,\mathcal{H}\}=0,\\
			\dot{a}&=\{a,\mathcal{H}\}=-\frac{\mathcal{N}p_a}{a},\\
			\dot{p_a}&=\{p_a,\mathcal{H}\}=-\frac{\mathcal{N}p^2_a}{2a^2}.
		\end{aligned}
	\end{align} 
	In the comoving gauge with $\mathcal{N}=1$, the dust degree of freedom is linearly related to the comoving time, $T(\tau)=\tau+C$, with $C$ being a constant of integration. The momentum conjugate to the dust degree of freedom $P_T$ is the constant of motion. The first two equations give rise to $\dot{a}^2+2a\ddot{a}=0\implies a(\tau)\propto\tau^{2/3}$, which is the standard solution of the Friedmann's equations with pressureless dust. Since the coordinate time is equal to the dust variable, the gauge-invariant relation between the scale factor and the dust variable is $a(T)\propto T^{2/3}$. Now that we have the phase space structure for this model, we can analyze various geometric quantities of relevance by expressing them as phase space functions and study their behavior in the quantum domain.
	\subsubsection{Hubble Parameter}
	\noindent The Hubble parameter for this model in terms of the phase space variables is given by
	\begin{align}
		\mathbb{H}=\frac{\dot{a}}{\mathcal{N}a}=-a^{-2}p_a.\label{HP_cl}
	\end{align}
	Classically, the Hubble parameter goes as $\mathbb{H}(\tau)=2/3\tau$, diverging at the singularity $\tau=0$.
	\subsubsection{Ricci Scalar}
	\noindent The canonical expression for $\dot{a}$ and $\ddot{a}$ is computed by using the defining equation for the momentum conjugate to the scale factor,
	\begin{align}
		\dot{a}&=-\frac{\;p_a\mathcal{N}}{a},\label{da}\\
		\ddot{a}&=-\left(\frac{\dot{p}_a\mathcal{N}}{a}+\frac{p_a\dot{\mathcal{N}}}{a}-\frac{p_a\mathcal{N}}{a^2}\dot{a}\right)\nonumber\\
		&=-\left(\frac{\{p_a,\mathcal{H}\}\mathcal{N}}{a}+\frac{p_a\dot{\mathcal{N}}}{a}+\frac{p^2_a\mathcal{N}^2}{a^3}\right).\label{dda}
	\end{align}
	The canonical expression of the Ricci scalar in Eq. \eqref{RiCl} in this case turns out to be
	\begin{align}
		\mathcal{R}=&\frac{6}{\mathcal{N}^2}\bigg[-\frac{\dot{\mathcal{N}}}{\mathcal{N}a}\left(-\frac{p_a\mathcal{N}}{a}\right)-\frac{1}{a}\bigg(\frac{\{p_a,\mathcal{H}\}\mathcal{N}}{a}+\frac{p_a\dot{\mathcal{N}}}{a}+\frac{p^2_a\mathcal{N}^2}{a^3}\bigg)\nonumber\\
		&+\frac{1}{a^2}\bigg(-\frac{p_a\mathcal{N}}{a}\bigg)^2\bigg]=-\frac{6\{p_a,\mathcal{H}\}}{\mathcal{N}a^2}.\label{Ricci}
	\end{align}
	In the comoving gauge, we have $\mathcal{N}=1$, and any other gauge choice is related to this gauge choice via
	\begin{align}
		\mathcal{R}=-\frac{6\{p_a,\mathsf{H}\}}{a^2}+6\frac{\partial\mathcal{N}}{\partial a}\frac{\mathsf{H}}{a^2}\approx -\frac{6\{p_a,\mathsf{H}\}}{a^2}.\label{Rgauge}
	\end{align}
	Thus the canonical expressions corresponding to different gauge choices are equal on the constraint surface. The on-shell expression (obtained by computing the Poisson bracket) for the Ricci scalar, therefore, is
	\begin{align}
		\mathcal{R}=\frac{3p_a^2}{a^4}.
	\end{align}
	For the dust dominated universe, the Ricci scalar behaves as $\mathcal{R}=4/3\tau^{2}$. Therefore, the flat-FLRW model with dust as the matter has a curvature singularity at $\tau=0$, when $a(\tau)\propto\tau^{2/3}\rightarrow 0$ and $\mathcal{R}\rightarrow\infty$.
	
	All the phase space functions that we have considered here do not commute with the Hamiltonian constraint and, in conventional terminology, are not Dirac observables. Following Kucha\v{r}'s prescription, we will consider these phase space functions as observables. Moreover, these observables are a product of scale factor and its conjugate momentum, and hence their quantum avatars suffer from the operator ordering ambiguity. Therefore, supplementing only the expectation value of the scale factor is not guaranteed to capture the full quantum behavior of these observables, as we shall see below.
	
	\subsection{Quantum Model}\label{QFBK}
	In the quantum realization of this model, the Brown-Kucha\v{r} dust provides the notion of time, and the degree of freedom associated with dust is the clock variable, whose rate of change is proportional to the flow of comoving time classically. In the quantum analysis, we will use the dust clock variable and comoving time interchangeably. The Wheeler-DeWitt equation for this model then takes the form,
	\begin{align}
		i\frac{\partial \Psi(a,\tau)}{\partial\tau}=\hat{H}&\Psi(a,\tau),\label{WdW}\\
		\hat{H}=\frac{1}{2}a^{-1+p+q}&\frac{d}{da}a^{-p}\frac{d}{da}a^{-q}.\label{Hamlitonian}
	\end{align}
	This model exhibits operator ordering ambiguity, and parameters $p$ and $q$ represent our freedom to choose operator ordering. This Eq. \eqref{WdW} has the form of Schr\"{o}dinger equation and the stationary states for this models are
	\begin{align}
		\begin{aligned}
			\phi_E^1(a)&=a^{\frac{1}{2}(1+p+2q)}J_{\frac{1}{3}|1+p|}\left(\frac{2}{3}\sqrt{2E}a^{\frac{3}{2}}\right)\\
			\phi_{-E}^1(a)&=a^{\frac{1}{2}(1+p+2q)}I_{\frac{1}{3}|1+p|}\left(\frac{2}{3}\sqrt{2E}a^{\frac{3}{2}}\right)\\
		\end{aligned}\\
		\begin{aligned}
			\phi_E^2(a)&=a^{\frac{1}{2}(1+p+2q)}Y_{\frac{1}{3}|1+p|}\left(\frac{2}{3}\sqrt{2E}a^{\frac{3}{2}}\right)\\
			\phi_{-E}^2(a)&=a^{\frac{1}{2}(1+p+2q)}K_{\frac{1}{3}|1+p|}\left(\frac{2}{3}\sqrt{2E}a^{\frac{3}{2}}\right)\\
		\end{aligned}\\
		\begin{aligned}
			\phi_0^1(a)&=a^q\quad,\quad\phi_0^2(a)=a^{1+p+q}.
		\end{aligned}
	\end{align}
	
	Here, $J_n,\;Y_n,\;K_n\;\;\text{and}\;I_n$ are the Bessel functions of the first and second kind. The eigenvalue of Hamiltonian can be interpreted as Misner-Sharp mass which is related on-shell to the energy density of dust\footnote{Misner-Sharpe mass for a spherically symmetric system $ds^2=g_{ab}(z)dz^adz^b+R^2(z)d\Omega^2$ is $M_{MS}=R(z)\left(1-g^{ab}\partial_aR(z)\partial_bR(z)\right)/2G$. For the case of FLRW model, the Misner-Sharp mass $M_{MS}=a\dot{a}^2r^3/2G=(4\pi r^3/3)\rho a^3=V_0a^3\rho$, is related to the mass of dust in the fiducial cell of volume $V_0$, which is a constant of motion. The gravitational Hamiltonian is given by $H=-(3V_0/8\pi G)a\dot{a}^2=-V_0\rho a^3=-M_{MS}$, therefore the Hamiltonian represents the energy associated with dust.}. We choose the Hilbert space $L^2(\mathbb{R}^+,a^{1-p-2q})$ that will make this Hamiltonian Hermitian,
	\begin{align}
		\braket{\psi|\chi}=\int_{0}^{\infty}da\;a^{1-p-2q}\psi^*(a)\chi(a).\label{inner}
	\end{align}
	The self-adjoint extensions and the spectrum of the Hamiltonian operator \eqref{Hamlitonian} are discussed in \cite{kiefer_singularity_2019}. Since $J_\nu$ functions have a closure relation,
	\begin{align}
		\int_{0}^{\infty}dx\;x\;J_\nu(ax)J_\nu(bx)=\frac{\delta(a-b)}{a},\;\text{for} \;\nu>-\frac{1}{2}.
	\end{align}
	The positive energy stationary states $\phi_{E}^1$ form an orthogonal set under the scalar product we have chosen, thus making them suitable for the construction of wave packets. 
	\begin{align}
		\braket{\widetilde{\phi}_{E}^1|\widetilde{\phi}_{E'}^1}=\delta\left(\sqrt{E}-\sqrt{E'}\right), \quad \tilde{\phi}_{E}^1=\frac{2}{\sqrt{3}}E^{\frac{1}{4}}\phi_{E}^1.
	\end{align}
	For the case of these positive energy modes, the behavior of the probability amplitude near singularity $a=0$ is
	\begin{align}
		a^{(1-p-2q)}|\widetilde{\phi}^1_E|^2\sim a^{2+|1+a|}\rightarrow 0.
	\end{align}
	Following DeWitt's criteria, the singularity is considered to be avoided for positive energy states and the wave packets constructed from it. The discussion on singularity resolution for other stationary states can be found in \cite{kiefer_singularity_2019}. From the positive energy modes, a unitarily evolving wave packet is constructed by choosing a normalized Poisson-like distribution 
	\begin{align}
		\psi(a,\tau)&=\int_{0}^{\infty}d\sqrt{E}\tilde{\phi}_E(a)e^{i E\tau}A(\sqrt{E}),\\
		A(\sqrt{E})&=\frac{\sqrt{2}\lambda^{\frac{1}{2}(\kappa+1)}}{\sqrt{\Gamma(\kappa+1
				)}}\sqrt{E}^{\kappa+\frac{1}{2}}e^{-\frac{\lambda}{2}E},\label{EnergyDis}
	\end{align}
	where $\kappa\geq0$ and $\lambda>0$ are real parameters with $\kappa$ being dimensionless and $\lambda$ has dimensions of length or inverse of energy. For the choice of distribution, the expectation value of Hamiltonian is inversely proportional to $\lambda$.
	\begin{align}
		\overline{E}=&\int_0^\infty d\sqrt E\; A(\sqrt{E})^2\;E=\frac{\kappa+1}{\lambda},\label{E}\\
		\Delta E=&\sqrt{\overline{E^2}-\overline{E}^2}=\frac{\sqrt{\kappa+1}}{\lambda}.\label{dE}
	\end{align}
    The distribution with well-defined energy, i.e., $\Delta E\ll\bar{E}$ corresponds to the limit $\kappa\rightarrow\infty$. With this choice of distribution, the wave packet takes the form,
    
	\begin{strip}	
	\begin{equation}
		\begin{aligned}
			\psi(a,\tau)=\sqrt{3}\left(\frac{\sqrt{2}}{3}\right)^{\frac{1}{3}|1+p|+1}&\frac{\Gamma\left(\frac{1}{6}|1+p|+\frac{\kappa}{2}+1\right)}{\sqrt{\Gamma(\kappa+1)}\Gamma\left(\frac{1}{3}|1+p|+1\right)}\frac{\lambda^{\frac{1}{2}(\kappa+1)}}{\left(\frac{\lambda}{2}-i\tau\right)^{\frac{1}{6}|1+p|+\frac{\kappa}{2}+1}}a^{\frac{1}{2}(1+p+|1+p|+2q)}\\
			&\;_1F_1\left(\frac{1}{6}|1+p|+\frac{\kappa}{2}+1;\frac{1}{3}|1+p|+1;-\frac{2a^3}{9\left(\frac{\lambda}{2}-i\tau\right)}\right).\label{gwp}
		\end{aligned}
	\end{equation}	
	\end{strip}
	
	To simplify the expression, one can take $\kappa=|1+p|/3$ \cite{kiefer_singularity_2019} with which the expression for the wave packet reduces to the form, 
	\begin{align}
		\psi(a,\tau)=\frac{\sqrt{3}a^{(1+p+|1+p|+2q)/2}}{\sqrt{\Gamma(\frac{1}{3}|1+p|+1)}}\left(\frac{\frac{\sqrt{2\lambda}}{3}}{\frac{\lambda}{2}-i\tau}\right)^{\frac{|1+p|}{3}+1}e^{-\frac{2a^3}{9\left(\frac{\lambda}{2}-i\tau\right)}}\label{wp}.
	\end{align}
	
	This simplification comes at a cost; the energy distribution, in this case, depends on the operator ordering parameter, which may become a point of reflection in the analysis later. In this case, the large ordering parameter $p$ corresponds to a sharply peaked distribution. This model avoids singularity according to DeWitt's criteria following \cite{kiefer_singularity_2019} and represents a bouncing universe that tunnels from the collapsing branch to the expanding branch.

	\section{Observables in the model with Brown-Kucha\v{r} dust}\label{Sec4}
	As discussed in Section \ref{Sec2}, we will not demand the observables to commute with the Hamiltonian constraint and will incorporate Brown-Kucha\v{r} model for dust as matter, where the dust proper time appears naturally in the quantum picture, thereby sidestepping the issue of frozen dynamics in QG models.
	
	In the quantum domain, we will be using the Hermitian extension of the observables as it ensures the reality of expectation values appearing in the model. Since this work does not involve studying the spectral properties of the operators corresponding to various phase space observables, we will adopt the viewpoint that Hermiticity is a sufficient requirement for an operator to be a quantum observable. In this section, we will write Hermitian extension of the phase space functions that are of particular importance in classical theory and compute the expectation values of these operators in the wave packet constructed in the previous section.
	
	The focus of this analysis is around the operator ordering ambiguity in the various observables, and in the last subsection \ref{OOAH}, we will address the case where the constraint on the ordering parameter $\kappa=|1+p|/3$ is relaxed. There are several physical prescriptions for choosing the ordering of Hamiltonian, e.g., the covariance of superspace leading to Laplace-Beltrami ordering \cite{dewitt_quantum_1967} or ordering used by Vilenkin \cite{PhysRevD.37.888}. For the case of observables, no such determination can be made a priori. The best one can hope is to do a comparative analysis of different ordering schemes, and check where the ambiguity plays a role, and look out for any unphysical inconsistencies.
	
	It is shown in \cite{PhysRevD.104.126027} that the parameter $q$ appears as a free parameter in theory, and we can work with the Hilbert space $L^2(\mathbb{R}^+,a^2da)$ following quantization on the half-line, which leads to the constraint on ordering parameters $p+2q+1=0$. The momentum operator Hermitian with this choice is $\hat{p}_a=-ia^{-1}\partial_aa$ and more discussion on the Hermiticity and self-adjointness of momentum operator on half-line can be found in \cite{PhysRevD.104.126027} and references therein. With this choice, the Hamiltonian operator and the wave packet \eqref{wp} take the form
	\begin{align}
		\hat{H}&=-\frac{1}{2}\hat{a}^{-q-1}\hat{p}_a\hat{a}^{2q+1}\hat{p}_a\hat{a}^{-q-1},\label{HO2}\\
		\psi(a,\tau)&=\frac{\sqrt{3}a^{|q|}}{\sqrt{\Gamma(\frac{2|q|+1}{3})}}\left(\frac{\frac{\sqrt{2\lambda}}{3}}{\frac{\lambda}{2}-i\tau}\right)^{\frac{2|q|}{3}+1}e^{-\frac{2a^3}{9\left(\frac{\lambda}{2}-i\tau\right)}}\label{mwp}.
	\end{align}
	Now, the idea is to write the Hermitian extension of the phase space functions corresponding to the geometric quantities that characterize a classical FLRW universe, such as the Hubble parameter, the Ricci scalar and higher curvature invariants derived in Subsection \ref{CFBK} and analyze their quantum behavior for the semiclassical\footnote{The wave packet is being referred to a semiclassical state as it is peaked on the classical trajectory.} wave packet in the Eq. \eqref{mwp}. For instance, the expectation value of the scale factor for the wave packet in \eqref{mwp} is,
	\begin{align}
		\bar{a}(\tau)&=\int_{0}^{\infty}da\;a^2\psi^*a\psi\nonumber\\
		&=\left(\frac{9(\lambda^2+4\tau^2)}{8\lambda}\right)^\frac{1}{3}\frac{\Gamma\left(\frac{2|q|}{3}+\frac{4}{3}\right)}{\Gamma\left(\frac{2|q|}{3}+1\right)}\label{sf}
	\end{align}
	From Eq. \eqref{sf}, we see that for large $|\tau|$, i.e., $\tau^2\gg\lambda^2$, the scale factor follows the classical trajectory $a(\tau)\propto\tau^{2/3}$ for the dust dominated universe, but for small $|\tau|$ the behavior differs. This model represents a bouncing universe where the scale factor has a global minimum at the classical singularity $\tau=0$. In this work, we will compute the expectation value of the various observables and compare them with the ``semiclassical" expression computed directly from the expectation value of the scale factor. This will be relevant for effective geometry, where the quantum corrections to observationally relevant objects, e.g., the power spectrum computed by substituting the quantum expectations of scale factor in the evolution equations of perturbations, e.g., the Mukhanov-Sasaki equation \cite{Mukhanov:1990me,10.1143/PTPS.78.1}.
	
	The interesting thing to note here is that the $q$ dependence of the expectation value of the scale factor is of the form $\bar{a}(\tau,\lambda,q)=f(\tau,\lambda)g(q)$. Since all observables in the dust dominated case depend on the scale factor via the terms of the form $\dot{a}/a$ or $\ddot{a}/a$, therefore, the semi-classical expressions corresponding to these observables will be independent of the parameter $q$. On a side note, as the energy distribution itself depends on the ordering parameter, the $q$ dependence of any observable in this choice is a combined effect of its dependence on the shape of energy distribution and the ordering of the Hamiltonian.

	\subsection{Hubble Parameter}
\begin{figure}
		\centering
		\includegraphics[width=\columnwidth]{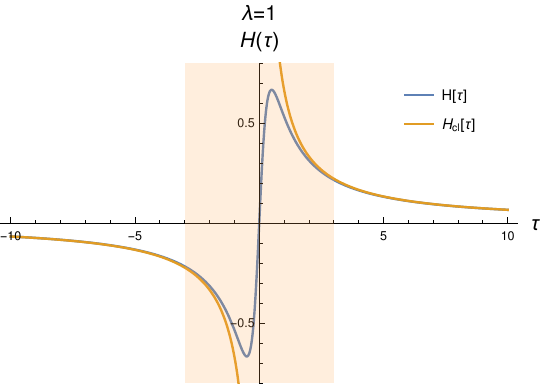}
		\caption{The expectation value of the Hubble Parameter is represented by the blue curve, the orange curve is the classical Hubble parameter, and the shaded region represents the regime where the quantum effects dominate.}
		\label{HP}
	\end{figure}

	The semiclassical expression of the Hubble parameter in Eq. \eqref{HP_cl}, computed from the expectation value of the scale factor \eqref{sf} is given by
	\begin{align}
		\mathbb{H}(\bar{a})=\frac{\dot{\bar{a}}}{\bar{a}}=\frac{8 \tau }{3 \left(\lambda ^2+4 \tau ^2\right)}.\label{schp}
	\end{align}
	In order to compare the quantum expectation w.r.t. the semiclassical expression, we will write symmetric operator orderings for the Hubble parameter. As the phase space expression of the Hubble parameter is a product of the scale factor and its conjugate momentum, its quantum counterpart exhibits the operator ordering ambiguity. Here, we will introduce two ordering schemes that we will follow throughout this work, first the trivial symmetric ordering as ordering 1 and a Weyl-like symmetric ordering as ordering 2.  
	\begin{align}
		F.O.1\rightarrow\hat{\mathbb{H}}_1&=-a^{-1}\hat{p}_aa^{-1},\label{HO1}\\
		F.O.2\rightarrow\hat{\mathbb{H}}_2&=-\frac{1}{2}\left(a^{n-2}\hat{p}_aa^{-n}+a^{-n}\hat{p}_aa^{n-2}\right)\label{HPO}.
	\end{align}
	It is a well-known result that for functions linear in either position or momentum, i.e., of the form $xp^n$ or $x^np$, the different ordering prescriptions give rise to the same differential operator \cite{Kerner_1970,e20110869}. This can be shown explicitly in this case,
	\begin{align}
		\hat{\mathbb{H}}_1\psi&=\hat{\mathbb{H}}_2\psi=i\; a^{-2}\frac{\partial\psi}{\partial a}=\hat{\mathbb{H}}\psi.\label{HBK}
	\end{align}
	These operators are Hermitian, provided the boundary term $\psi^*\chi\big|^\infty_0$ vanishes, which is the case for the set of wave packets \eqref{mwp}, provided $q\neq 0$. The expectation value of the Hubble parameter for the wave packet \eqref{mwp} is,
	\begin{align}
		\overline{\mathbb{H}}(\tau)=&\braket{\psi|\hat{\mathbb{H}}|\psi}=i\int_{0}^{\infty}\psi^*(a,\tau)\frac{\partial\psi(a,\tau)}{\partial a}\;da\nonumber\\
		=&\frac{8\tau}{3(\lambda^2+4\tau^2)}.
	\end{align}
	Interestingly, the expectation value of the Hubble parameter matches the semiclassical expression in \eqref{schp}. Therefore, in this case, the effective geometry approach is well-justified as the semiclassical expression completely captures the quantum gravity effects. Another thing to note is that the expectation value is independent of the parameter $q$, which is the ordering parameter of the Hamiltonian and the parameter that describes the shape of the energy distribution. In the large $|\tau|$ limit, i.e. $\tau^2\gg\lambda^2$, we recover the classical expression of the Hubble parameter
	\begin{align}
		\overline{\mathbb{H}}(\tau)\big|_{\tau^2\gg\lambda^2}=\frac{2}{3\tau}.
	\end{align}
	The expectation value of the Hubble parameter is plotted in Fig. \ref{HP}. 
	The Hubble parameter $\overline{\mathbb{H}}(\tau)$ has a global maximum at $\tau=\lambda/2$ and a global minimum at $\tau=-\lambda/2$. At the point of classical singularity, the Hubble parameter vanishes, and the quantum effects regularize the divergent classical Hubble parameter, thereby representing the bouncing universe. Early on in the collapsing branch, the Hubble parameter decreases and follows the classical behavior. As the system approaches $\tau=-\lambda/2$, the quantum effects kick in, and it deviates from the classical trajectory with the Hubble parameter acquiring a minimum at $\tau=-\lambda/2$. Thereon, the Hubble parameter starts increasing, vanishes at $\tau=0$, and just before $\tau=\lambda/2$, it again turns around and acquires a maximum at $\tau=\lambda/2$ then starts decreasing and follows the classical behavior for the late time in the expanding branch.
 
	The differential operators corresponding to the square of the Hubble parameter operator are again the same for both ordering choices in \eqref{HPO}.
	\begin{align}
		\hat{\mathbb{H}}^2_1\psi=-a^{-2}\partial_a(a^{-2}\partial_a\psi)=\hat{\mathbb{H}}^2_2\psi.
	\end{align}
	However, this analysis can be generalized by writing the symmetric orderings corresponding to the phase space function that represents the square of the Hubble parameter $p_a^2a^{-4}$,
	\begin{align}
		\widehat{\mathbb{H}^2_1}&=a^{-j}\hat{p}_aa^{2j-4}\hat{p}_aa^{-j}\label{Hsq1},\\
		\widehat{\mathbb{H}^2_2}&=\frac{1}{2} \left(a^{-j}\hat{p}_aa^{-k}\hat{p}_aa^{j+k-4}+ a^{j+k-4}\hat{p}_aa^{-k}\hat{p}_aa^{-j}\right)\label{Hsq2}.
	\end{align}
	Here parameters $j$ and $k$ encapsulate the operator ordering ambiguity in the square of the Hubble parameter. The choice $j=1$ in the case of the first ordering gives the square of the Hubble parameter operator in \eqref{HPO}. Here, the Hermiticity of these operators requires the boundary term to vanish
	\begin{align}
		\bigg[a^{-2}\bigg(\psi^*\frac{\partial\chi}{\partial a}-\frac{\partial\psi^*}{\partial a}\chi\bigg)\bigg]^\infty_0\longrightarrow 0,
	\end{align}
	which holds in the case when $|q|>3/2$ for wave packet under consideration in \eqref{mwp}. The expectation value of the square of the Hubble parameter for such states is
	\begin{align}
		\overline{\mathbb{H}_1^2}&=\overline{\mathbb{H}}^2\left(1+\frac{3|q|+2 (j-4) (j-1)}{36|q| (2|q|-3) }\frac{\lambda^2}{\tau^2}\right),\\
		\overline{\mathbb{H}^2_2}&=\overline{\mathbb{H}}^2\left(1+\frac{3|q|-2 j (j+k-4)+5 k-12}{36|q| (2|q|-3) }\frac{\lambda^2}{\tau^2}\right).
	\end{align}
	For the second ordering, we see the square of the Hubble parameter acquires negative values\footnote{On the face of it, this result is troubling, the expectation value of the square of a Hermitian operator $\hat{O}$ is always positive as $\braket{\psi|\hat{O}^2|\psi}=\sum_{n}\braket{\psi|\hat{O}|n}\braket{n|\hat{O}|\psi}=\sum_{n}|\braket{\psi|\hat{O}|n}|^2>0$. But notably, a Weyl-like ordered operator is not the square of a Hermitian operator, and therefore the negative expectation value do not raise any logical fallacy in the quantum model.} for small $|q|$ which we will later see is the theme for this class of orderings. 
	\begin{figure}[H]
		\centering
		\includegraphics[width=\columnwidth]{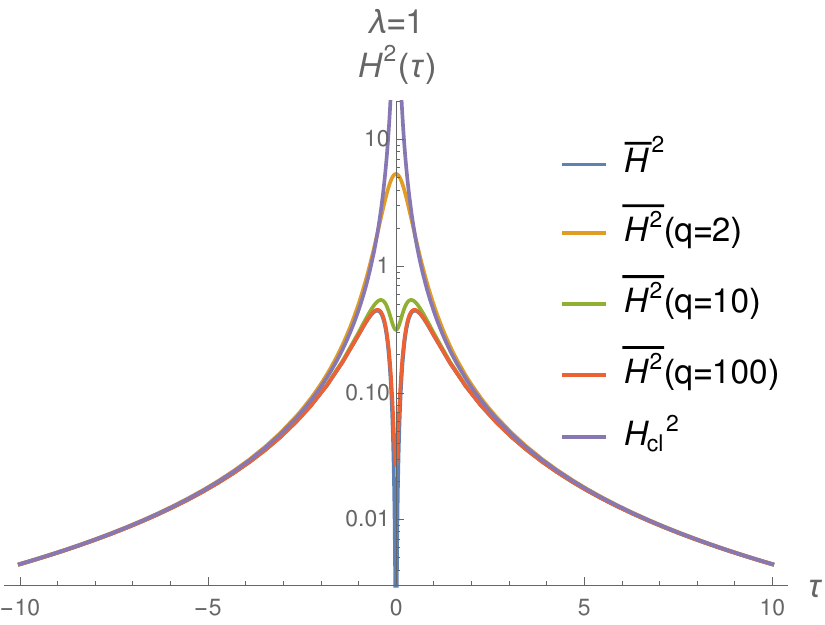}
		\caption{Expectation value of the square of the Hubble parameter for the case of $j=1$ is plotted along with the square of the expectation value of the Hubble parameter.}
		\label{HPsq}
	\end{figure}
	
	In the large $q$ limit, the expectation value of the square of the Hubble parameter closely follows the square of the expectation value of the Hubble parameter (which is the semiclassical expression for the square of the Hubble parameter), as seen in Fig. \ref{HPsq}. For small $q$, the function has a global maximum at the classical singularity, whereas, for large $q$, the function has a local minimum at the classical singularity and global maxima at $\tau\approx\pm\lambda/2$. The key finding of this subsection is that the expectation value of the Hubble parameter matches its semiclassical expression. The analysis of the square of the Hubble parameter hints that in the large $q$ regime, which implies a sharply peaked energy distribution, the ordering of the square of the Hubble parameter is irrelevant, and the expectation value of the square of the Hubble parameter correlates well with its semiclassical counterpart.
	
	\subsection{Ricci Scalar}\label{RicciBK}
	In the study of Friedmann universes, the Ricci scalar is one of the most prominent geometric quantity that appears in the dynamical equations for the non-minimal coupling case \cite{Birrell:1982ix}. First, we compute the Ricci scalar from the expectation value of scale factor given in \eqref{sf}, i.e., the semiclassical expression of the Ricci scalar
	\begin{align}
		\mathcal{R}(\bar{a})=6\left[\frac{\ddot{\bar{a}}}{\bar{a}}+\left(\frac{\dot{\bar{a}}}{\bar{a}}\right)^2\right]=\frac{16 \left(3 \lambda ^2+4 \tau ^2\right)}{3 \left(\lambda ^2+4 \tau ^2\right)^2}.\label{R1SC}
	\end{align}
	This semiclassical expression for Ricci scalar represents the regularized function with a maximum at the origin and follows the classical behavior $\mathcal{R}(\tau)\rightarrow4/3\tau^2$ in the large $\tau$ regime. Near bounce, the quantum effects in this approach are accounted for via the parameter $\lambda$, which is inversely proportional to the mean energy \eqref{E}.
		\begin{figure*}
		\centering
		\includegraphics[width=\textwidth]{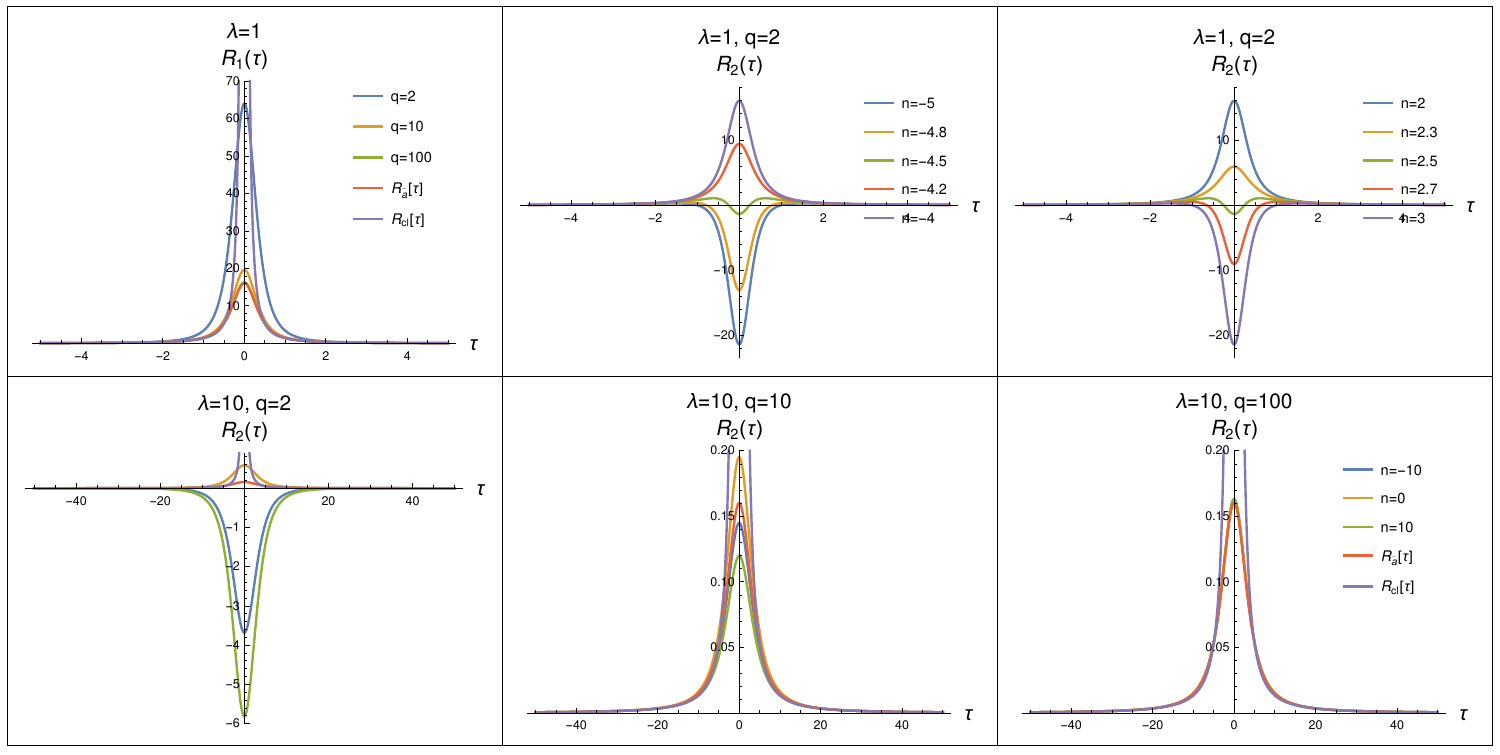}
		\caption{ Expectation value of Ricci scalar for both orderings \eqref{RBJO} and \eqref{RWO}.  The first snap in the first row has the expectation value of the Ricci scalar with the first ordering choice for different values of the parameter $q$. The next two snaps contain the expectation value of the Ricci scalar for the second choice of ordering for a fixed value of $q$ and different values of $n$. The second row contains the expectation value of the Ricci scalar ordered according to the second scheme for different values of $q$.}\label{RiP}
	\end{figure*}
	
	We are interested in writing the Hermitian extension of the Ricci scalar in the Hilbert space under consideration. Since the Ricci scalar is a product of the phase space variables and exhibits operator ordering ambiguity, we will write general operator orderings that will make the operator that corresponds to the phase space function given in Eq. \eqref{Ricci} Hermitian with the given measure following the symmetrization schemes as in Eqs. \eqref{HO1} and \eqref{HPO}
	\begin{align}
		\widehat{\mathcal{R}}_{1}=&6i\;\hat{a}^{-1}[\hat{p}_a,\hat{\mathsf{H}}]\hat{a}^{-1},\label{RBJO}\\
		\widehat{\mathcal{R}}_{2}=&3i\;\left(\hat{a}^{-n-2}[\hat{p}_a,\hat{\mathsf{H}}]\hat{a}^{n}+\hat{a}^{n}[\hat{p}_a,\hat{\mathsf{H}}]\hat{a}^{-n-2}\right)\label{RWO}.
	\end{align}
	Here the parameter $n$ encapsulates the freedom we have to choose the operator ordering. The commutator between the momentum operator and Hamiltonian operator \eqref{HO2} takes the form,	
	\begin{align}
		[\hat{p}_a,\hat{\mathsf{H}}]=&-\frac{1}{2}[\hat{p}_a,\hat{a}^{-q-1}\hat{p}_a\hat{a}^{2q+1}\hat{p}_a\hat{a}^{-q-1}]\nonumber\\
		=&\frac{i}{2}\bigg(-(q+1)\big(\hat{a}^{-q-2}\hat{p}_a\hat{a}^{2q+1}\hat{p}_a\hat{a}^{-q-1}+\hat{a}^{-q-1}\times\nonumber\\
            &\quad\hat{a}^{2q+1}\hat{p}_a\hat{a}^{-q-2}\big)+(2q+1)\hat{a}^{-q-1}\hat{p}_a\hat{a}^{2q}\hat{p}_a\hat{a}^{-q-1}\bigg).\label{Comm}
	\end{align}
	The differential operators corresponding to the Ricci scalar operator for the two orderings come out to be, 
	\begin{align}
		\hat{\mathcal{R}}_{1}=&3a^{-6}\left(3(q^2-1)+2a\partial_a-a^2\partial^2_a\right),\label{RBJDO}\\
		\hat{\mathcal{R}}_2=&3a^{-6}\left((3q^2-n(n+2)-4)+2a\partial_a-a^2\partial^2_a\right)\label{RWDO}.
	\end{align}
	These operators are Hermitian provided the states satisfy the boundary condition
	\begin{align}
		-3\bigg[a^{-2}\bigg(\psi^*\frac{\partial\chi}{\partial a}-\frac{\partial\psi^*}{\partial a}\chi\bigg)\bigg]^\infty_0\longrightarrow 0,
	\end{align}
	which is the case for the wave packet in consideration, provided $|q|>3/2$. Thus, the expectation value of the Ricci scalar operator with first ordering in the wave packet \eqref{mwp} is
	\begin{align}
		\overline{\mathcal{R}}_{1}(\tau)=\frac{16 \left(3 \lambda ^2 \left(|q|(1+2 |q|)-2\right)+4 |q| (2 |q|-3) \tau ^2\right)}{3 |q| (2 |q|-3) \left(\lambda ^2+4 \tau ^2\right)^2}.\label{RBJE}
	\end{align}
	For the second ordering choice of Ricci scalar, the expectation value is
	\begin{align}
		\overline{\mathcal{R}}_2(\tau)=&\frac{16}{3 |q| (2 |q|-3) \left(\lambda ^2+4 \tau ^2\right)^2} \big(\lambda ^2 (-2 n (n+2)\nonumber\\
		&+3 |q| (2 |q|+1)-8)+4 (2 |q|-3) |q| \tau ^2\big).\label{RWE}
	\end{align}
	We see the expectation value for both cases is a well-behaved regular function in the domain of parameters that ensures the Hermiticity. The Hermiticity of the operator and regularity of the expectation value does not have any direct correlation, and the Hermiticity constraint appears as {\it deus ex machina} that saves the model from possible divergences. In Appendices \ref{Hermiticity} and \ref{Regularity}, we have derived the conditions for the hermiticity of the Ricci scalar operator and the regularity of its expectation values, among others, and shown that their domain of applicability matches. 
	
	In this case as well, early in the collapsing regime or late in the expanding regime, i.e., $\tau^2\gg\lambda^2$, we recover the classical expression for the Ricci scalar irrespective of the operator ordering chosen,
	\begin{align}
		\overline{\mathcal{R}}_{1,2}(\tau)\big|_{\tau^2\gg\lambda^2}=\frac{4}{3\tau^2}.
	\end{align}
	Therefore, this quantum gravity analysis predicts a regularized Ricci scalar, which follows the classical behavior far away from the region of the classical singularity, where the quantum gravity effects are expected to be prominent. The expressions for various ordering merge to those of semiclassical one for $q\rightarrow\infty$, a sharply peaked energy distribution.
	\begin{align}
		\overline{\mathcal{R}}_{1,2}(\tau)\bigg|_{q\rightarrow\infty}=\frac{16 \left(3 \lambda ^2+4 \tau ^2\right)}{3 \left(\lambda ^2+4 \tau ^2\right)^2}=\mathcal{R}(\bar{a}).
	\end{align}
	Thus, we see that the semiclassical expression is a limiting case, and the quantum expectation, in general, is different for finite $q$. At the location of the classical singularity, the expectation value is always positive for the first ordering in the allowed $ q$ range. The operator with Weyl-like ordering can have a negative expectation value (which is classically forbidden) that depends on the value of the parameters $n$ and $q$. Moreover, the expectation value always has a maximum at classical singularity for first ordering while for Weyl-like ordering, the expectation value can have a minimum as well as maximum depending again on the parameters $n$ and $q$.
	
	We have plotted the expectation value of the Ricci scalar for the two orderings in Fig. \ref{RiP}. In the case of the operator ordered with the first scheme, the expectation value has a global maximum at the classical singularity for allowed parameter values, and it matches the semiclassical expression for large $q$. However, the case of the operator with the second ordering choice is more interesting. The function can have a maximum or a minimum at the classical singularity and can also attain negative values as well. The trend observed is, for fixed $q$ and $|n|\gg |q|$, the Ricci scalar has a minimum with negative amplitude at the singularity, and it increases as $|\tau|$ increases. At finite time, the Ricci scalar becomes positive, attains a maximum, and then starts decreasing and matches the classical behavior. There exists a window where $O(n)\approx O(q)$, where the crossover happens, from the profile of Ricci scalar with a global minimum at the origin to the profile with a global maximum at the origin. This is illustrated in the last two plots in the first row of Fig. \ref{RiP}. For the case where $|q|\gg|n|$, the expectation value for all the operator orderings merge to the same profile, i.e., the semiclassical expression. The standard deviation of the Ricci scalar $\delta\mathcal{R}^2=\overline{\mathcal{R}^2}-\overline{\mathcal{R}}^2$ for the two operator orderings is given by
	\begin{figure}[H]
		\centering
		\includegraphics[width=\columnwidth]{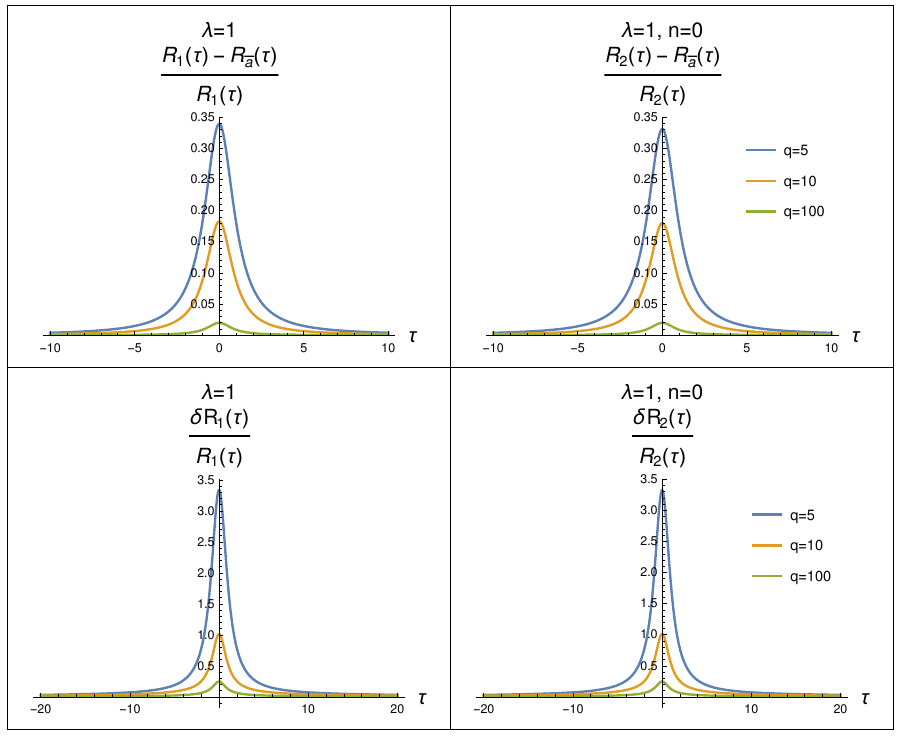}
		\caption{In the first row, we have a fractional change in the expectation value of different operator orderings of the Ricci scalar as compared to the semiclassical expression for the Ricci scalar. In the second row, we have plotted the relative standard deviation in the Ricci scalar.}\label{SDRi}
	\end{figure}
    \begin{strip}
	\begin{align}
		\delta\mathcal{R}_1^2&=\frac{1024 \lambda ^2 \left(4 (|q|-3) |q|^2 (2 |q|-9) (2 |q|-3) \tau ^2+3 \lambda ^2 (|q| (|q| (|q| (4 |q| (3 |q|-8)-27)+60)+12)-27)\right)}{3 (2|q|-3)^2 (|q|-3) |q|^2 (2 |q|-9) \left(\lambda ^2+4 \tau ^2\right)^4},\\
		\delta\mathcal{R}_2^2&=\frac{1024 \lambda ^2 }{3 (3-2 |q|)^2 (|q|-3) |q|^2 (2 |q|-9) \left(\lambda ^2+4 \tau ^2\right)^4}\biggr(\lambda ^2 \biggr(-3 (8 n (n+2)+35) |q|^3+60 (n (n+2)+4) |q|^2\nonumber\\
		&\qquad+4 (n (n+2)+4)^2 |q|-9 (n (n+2)+4)^2+36 |q|^5-96 |q|^4\biggr)+4 (|q|-3) |q|^2 (2 |q|-9) (2 |q|-3) \tau ^2\biggr).
	\end{align}
    \end{strip}
	
	In the first row of Fig. \ref{SDRi}, we have plotted the fractional change in the expectation value of the Ricci scalar as compared to the Ricci scalar computed from the expectation value of the scale factor for both ordering choices. As we continue to increase the parameter $q$, the fractional change continues to decrease, as expected. In the second row of Fig. \ref{SDRi}, we have plotted the relative standard deviation in the Ricci scalar as a function of time for both orderings. Here, we also notice that the quantum fluctuations are small for a large $q$ parameter. Moreover, the relative standard deviation in the Ricci scalar overshoots the fractional change in the expectation value of the Ricci scalar at all times for both orderings. This means that for large $q$, we can trust the semiclassical expressions even near the classical singularity where the quantum effects dominate. We see even for small $q$, the fractional change in the expectation of the Ricci scalar as compared to its semiclassical counterpart does not exceed 35\%, and it decreases as $q$ increases. Therefore, the effective geometry approach does not receive significant corrections for the case of the non-minimally coupled scalar field. 
	
	\subsection{Operator Ordering Ambiguity in the Hamiltonian constraint}\label{OOAH}
	For the case of the quantum FLRW model with Brown-Kucha\v{r} dust, the simplification of the functional form of wave packet \eqref{wp} comes at the cost of making the distribution parameter `$\kappa$' a function of operator ordering parameter. This would mean that the operator ordering ambiguity in the Hamiltonian constraint is harder to address in full generality. We will circumvent this issue following the approach in \cite{PhysRevD.104.126027}, where we simplify the expression for the wave packet by fixing the distribution parameters and have different ordering parameters leading to the wave packets	
	\begin{align}
		&\quad\kappa=4,\;\lambda=1,\;p=5,\;\text{and }q=-3\nonumber\\
            &\psi_\text{I}(a,\tau)=\frac{16a^3(27-4a^3-54 i \tau)}{243(1-2 i \tau)^5}e^{-\frac{2a^3}{9\left(\frac{1}{2}- i \tau\right)}},\label{wp1}
        \end{align}
        \begin{align}
		&\quad\kappa=4,\;\lambda=1,\;p=11,\;\text{and }q=-6\nonumber\\
            &\psi_{\text{II}}(a,\tau)=\frac{64a^6}{243(1-2 i \tau)^5}e^{-\frac{2a^3}{9\left(\frac{1}{2}- i \tau\right)}}\label{wp2}.
	\end{align}
	The operator ordering ambiguity in the Hamiltonian can be addressed in a restricted sense by comparing the expectation value of the observables in these wave packets. The expectation value of the scale factor is given by
	\begin{align}
		\overline{a}_\text{I}(\tau)&=\frac{\left(260 \tau ^2+47\right) \Gamma \left(\frac{13}{3}\right)}{240 \sqrt[3]{3} \left(4 \tau ^2+1\right)^{2/3}},\\
		\overline{a}_{\text{II}}(\tau)&=\frac{\sqrt[3]{4 \tau ^2+1} \Gamma \left(\frac{16}{3}\right)}{16 \sqrt[3]{3}}.
	\end{align}
	Expressions for the Hubble parameter and the Ricci scalar computed from the scale factor expectation value are
	\begin{align}
		\mathbb{H}(\overline{a}_\text{I})&=\frac{8 \tau  \left(260 \tau ^2+101\right)}{\left(12 \tau ^2+3\right) \left(260 \tau ^2+47\right)},\\
		\mathbb{H}(\overline{a}_\text{II})&=\frac{8\tau}{3(1+4\tau^2)},\\
		\mathcal{R}(\overline{a}_\text{I})&=\frac{16 \left(5200 \left(52 \tau ^2+47\right) \tau ^4+137452 \tau ^2+14241\right)}{3 \left(1040 \tau ^4+448 \tau ^2+47\right)^2},\\
		\mathcal{R}(\overline{a}_\text{II})&=\frac{16 \left(4 \tau ^2+3\right)}{3 \left(4 \tau ^2+1\right)^2}.
	\end{align}
	The boundary term $\psi^*\chi\big|_0^\infty$ vanishes for the wave packets in Eq. \eqref{wp1} and \eqref{wp2} and the Hubble parameter is Hermitian. The expectation value of the Hubble parameter is given by
	\begin{align}
		\overline{\mathbb{H}}^{}_{\text{I}}(\tau)&=\frac{16(\tau+2\tau^3)}{3(1+4\tau^2)^2},\\
		\overline{\mathbb{H}}^{}_{\text{II}}(\tau)&=\frac{8\tau}{3(1+4\tau^2)}.
	\end{align}
	The locations of the extrema of both expressions do not match; for the first expression, the extrema are located at $\tau\approx\pm 0.375$ whereas for the second case, the extrema are at $\tau=\pm 1/2$. Furthermore, the profile for the second case is completely enveloped by the profile for the first case. The extrema in the first case are closer to the singularity and are greater in magnitude as compared to the second case.
	
	\begin{figure*}
		\centering
		\includegraphics[width=\textwidth]{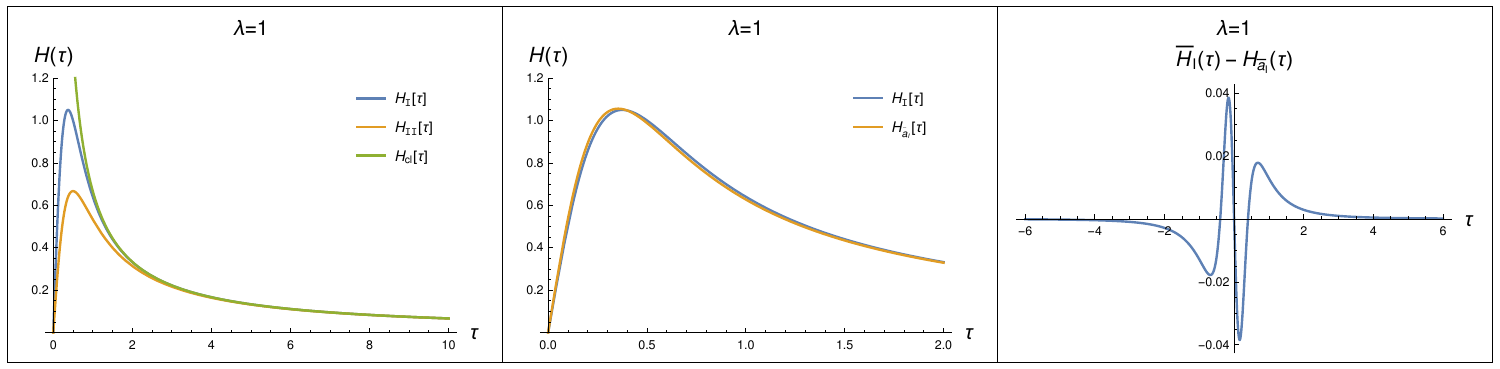}\\\vspace*{-0.2cm}
		\includegraphics[width=\textwidth]{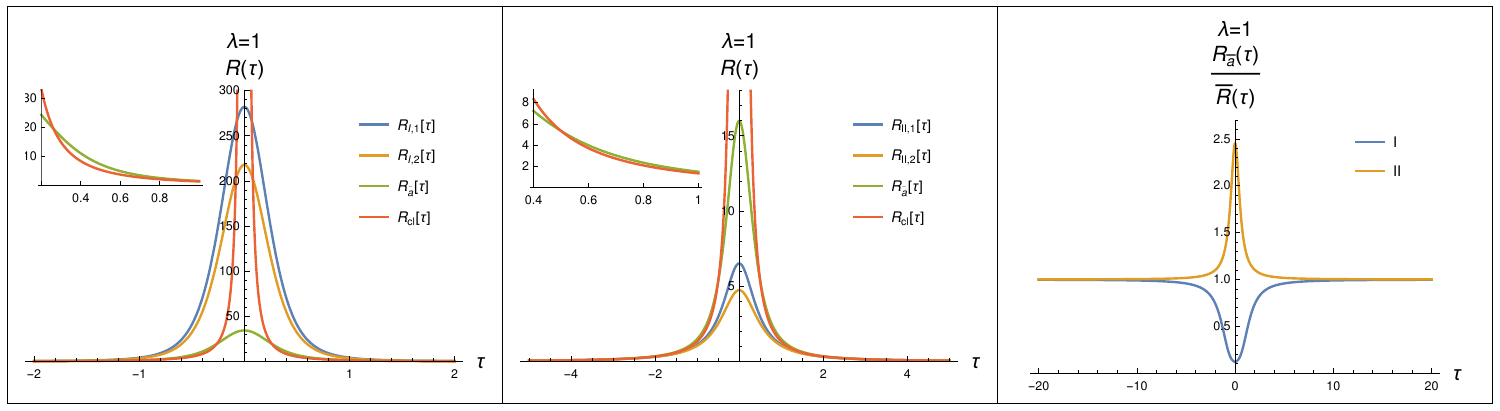}	
		\caption{In the first row, we have the expectation value of the Hubble parameter for both wave packets. For the first wave packet, the expectation value along with the semiclassical expression and the difference between the expectation value of the Hubble parameter and the semiclassical expression, respectively. The expectation values of the different orderings of Ricci scalar with the two wave packets under consideration are plotted in the first two sub-figures of second row. In the third sub-figure, we have plotted the ratio of the Ricci scalar computed from scale factor expectation and the expectation value of the Ricci scalar for both cases.}
		\label{RGWP}
	\end{figure*}
	
	For the wave packet in Eq. \eqref{wp2}, the expectation value of the Hubble parameter matches the semiclassical expression, whereas it is not the case for the wave packet in \eqref{wp1}. This is an indication that the expectation value of the Hubble parameter depends on the operator ordering chosen for the Hamiltonian, and the exact matching with the semiclassical expression may not always be the case, as is seen in Fig. \ref{RGWP}. However, the difference is small, and it has a global minimum and maximum close to the singularity at the $\tau\approx \pm 0.16$, which are sandwiched between the local minimum and maximum at $\tau\approx \mp 0.68$, and it vanishes away from the singularity. 
	
	We also compare the expectation value of the Ricci scalar operator in this case, as the Hermiticity condition is satisfied for both wave packets in Eq. \eqref{wp1} and \eqref{wp2}. In the case of the first ordering scheme of the Ricci scalar in Eq. \eqref{RBJO}, the expectation value takes the form
	\begin{align}
		\overline{\mathcal{R}}^\text{1}_{\text{I}}(\tau)&=\frac{64 \left(36 \tau ^4+65 \tau ^2+119\right)}{27 \left(4 \tau ^2+1\right)^3},\\
		\overline{\mathcal{R}}^\text{1}_{\text{II}}(\tau)&=\frac{16 \left(36 \tau ^2+11\right)}{27 \left(4 \tau ^2+1\right)^2}.
	\end{align}
	Whereas for the case of second ordering in Eq. \eqref{RWO}, the expectation value is
	\begin{align}
		\overline{\mathcal{R}}^\text{2}_{\text{I}}(\tau)=&\frac{16 }{81 \left(4 \tau ^2+1\right)^3}\big(-4 n (n+2) \tau ^2-13 n (n+2)\nonumber\\
		&+432 \tau ^4+776 \tau ^2+1415\big),\\
		\overline{\mathcal{R}}^\text{2}_{\text{II}}(\tau)=&\frac{16 \left(4 \left(27 \tau ^2+8\right)-n (n+2)\right)}{81 \left(4 \tau ^2+1\right)^2}.
	\end{align}
	
	For the Weyl-like ordered Ricci scalar operator, the expectation value can take negative values, and we have plotted for the case where the Ricci scalar is strictly positive ($n=4$ and $n=2$ respectively) in Fig. \ref{RGWP}. For the wave packet in \eqref{wp1}, the expectation value overshoots the classical value and joins the tail from the above, whereas it is the other way around for the wave packet in \eqref{wp2}. The striking difference is in how these expectation values relate to the semiclassical expression. For the first case, the expectation value overshoots the semiclassical expression, and it is the other way around for the second case. Thus, as is seen for the case of the Hubble parameter, the expectation value of the Ricci scalar follows pretty much the same trend but its relation to semiclassical expression changes drastically when we change the ordering of the Hamiltonian operator. Therefore, in conclusion, one has to be careful while using the effective geometry approach, as different ordering schemes of Hamiltonian may lead to inconsistency in the semiclassical analysis.
	
	For the case of the FLRW model with Brown-Kucha\v{r} dust clock, the main findings can be summarized as follows. The model shows robust singularity resolution, where the expectation of the observables that mark the singularity in a classical model has regular expressions with appropriate behavior away from the singularity. The operator ordering ambiguity is relevant only near the classical singularity and has no signature away from the singularity, reaffirming the behavior reported in a previous work \cite{PhysRevD.104.126027}. Moreover, in the limit of a sharply peaked distribution, the operator ordering of observables is not relevant. The applicability of the effective geometry approach is addressed in this model. It is found that in a certain class of orderings of the Hamiltonian, the use of the semiclassical expression is well justified as it matches the expectation of the Hubble parameter, although a different choice of ordering leads to the case where these expressions do not match. For observables other than the Hubble parameter, these expressions do not match in general, but it is observed that the semiclassical expression is the limiting case of the expectation of these observables when we take the limit $q\rightarrow\infty$.  Therefore, the semiclassical expressions can be trusted for the sharply peaked distribution. On the other hand, for small $q$, there are appreciable departures from the effective geometry approach, particularly for small $\tau$. Although there are no significant departures from semiclassical expression at large $\tau$, it was shown in \cite{PhysRevD.104.126027} that in the LTB collapse model, the post-bounce outgoing modes do contain signatures of the ordering parameter in the infrared regime, even at late time. A similar analysis in this context is worthy of inspection.
	
	\section{FLRW Model with Cosmological constant}\label{Sec5}
	In this section, we will investigate the same questions in the most trivial generalization of the previous case, the cosmological constant driven universe. Here, we will consider the unimodular formulation of gravity, where the action is invariant under the coordinate transformations that leave the volume form $\sqrt{-g}$ invariant. In this case, the cosmological constant is a dynamical variable, and its conjugate momentum is the clock variable \cite{Henneaux:1989zc,Unruh:1988in,Kuchar:1991xd,Smolin:2009ti}. The accepted point of view is that the cosmological constant is not a constant of nature but a constant of motion that fixes the initial data \cite{Gielen:2020abd}. The aim is to check the robustness and model independence of the earlier analysis and see if the results hold true in this setting. We will repeat the same exercise, and we will work with the ordering of the Hamiltonian similar to the previous case. In this section, we will present a classical and quantum analysis of this model.
	
	\subsection{Classical Model}
	
	\noindent 
	The action for a parameterized version of the unimodular gravity is obtained by introducing auxiliary field $T$ \cite{Smolin:2009ti}, 
	\begin{align}
		S=\frac{1}{2\kappa}\int d^4x\left[\sqrt{-g}(R-2\Lambda)+\Lambda\partial_\mu T^\mu\right].
	\end{align}
	The Hamiltonian constraint for the flat-FLRW model with cosmological constant is
	\begin{align}
		\mathcal{H}=\mathcal{N}\left[-\frac{p_a^2}{2a}+a^3\Lambda\right],
	\end{align}
	where $\Lambda$ is not a constant anymore, and its conjugate momentum $T$ is the clock variable. With the gauge choice $\mathcal{N}=a^{-3}$, we have $\dot{T}=1$, and the clock variable is linearly related to the coordinate time. The equations of motion with this gauge choice are
	\begin{align}
		\dot{T}=1\qquad&\&\qquad\dot{\Lambda}^{}=0\\
		\frac{\ddot{a}}{a}+2\left(\frac{\dot{a}}{a}\right)^2=0,\qquad&\&\qquad a^{4}\dot{a}^2=2\Lambda.
	\end{align}
	Classically the scale factor in this gauge behaves as
	\begin{align}
		a(\tau)=\left(18\Lambda\tau^2\right)^{\frac{1}{6}}.
	\end{align}
	With Hubble parameter and Ricci scalar in Eq. \eqref{RiCl} given by
	\begin{align}
		\mathbb{H}=&\frac{\dot{a}}{a\mathcal{N}}=a^{2}\dot{a}=\sqrt{2\Lambda},\label{HPSFS}\\
		\mathcal{R}=&6a^{6}\left(\frac{\ddot{a}}{a}+4\left(\frac{\dot{a}}{a}\right)^2\right)=24\Lambda=12\mathbb{H}^2.\label{RSSFS}
	\end{align}
	The classical model exhibits a coordinate singularity at $\tau=0$ where both the Hubble parameter and the Ricci scalar are finite. This coordinate singularity will disappear for an appropriate choice of coordinates, e.g., the standard cosmic time gauge\footnote{The comoving gauge here will be defined according to the observer comoving with the fluid and the $\mathcal{N}=1$ choice is identified with the cosmic time gauge. For the case of dust as fluid, the comoving time and cosmic time match.} where the lapse function is $\mathcal{N}=1$, and the scale factor vanishes at the infinity of cosmic time. The phase space expression for the Hubble parameter in this model is
	\begin{align}
		\mathbb{H}=\frac{\dot{a}}{\mathcal{N}a}=-a^{-2}p_a.\label{HPS}
	\end{align}
	The canonical expression for the Ricci scalar derived in section \ref{Sec2} is lapse-choice independent. In this case, using Eq. \eqref{Ricci} and with gauge $\mathcal{N}=a^{-3}$, we have
	\begin{align}
		\mathcal{R}=-6a\{p_a,\mathcal{H}\}.\label{ROS}
	\end{align}
	Other gauge choices are equal on the constraint surface, as seen in Eq. \eqref{Rgauge}. These observables are a product of the scale factor, and their conjugate momentum and their quantum counterparts will be non-trivial. As is done previously, we will use momentum conjugate to cosmological constant $T$ and coordinate time $\tau$ interchangeably.
	
	\subsection{Quantum Model}
	
	The Wheeler-DeWitt equation for the flat-FLRW model with the perfect fluid is,
	\begin{align}
		\frac{1}{2}a^{-4+p+q}\frac{\partial}{\partial a}a^{-p}\frac{\partial}{\partial a}a^{-q}\Psi=i\frac{\partial\Psi}{\partial\tau}.\label{HSO}
	\end{align}
	For the Hamiltonian operator to be Hermitian, the inner product is chosen as
	\begin{align}
		\braket{\Phi|\Psi}=\int_0^\infty\Phi^*\Psi a^{4-p-2q}da
	\end{align}
	With this choice of the Hilbert space $L^2(\mathbb{R}^+,a^{4-p-2q}da)$, the Hermitian representation of the momentum operator is given by
	\begin{align}
		\hat{p}_a=-ia^{-\frac{(4-p-2q)}{2}}\frac{\partial}{\partial a}a^{\frac{(4-p-2q)}{2}}.
	\end{align}
	The ordering for the Hamiltonian operator corresponding to this representation of the momentum operator is
	\begin{align}
		\hat{H}=-\frac{1}{2}a^{-2+\frac{p}{2}}\hat{p}_aa^{-p}\hat{p}_aa^{-2+\frac{p}{2}}.
	\end{align}
	The solution of the WDW equation is obtained via the separation ansatz, and the eigenfunctions with the positive cosmological constant are,
	\begin{align}
		\begin{aligned}
			\Psi_\Lambda^1(a,\tau)=a^{\frac{1+p+2q}{2}}\;J_{\frac{|1+p|}{6}}\left(\frac{\sqrt{8\Lambda}}{6}a^{3}\right),\\
			\Psi_\Lambda^2(a,\tau)=a^{\frac{1+p+2q}{2}}\;Y_{\frac{|1+p|}{6}}\left(\frac{\sqrt{8\Lambda}}{6}a^{3}\right).
		\end{aligned}
	\end{align}
	The model is singularity-free according to DeWitt's criteria as the probability amplitude associated with the $\psi_E^1$ states vanishes $a^{4-p-2q}|\psi_E^1|^2\rightarrow a^{5+\frac{|1+p|}{2}}\rightarrow 0$. The Hamiltonian operator in Eq. \eqref{HSO} is essentially self-adjoint if $|1+p|>6$, and it admits infinite self-adjoint extensions when $|1+p|\leq 6$, following \cite{kiefer_singularity_2019}. The normalized wave packet constructed from $\psi_\Lambda^1$ states with Poisson-like distribution in Eq. \eqref{EnergyDis} takes the form

 \newpage
 
	\begin{strip} 
	\begin{align}
		\Psi(a,\tau)&=\int_0^\infty A\left(\sqrt{\Lambda}\right)e^{i\Lambda\tau}\Psi^1_\Lambda(a,\tau) d\Lambda,\nonumber\\
		&=\sqrt{\frac{6}{\Gamma \left(\frac{| p+1| }{6}+1\right)}} a^{\frac{1}{2} (| p+1| +p+2 q+1)} \exp \left(-\frac{a^6}{18 \left(\frac{\lambda }{2}-i \tau \right)}\right) \left(\frac{\sqrt{2 \lambda }}{6 \left(\frac{\lambda }{2}-i \tau \right)}\right)^{\frac{| p+1| }{6}+1}.\label{wpS}
	\end{align}
	\end{strip}
	In this case, the distribution parameter is chosen as $\kappa=|1+p|/6$ to simplify the wave packet. In this case, as well, the parameter $q$ will appear as a free parameter in the model. Now the stage is set to investigate the status of observables in this quantum model, as we did for the dust dominated universe. 
	
	\section{Observables in the Cosmological constant driven universe}\label{Sec6}
	The classical dynamics of this model implies that the curvature invariant is finite, even though the scale factor vanishes at a finite coordinate time, indicating that it is just a ``coordinate" singularity. Therefore, it will be interesting to study the quantization of this ``singularity-free" classical model and see how the boundary conditions, that are required for the unitarity modify the dynamics in this quantum model \cite{Gielen:2021igw,Gielen:2022tzi,Gielen:2020abd}. The discussion on a general ordering scheme of the Hamiltonian with the arbitrary equation of state parameter will be presented elsewhere. The expectation value of the scale factor with wave packet in Eq. \eqref{wpS} is
	\begin{align}
		\bar{a}(\tau)= \left(\frac{9(4 \tau ^2+\lambda^2)}{2\lambda }\right)^{\frac{1}{6}} \frac{\Gamma \left(\frac{| p+1| +7}{6}\right)}{\sqrt[6]{2} \Gamma \left(\frac{| p+1| }{6}+1\right)}.\label{sfS}
	\end{align}
	Instead of vanishing, the scale factor acquires a finite minimum at the point of coordinate singularity, thereby representing a bouncing cosmological model. Again, as before, the late time in the expanding (early time in collapsing) regime, i.e., when $\tau^2\gg\lambda^2$, the scale factor behaves as
	\begin{align}
		\bar{a}(\tau)\big|_{\tau^2\gg\lambda^2}\rightarrow\frac{\Gamma \left(\frac{| p+1| +7}{6}\right)}{\Gamma \left(\frac{| p+1| }{6}+1\right)}\left(\frac{18\tau^2}{\lambda}\right)^{\frac{1}{6}}\label{sfSex}
	\end{align}
	following the classical trajectory. Classically, the Hubble parameter has a step function-like discontinuity and is negative for $\tau<0$ and positive for $\tau>0$, whereas the Ricci scalar is constant throughout. The classically anticipated values of these observables for the universe that is represented by the wave packet in Eq. \eqref{sfS} are discussed in Appendix \ref{AppSchCl}. Again, we will compare the semiclassical expressions of the Hubble parameter and Ricci scalar computed from this expectation value of the scale factor with the expectation value of these observables. 
	
	\subsection{Hubble Parameter}\label{HPSsubs}
	\begin{figure*}
		\centering
		\includegraphics[width=\textwidth]{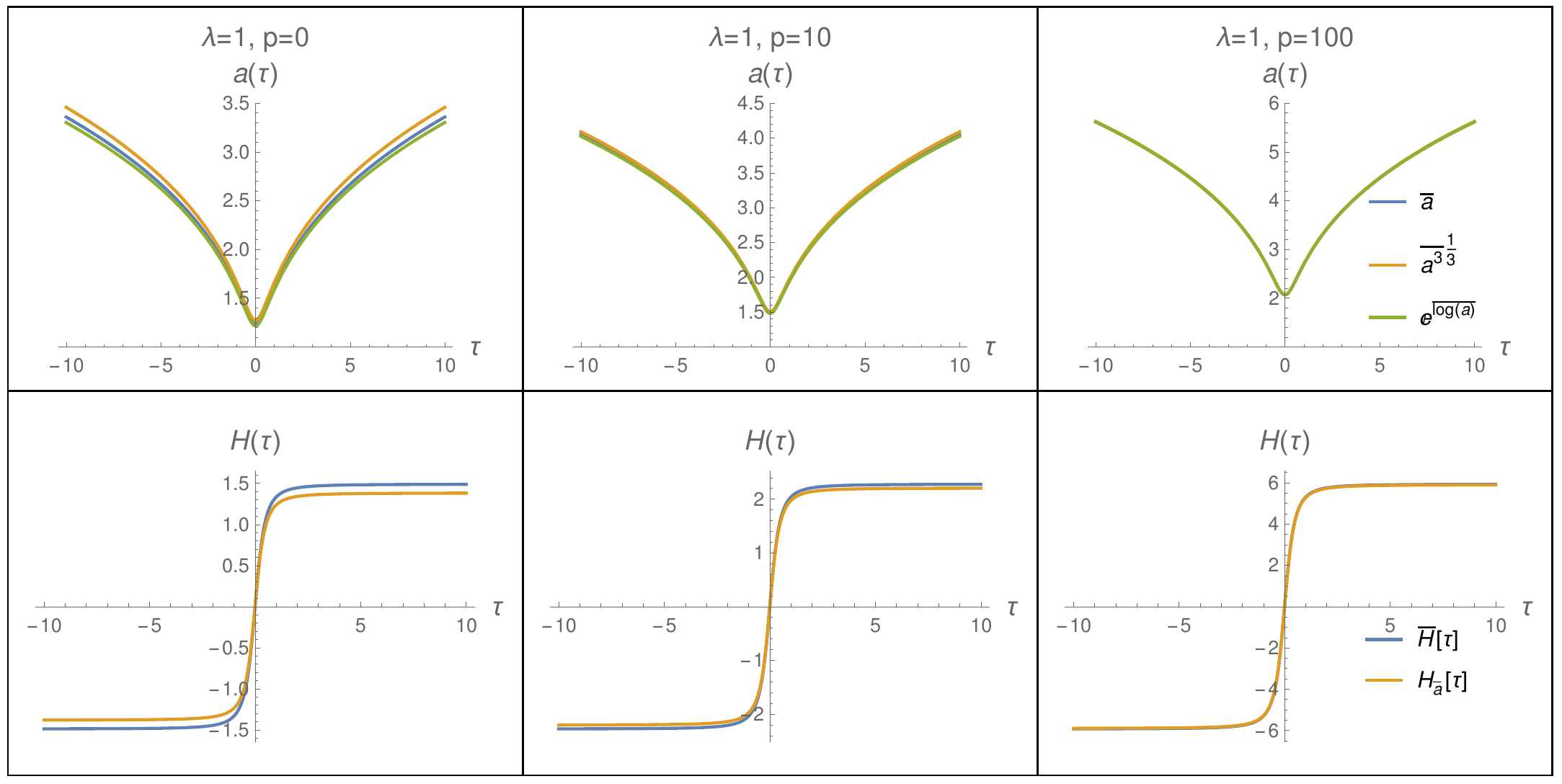}
		\caption{In the first row, we have plotted the expectation of the scale factor and the cube root of the expectation value of the volume variable. In the second row, we plotted the expectation value of the Hubble parameter along with its semiclassical expression. Different notions of the observables match in the limit of a sharply peaked distribution, $|p|\rightarrow\infty$.}\label{HSF}
	\end{figure*}
	\noindent The Hubble parameter in Eq. \eqref{HPSFS} computed from the expectation value of the scale factor \eqref{sfS} turns out to be
	\begin{align}
		\mathbb{H}(\bar{a})=&\dot{\bar{a}}\bar{a}^2=\frac{2 \sqrt{2} \tau}{\sqrt{\lambda ^3+4 \lambda  \tau ^2}}\frac{\Gamma \left(\frac{| p+1| +7}{6}\right)^3}{\Gamma \left(\frac{| p+1| }{6}+1\right)^3}.
	\end{align}
	The semiclassical expression asymptotes to a constant negative value early in the collapsing branch $\tau<-\lambda$ and to a positive value late in the expanding branch $\tau>\lambda$, with a smooth transition from the collapsing branch to the expanding branch, representing a bounce. At the leading order, the classical step function-like behavior of the Hubble parameter is recovered,
	\begin{align}
		\mathbb{H}(\bar{a})\big|_{\tau^2\gg\lambda^2}=\sqrt{\frac{2}{\lambda}}\frac{\Gamma \left(\frac{| p+1| +7}{6}\right)^3}{\Gamma \left(\frac{| p+1| }{6}+1\right)^3}\frac{|\tau|}{\tau}.
	\end{align}
	The symmetric operator orderings of the Hubble parameter in Eq. \eqref{HPS}, following the prescription of ordering used earlier, are 
	\begin{align}
		\hat{\mathbb{H}}_1&=-a^{-1}\hat{p}_aa^{-1},\label{HPS1}\\
		\hat{\mathbb{H}}_2&=-\frac{1}{2}\left(a^{-2+n}\hat{p}_aa^{-n}+a^{-n}\hat{p}_aa^{-2+n}\label{HPS2}\right).
	\end{align}
	Again, both orderings for the Hubble parameter operator lead to the same differential operator,
	\begin{align}
		\hat{\mathbb{H}}_1\psi=\hat{\mathbb{H}}_2\psi=\frac{i}{a^2}\left(\frac{\partial\psi}{\partial a}-\frac{p+2q-2}{2a}\psi\right)=\hat{\mathbb{H}}\psi.\label{HSFeq}
	\end{align} 
	The Hermiticity of this operator requires the vanishing of the boundary term, 
	\begin{align}
		\left[a^{2-p-2q}\psi^*\chi\right]_0^\infty\rightarrow0,
	\end{align}
	which is satisfied for the case of the wave packet in Eq. \eqref{wpS}. The expectation value of the Hubble parameter for the wave packet in Eq. \eqref{wpS} is,
	\begin{align}
		\overline{\mathbb{H}}(\tau)=\frac{2 \sqrt{2} \tau }{\sqrt{\lambda ^3+4 \lambda  \tau ^2}}\frac{\Gamma \left(\frac{1}{6} (| p+1| +9)\right)}{\Gamma \left(\frac{| p+1| }{6}+1\right)}.\label{HPES}
	\end{align}
	At the leading order, the expectation value of the Hubble parameter asymptotes to a constant value given by 
	\begin{align}
		\overline{\mathbb{H}}(\tau)\big|_{\tau^2\gg\lambda^2}=\sqrt{\frac{2}{\lambda}}\frac{\Gamma \left(\frac{1}{6} (| p+1| +9)\right)}{\Gamma \left(\frac{| p+1| }{6}+1\right)}\frac{|\tau|}{\tau}.\label{HPex}
	\end{align}

	We see that the expectation value of the Hubble parameter does not agree with its semiclassical expression in the classical regime ($\tau^2\gg\lambda^2$). The possible origin of this disagreement is discussed in Appendix \ref{AppSchCl}.
	
	The expectation value of the Hubble parameter is plotted in Fig. \ref{HSF} along with its semiclassical expression. Classically, the Hubble parameter has a behavior like that of a step function with a constant negative value in the contracting branch and a constant positive value in the expanding branch, with the transition from collapsing to expanding branch forbidden. Both the semiclassical expression and the quantum expectation follow the same generic trend. The Hubble parameter, in this case, is a continuous generalization of the step function and has a smooth transition from a constant negative value for $\tau\ll-\lambda$ to a constant positive value for $\tau\gg\lambda$ with the Hubble parameter passing through the origin at $\tau=0$. This represents a quantum tunneling of the universe from a collapsing branch to an expanding branch, i.e., a bouncing universe. In the large $p$ limit, the expectation value of the Hubble parameter matches its semiclassical counterpart
	\begin{align}
		\lim_{p\rightarrow\infty}\frac{\overline{\mathbb{H}}(\tau)}{\mathbb{H}(\overline{a})}=1.
	\end{align} 
	The expectation value of the square of the Hubble parameter is given by
	\begin{align}
		\overline{\mathbb{H}^2}(\tau)=\braket{\psi|\hat{\mathbb{H}}^2|\psi}=\frac{2}{\lambda}+\frac{9 \lambda ^2+4 (p+1)^2 \tau ^2}{3 \lambda  | p+1|  \left(\lambda ^2+4 \tau ^2\right)},\label{HPsqE}
	\end{align}
	with the Hermiticity constraint being $|1+p|\neq0.$ We are interested in looking at the large $p$ limit of the standard deviation
	\begin{align}
		\lim_{p\rightarrow\infty}\left(\braket{\psi|\hat{\mathbb{H}}^2|\psi}-\braket{\psi|\hat{\mathbb{H}}|\psi}^2\right)=\frac{2 \left(\lambda ^2+\tau ^2\right)}{\lambda ^3+4 \lambda  \tau ^2},
	\end{align}
	which settles at $1/2\lambda$ in the large $\tau$ limit. Since the expectation value of the Hubble parameter diverges in the large $p$ and large $\tau$ limits, therefore relative standard deviation will vanish in that limit.
	
	\begin{figure*}
		\centering
		\includegraphics[width=\textwidth]{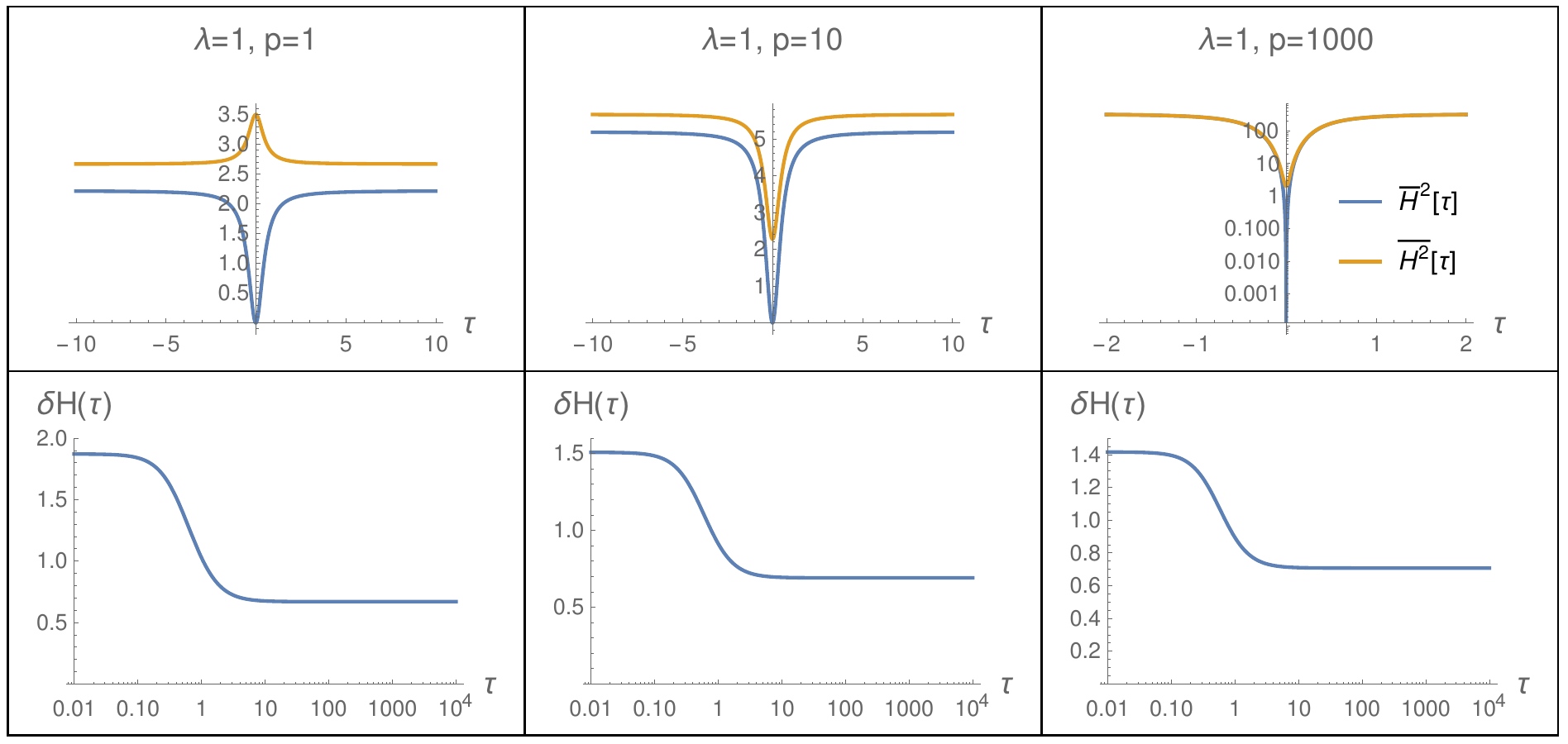}
		\caption{We have plotted the square of the expectation value of the Hubble parameter along with the expectation value of the square of the Hubble parameter. In this case, the two profiles match for large $|p|$ and in the large $|\tau|$ limit.}\label{HSqF}
	\end{figure*}
	
	As is apparent from the plots in Fig. \ref{HSqF}, $\overline{\mathbb{H}^2}$ and $\overline{\mathbb{H}}^2$ match in the large $p$ and large $\tau$ limits for all practical purposes. However, these two differ by the same amount $\sim 1/2\lambda$ for all $p$ and large $\tau$, with the amplitude of the Hubble parameter increasing as $p$ increases. In the second row of Fig. \ref{HSqF}, we have plotted the standard deviation in the Hubble parameter that shows a peculiar feature. It has a maximum at $\tau=0$, but instead of decaying for large $|\tau|$, it settles at a constant value that remains the same for different choices of parameter $p$. Therefore, even for sharply peaked distributions, the standard deviation in the Hubble parameter remains finite at late times.

	\subsection{Ricci Scalar}
	\begin{figure*}
		\centering
		\includegraphics[scale=0.7]{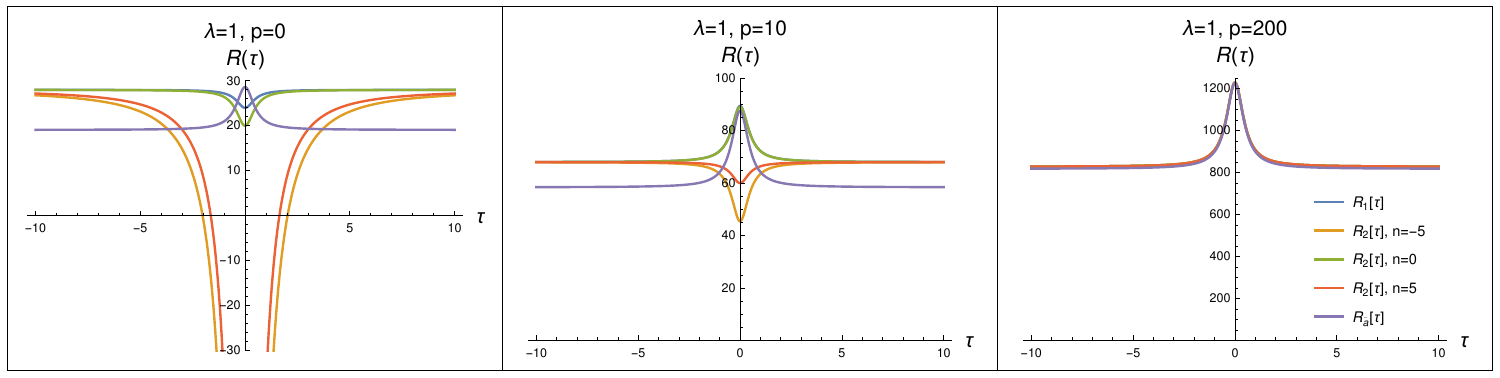}
		\caption{Expectation value of Ricci scalar for different operator ordering along with its semiclassical counterpart.}\label{RSF}
	\end{figure*}
	\noindent The semiclassical expression for the Ricci scalar is obtained as
	\begin{align}
		\mathcal{R}(\bar{a})=\frac{12 \left(3 \lambda ^2+8 \tau ^2\right)}{\left(\lambda ^3+4 \lambda  \tau ^2\right)}\frac{\Gamma \left(\frac{1}{6} (| p+1| +7)\right)^6}{\Gamma \left(\frac{| p+1| }{6}+1\right)^6}.
	\end{align}
	At the leading order, the semiclassical expression of the Ricci scalar settles at the value
	\begin{align}
		\mathcal{R}(\bar{a})\big|_{\tau^2\gg\lambda^2}=\frac{24 \Gamma \left(\frac{1}{6} (| p+1| +7)\right)^6}{\lambda  \Gamma \left(\frac{| p+1| }{6}+1\right)^6},
	\end{align}
	in the classical regime. Moreover, the semiclassical expression follows the classical relation between the Ricci scalar and the Hubble parameter for large $\tau$
	\begin{align}
		\mathcal{R}(\bar{a})=12\mathbb{H}(\bar{a})^2+\frac{36 \lambda}{\left(\lambda ^2+4 \tau ^2\right) }\frac{\Gamma \left(\frac{1}{6} (| p+1| +7)\right)^6}{\Gamma \left(\frac{| p+1| }{6}+1\right)^6}.
	\end{align}
	
	In order to analyze the quantum behavior of the Ricci scalar operator, we will again write the symmetric operator orderings corresponding to the phase space function in Eq. \eqref{ROS} that is Hermitian with the given measure, following the ordering scheme introduced in Eqs. \eqref{HO1} and \eqref{HPO},
	\begin{align}
		\hat{\mathcal{R}}_1&=6i\hat{a}^{\frac{1}{2}}[\hat{p}_a,\hat{\mathcal{H}}]\hat{a}^{\frac{1}{2}},\\
		\hat{\mathcal{R}}_2&=3i\left(\hat{a}^{1-n}[\hat{p}_a,\hat{\mathcal{H}}]\hat{a}^n+\hat{a}^n[\hat{p}_a,\hat{\mathcal{H}}]\hat{a}^{1-n}\right).
	\end{align}
	The differential operators corresponding to these orderings turn out to be
	\begin{align}
		\hat{\mathcal{R}}_1&=\frac{3}{2a^{6}} \big(\left(p^2+p (2-8 q)-2 (2 q+1)^2\right)\nonumber\\
		&\qquad\qquad+8 a \left((p+2 q) \partial_a-a \partial_a^2\right)\big),\label{RSDO}\\
		\hat{\mathcal{R}}_2&=\hat{\mathcal{R}}_1-3 (1-2 n)^2a^{-6}\label{RSWDO}.
	\end{align}
	The Hermiticity analysis of the Ricci scalar operator for both orderings yields the boundary term, 
	\begin{align}
		\bigg[a^{-p-2q}\left(\psi^*\partial_a\chi-\chi\partial_a\psi^*\right)\bigg]_0^\infty
	\end{align}
	which goes to zero for the wave packets under consideration \eqref{wpS}, provided $p\neq-1$. In this case, the expectation value of the Ricci scalar for both ordering schemes with the wave packet in Eq. \eqref{wpS} is given by
	\begin{align}
		\overline{\mathcal{R}}_1(\tau)&=\frac{24}{\lambda }+\frac{6 \lambda ^2 p (p+2)+16 (p+1)^2 \tau ^2}{\lambda  | p+1|  \left(\lambda ^2+4 \tau ^2\right)},\label{RSex1}\\
		\overline{\mathcal{R}}_2(\tau)
		&=\overline{\mathcal{R}}_1(\tau)-\frac{4 \lambda  (1-2 n)^2}{| p+1|  \left(\lambda ^2+4 \tau ^2\right)}.\label{RSex2}
	\end{align}
	In this case as well, the expectation value of the Ricci scalar is regular except for at $p=-1$, excluded by the Hermiticity consideration. The Ricci scalar settles at a constant value for large $\tau$, which is different from the value that the semiclassical expression settles at. The quantum imprints on the Ricci scalar are pronounced near the coordinate singularity, where the universe tunnels from a collapsing branch to an expanding branch, similar to the case of semiclassical expression. In the limit of the sharply peaked trajectory, i.e., large $p$, different expressions asymptotes to the same profile given by
	\begin{align}
            \overline{\mathcal{R}}_{1/2}(\tau)\big|_{p\rightarrow\infty}=\frac{2 p \left(3 \lambda ^2+8 \tau ^2\right)}{\lambda  \left(\lambda ^2+4 \tau ^2\right)}=\mathcal{R}(\bar{a})\big|_{p\rightarrow\infty}.
	\end{align}
	
	In Fig. \ref{RSF}, we have plotted the Ricci scalar expectation value for different ordering choices along with its semiclassical expression. For small $p$, the quantum expectation and semiclassical expression do not agree, even in the ``classical" regime, i.e., when $\tau^2\gg \lambda^2$. Moreover, the nature of the extrema is also different for small $p$. The various profiles merge onto a single profile in the case of large $p$. Therefore, the semiclassical expression can be trusted, and the ordering ambiguity is not relevant for a sharply peaked distribution. 
	
	Next, we will check whether the quantum expectations respect the classical relation between the Ricci scalar and the Hubble parameter, $\mathcal{R}=12\mathbb{H}^2$. The expression for the expectation value of the Hubble parameter is given in Eq. \eqref{HPES}, and this relation is not satisfied even in the ``classical" regime. The expectation value of the square of the Hubble parameter is given in Eq. \eqref{HPsqE} and the quantum expectation values satisfy the classical relation at the leading order, i.e., for $\tau^2\gg\lambda^2$, irrespective of the ordering of the Ricci scalar operator.
	\begin{align}
		\overline{\mathcal{R}}_{1/2}(\tau)\bigg|_{\tau^2\gg\lambda^2}=12\overline{\mathbb{H}^2}(\tau)\bigg|_{\tau^2\gg\lambda^2}+O(\tau^{-2}).
	\end{align}
	The expectation value of the square of the Ricci scalar for both orderings is given by
	\begin{strip}
	\begin{align}
		\overline{\mathcal{R}_1^2}(\tau)&=\frac{4 }{(| p+1| -6) | p+1|  \left(\lambda ^3+4 \lambda  \tau ^2\right)^2}\bigg(48 \lambda ^2 \tau ^2 \left(6 (p (p+2)-46) | p+1| +(p-4) (p+6) (p+1)^2\right)+9 \lambda^4 \\
		&\big(8 (p-4) (p+6) | p+1| +(p-6) p (p+2) (p+8)+48\big)+64 (p-5) (p+7) \tau ^4 \left(12 | p+1| +(p+1)^2\right)\bigg)\\
		\overline{\mathcal{R}_2^2}(\tau)&=\overline{\mathcal{R}_1^2}(\tau)+\frac{16 (1-2 n)^2 \left(-12 | p+1|  \left(\lambda ^2-4 \tau ^2\right)+\lambda ^2 (4 (n-1) n-3 p (p+2)+145)-8 (p+1)^2 \tau ^2\right)}{(| p+1| -6) | p+1|  \left(\lambda ^2+4 \tau ^2\right)^2}
	\end{align}
	At the leading order $\tau^2\gg\lambda^2$, the relative standard deviation in the Ricci scalar settles at
	\begin{align}
		\frac{\delta\mathcal{R}_1}{\overline{\mathcal{R}}_1}&=\sqrt{\frac{6 \left(6 | p+1|  \left((p+1)^2-| p+1|  (| p+1| +6)\right)+(p+1)^4\right)}{(| p+1| -6) \left(6 | p+1| +(p+1)^2\right)^2}}+O(\tau^{-2})= \frac{\delta\mathcal{R}_2}{\overline{\mathcal{R}}_2},
	\end{align}
	\end{strip}
	which vanishes for $|p|\rightarrow\infty$ limit. Similar to what we have observed in the case of dust dominated universe, the quantum fluctuations in the Ricci scalar are small for a sharply peaked distribution, and the semiclassical expressions match in this regime. However, the quantum fluctuations saturate to a finite value instead of decaying at the late time in the expanding branch for all ordering parameter values, as is seen for the Hubble parameter. 
	\begin{figure*}
		\centering
		\includegraphics[width=0.7\textwidth]{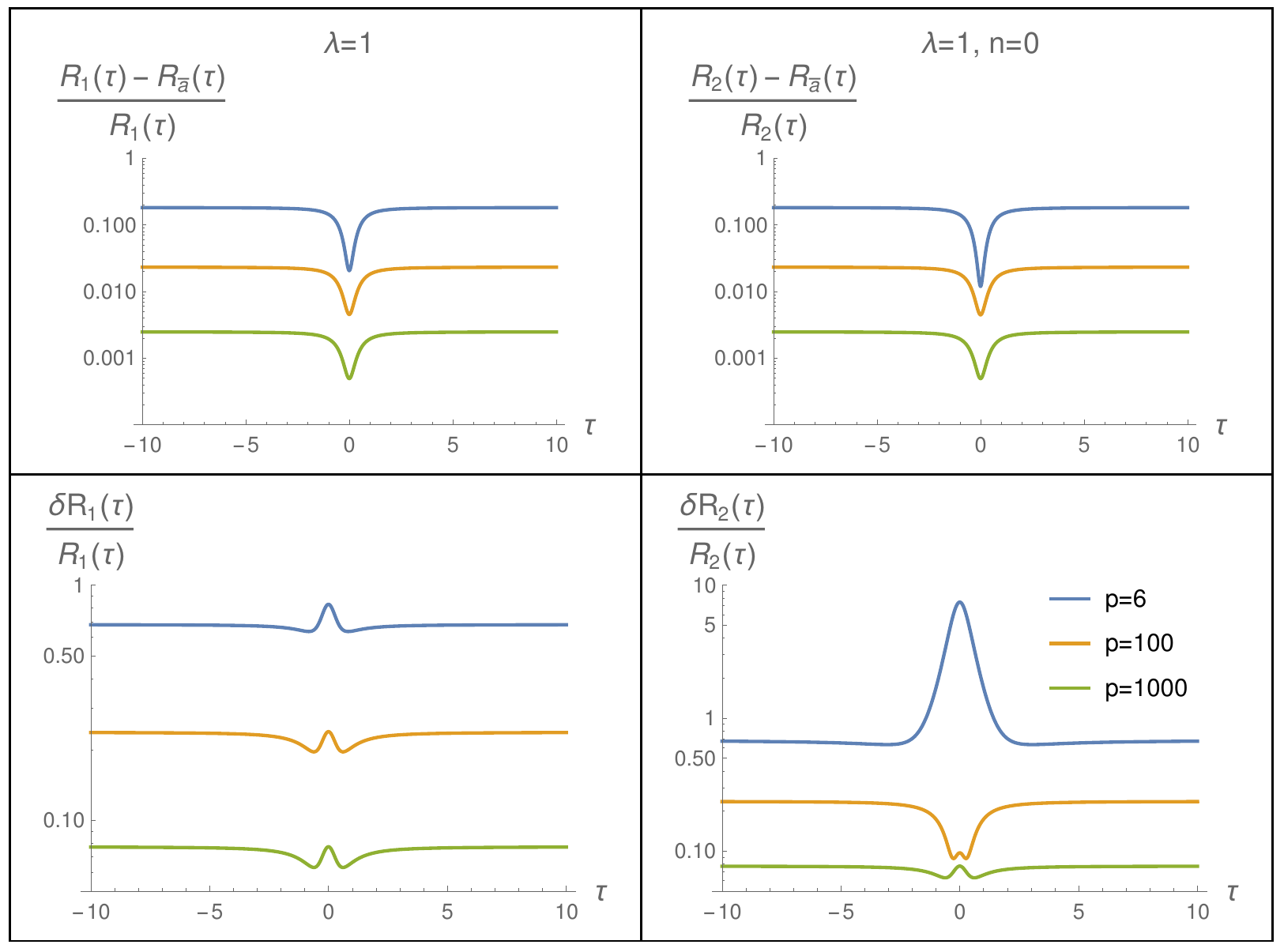}
		\caption{In the first row, we have plotted the fractional change in Ricci scalar expectation value as compared to its semiclassical counterpart for both ordering schemes. In the second row, we have plotted the relative standard deviation in the Ricci scalar for both ordering schemes.}\label{RSqF}
	\end{figure*}
	
	The results for the cosmological constant driven universe can be summarized as follows. The quantum model admits the resolution of the coordinate singularity and represents a bouncing universe that tunnels from a collapsing branch to an expanding branch. This behavior is apparent from the expectation value of the scale factor and the Hubble parameter, where the scale factor has a global minimum at the point of coordinate singularity, and the Hubble parameter has a smooth transition from negative to positive values. Due to quantum effects, the Ricci scalar gets disturbed from its constant value near the tunneling point. Furthermore, the operator ordering of the observables is relevant only near the tunneling point, and the expectation value of the Ricci scalar settles at the same value for different orderings late in the expanding phase. 
	
	As far as the applicability of the effective geometry approach is concerned, the results are in the same spirit as is the case for dust dominated universe. Again, the shape parameter (that is identified with the ordering parameter) acts as the control parameter, and the usage of effective geometry approach is well justified in the limit of a sharply peaked distribution, as conjectured in the \cite{Ashtekar:2009mb}.
	
	\section{Conclusions}\label{Conclusion}
	
	We have investigated the three questions discussed in the introduction in the context of dust dominated and the cosmological constant-dominated flat-FLRW universes: (i) to check the robustness of DeWitt's criteria of singularity resolution, (ii) the status of operator ordering ambiguity in this quantum model, and (iii) the domain of validity of the effective geometry approach, where the expectation of certain quantity is used for characterizing all quantum corrections. For the canonical system corresponding to the aforementioned models, we obtained the phase space expressions for the observables relevant to the analysis, e.g., the Hubble parameter and the Ricci scalar. Apart from the fact that these observables mark the existence of the singularity in the classical picture, they are also involved in the semiclassical analysis. Therefore, studying these observables in the quantum picture allows us to address all three questions at hand. Furthermore, as the quantum model has unitary evolution with respect to the fluid degree of freedom, the expectation values of gravitational observables in the quantum model are with respect to the fluid variable and therefore are time reparameterization invariant, i.e., gauge invariant.
	
	We have addressed the robustness of the singularity resolution in these quantum models by showing that the expectation values of the operators associated with curvature invariants are regular functions that follow the classical behavior away from the classical singularity and remain finite at the location of singularity. In the case of the Hubble parameter, the general trend is that the expectation value has a minimum in the collapsing branch and a maximum in the expanding branch, while it vanishes at the singularity. The location and width of these extrema depend upon the energy density of the fluid, and the extrema are sharp and closely spaced for highly energetic fluid. For the case of Ricci scalar in the dust dominated universe, the expectation value has a maximum at the singularity for a certain class of orderings and a minimum of negative magnitude sandwiched between maxima. Away from the singularity, the expectation value asymptotes to the classical trajectory for all orderings. Therefore, we have demonstrated the robustness of the singularity resolution in the FLRW model with a dust clock. In the case of the cosmological constant driven universe, the classical model has two disjoint branches, a collapsing branch labeled by a constant negative Hubble parameter and an expanding branch labeled by a constant positive Hubble parameter. The quantum model predicts a quantum tunneling from a collapsing branch to an expanding branch. Classically, the Ricci scalar is constant throughout, and the model has no curvature singularity. The Hubble parameter has a step function-like discontinuity at the coordinate singularity, and its quantum expectation is a smooth approximation of the step function behavior representing a bouncing cosmological model. Even though there is no divergence or discontinuity in the classical Ricci scalar, the quantum expectation of the Ricci scalar still gets modified from the classical value due to the quantum tunneling from the collapsing branch to the expanding branch.
	
	Another question of interest in this model is the operator ordering ambiguity. The phase space expressions of all observables under consideration are in the product form of the scale factor and its conjugate momentum. It is apparent that there will be operator ordering ambiguity since multiple ordering choices are available for the same observable. In this work, we write two classes of ordering to symmetrize operators: trivial symmetric ordering and Weyl-like ordering. For the case of the Hubble parameter, it is observed that both ordering prescriptions give rise to the same differential operator, following a generic result that various ordering prescriptions lead to the same operator for functions linear in either position or momentum. Therefore, there is no operator ordering ambiguity at the level of the Hubble parameter operator. For the case of observables that are quadratic or have a higher power in momentum, the key finding of this analysis is that the operator ordering ambiguity plays a role only near the classical singularity, the regime where quantum effects dominate. There is no strong signature of ordering ambiguity away from the singularity, and various orderings converge to the classical behavior. The regulated expectation value of various observables is highly sensitive to the ordering scheme as observed,  e.g., for the square of the Hubble parameter, Ricci scalar, and other curvature invariants. The parameters, $q$ for dust dominated and $p$ for the cosmological constant driven universe, appear as a control parameter that is related to the ordering of the Hamiltonian as well as the shape of the distribution. Different orderings of the observables merge to the same expectation in the limit of sharply peaked distribution, i.e., $q,\;p\rightarrow\infty$.
	
	Lastly, we have investigated the applicability of the effective geometry approach in this setting. The semiclassical expressions for the observables are computed from the expectation value of the scale factor. We compared these semiclassical expressions against the expectation value of these observables. In the case of dust dominated universe, the general trend observed is that the semiclassical expression and quantum expectation match in the classical domain, i.e., $\tau^2\gg\lambda^2$, and the difference is pronounced only near the classical singularity. The Hubble parameter is one of the most important objects in this regard, as it appears in almost all dynamical equations for perturbations and leaves a direct imprint on the physical observations \cite{Mukhanov:1990me}. The semiclassical expression matches the expectation value of the Hubble parameter for a certain choice of the ordering of the Hamiltonian. However, for a different choice of the ordering of Hamiltonian and distribution parameters, these expressions do not match, although the difference between them remains substantially small. The Ricci scalar also plays a crucial role in the dynamics of a non-minimally coupled scalar field at the semiclassical level. In the case of dust dominated universe, the semiclassical expression for Ricci scalar does not match the expectation value in general. In the limit of the control parameter $q\rightarrow\infty$, different orderings of observables merge to the same profile, which in fact is the semiclassical expression for the Ricci scalar. The same behavior is true for other curvature invariants as well. Furthermore, the fractional change in the expectation of the Ricci scalar as compared to the semiclassical expression does not exceed 35\%, with the limiting case being for small $q$ and at the singularity. Therefore, the semiclassical expression for the Ricci scalar can be trusted for the case when the distribution is sharply peaked.
	
	The case of quantum dynamics of the cosmological constant driven universe is somewhat different. The major deviation from the dust dominated case comes from the comparison of the semiclassical expressions and respective quantum expectations for the Hubble parameter and the Ricci scalar. The semiclassical expression and quantum expectation values for both the Hubble parameter and the Ricci scalar asymptote to different constant values away from the bounce point. However, again in the limit of the control parameter $p\rightarrow\infty$, the different profiles for the expectation value of the observables merge onto the semiclassical profile. Therefore, the use of the effective geometry approach is well motivated in the assumption of a sharply peaked distribution.
	
	The matching of the semiclassical expressions with quantum expectations in the models under consideration is ordering and state-dependent. The conjecture of a state sharply peaked on the classical trajectory, proposed in \cite{Ashtekar:2009mb}, is crucial for the applicability of the effective geometry approach, which we have shown to be applicable in this setting. For a general state, this approximation of the quantum corrected spacetime breaks down, and hence, one has to be careful while using the effective geometry in the semiclassical analysis. For the cosmological constant driven universe, the disagreement between the semiclassical expressions and the quantum expectations is most pronounced at the late time in the expanding branch, and the quantum fluctuations are finite in this regime, hinting at quantum effects surviving at late times. There have been recent studies that also indicate significant quantum effects at late times in matter and dark energy-dominated universes, e.g., possible quantum effects at the transition from cosmological deceleration to acceleration in \cite{Alexandre:2022ijm}, the quantum fluctuations survive at the late time leading to a large backreaction \cite{Dhanuka:2020yxp} and enhanced quantum correlations for a nearly matter-dominated universe in \cite{Dhanuka:2022ggi}. The observational signature of the operator ordering ambiguity in the late time universe will be pursued in a future work.
	
	\section{Acknowledgments}
	HSS would like to acknowledge the financial support from the University Grants Commission, Government of India, in the form of Junior Research Fellowship (UGC-CSIR JRF/Dec-2016/503905). Research of KL is partially supported by the Department of Science and Technology (DST) of the Government of India through a research grant under INSPIRE Faculty Award (DST/INSPIRE/04 /2016/000571). HSS would like to thank S. Shankaranarayanan and Sumanta Chakraborty for their helpful comments.
	
	\appendix

	\section{Hermiticity of operators}\label{Hermiticity}
	The self-adjointness of the Hamiltonian operator \eqref{Hamlitonian} is discussed in \cite{kiefer_singularity_2019} whereas the self-adjointness of the Hamiltonian in Eq. \eqref{HSO} follows along the same line. In this section, we will discuss the Hermiticity of other operators that appears in the main text.
	
	\subsection{Hubble parameter: Dust dominated universe}
	
	\noindent For the case of Brown-Kucha\v{r} dust, the Hermiticity of the Hubble parameter operator in Eq. \eqref{HBK} implies
	\begin{align}
		\braket{\psi|\hat{\mathbb{H}}|\chi}&=\int_0^\infty da\;a^2\;\psi^*\hat{\mathbb{H}}\chi=i\int_{0}^{\infty}da\;\psi^*\frac{\partial\chi}{\partial a}\nonumber\\
		&=i\left(\psi^*\chi\bigg|^\infty_0-\int_{0}^{\infty}da\;\frac{\partial\psi^*} {\partial a}\chi\right)\nonumber\\
		&=i\bigg[\psi^*\chi\bigg]^\infty_0+\braket{\hat{\mathbb{H}}\psi|\chi}.
	\end{align}
	The Hermiticity of the operator requires the vanishing of the boundary term in the square bracket. This term vanishes for the case where the wavefunctions vanish at $a\rightarrow0$ and $a\rightarrow\infty$. For the wave packets under consideration in Eq. \eqref{mwp}, the boundary condition is satisfied, provided $q\neq0$. 
	\subsection{Hubble parameter: Cosmological constant driven universe}
	\noindent Similarly, for the case of Schutz fluid, the Hermiticty of Hubble parameter operator in Eq. \eqref{HSFeq} implies
	\begin{align}
		\braket{\psi|\widehat{\mathbb{H}}|\chi}&=\int_0^\infty da\;a^{4-p-2q}\;\psi^*\widehat{\mathbb{H}}\chi\nonumber\\
		&=i\int_{0}^{\infty}da\;a^{2-p-2q}\psi^*\left(\frac{\partial\chi}{\partial a}-\frac{p+2q-2}{a}\chi\right)\nonumber\\
		&=i\left[a^{2-p-2q}\psi^*\chi\right]_0^\infty+\braket{\widehat{\mathbb{H}}\psi|\chi}.
	\end{align}
	The Hubble parameter operator is Hermitian, provided the boundary term in the square bracket vanishes. For the case of the wave packets in Eq. \eqref{wpS}, at the lower limit $a\rightarrow 0$, the boundary term goes as $a^{|p+1|+3}$ and therefore vanishes for all $p$. Whereas at the upper limit $a\rightarrow \infty$, the exponential term will kill off the boundary contribution. Therefore, the Hubble parameter is Hermitian for the set of wave packets under consideration. 
	\subsection{Square of Hubble parameter and Ricci scalar: Dust dominated universe}
	\noindent The differential operators corresponding to the square of the Hubble parameter operator in Eqs. \eqref{Hsq1}, \eqref{Hsq2} is
	\begin{align}
		\widehat{\mathcal{H}}^2_1=&a^{-6}\left((j-1)(j-4)+2a\partial_a-a^2\partial^2_a\right),\\
		\widehat{\mathcal{H}}^2_2=&a^{-6}\left(2 j (j+k-4)-5 k+12+2a\partial_a-a^2\partial^2_a\right).
	\end{align}
	We see that the difference comes from the term with no derivative and the terms with a derivative match with the terms for the Ricci scalar operator in Eqs. \eqref{RBJDO} and \eqref{RWDO}. The reason for this happenstance is that the two observables in question have similar phase space expression on-shell $\propto p_a^2a^{-4}$. We can write a general operator that represents all four cases via
	\begin{align}
		\widehat{\mathcal{O}}=\alpha a^{-6}\left(\beta+2a\partial_a-a^2\partial^2_a\right).
	\end{align}
	The Hermiticity of this operator implies
	\begin{align}
		\braket{\psi|\widehat{\mathcal{O}}|\chi}=&\int_{0}^\infty da\;a^2\psi^*\widehat{\mathcal{O}}\chi\nonumber\\
		=&\alpha\int_{0}^\infty da\;a^{-4}\psi^*\left(\beta\chi+2a\partial_a\chi-a^2\partial^2_a\chi\right),\nonumber\\
		=&\alpha\bigg[a^{-2}\bigg(\psi^*\frac{\partial\chi}{\partial a}-\frac{\partial\psi^*}{\partial a}\chi\bigg)\bigg]^\infty_0+\braket{\hat{\mathcal{O}}\psi|\chi}.
	\end{align}
	The boundary conditions required for the Hermiticity of these operators are the same. For operators to be Hermitian, these boundary terms have to vanish. For a function with asymptotic behavior $\psi(a)\rightarrow a^k$ as for $a\rightarrow 0$ and $\psi(a)\rightarrow a^{k'}$ as for $a\rightarrow \infty$, the boundary term behaves as
	\begin{align}
		&\lim_{a\rightarrow 0}\bigg[a^{-2}\bigg(\psi^*\frac{\partial\chi}{\partial a}-\frac{\partial\psi^*}{\partial a}\chi\bigg)\bigg]=\lim_{a\rightarrow 0}a^{2k-3}=0,\;\text{if}\; k>\frac{3}{2},\\
		&\lim_{a\rightarrow \infty}\bigg[a^{-2}\bigg(\psi^*\frac{\partial\chi}{\partial a}-\frac{\partial\psi^*}{\partial a}\chi\bigg)\bigg]=\lim_{a\rightarrow \infty}a^{2k'-3}=0,\;\text{if}\; k'<\frac{3}{2}.
	\end{align}
	For the set of wave packets, the boundary term vanishes at the upper limit for all parameter values, while at the lower limit, the boundary term vanishes when $|q|>3/2$. 
	
	\subsection{Square of Hubble parameter and Ricci scalar: Cosmological constant driven universe}
	The differential operators correspond to different ordering in Eqs. \eqref{RSDO} and \eqref{RSWDO} differ only at the level of the term without derivative. The general operator can be cast in the form
	\begin{align}
		\widehat{\mathcal{O}}=a^{-6} \left(\beta(p,n)+12a((p+2q)\partial_a-a\partial^2_a)\right)
	\end{align}
	The Hermiticity of this operator implies
	\begin{align}
		\braket{\psi|\hat{\mathcal{O}}|\chi}=&\int_{0}^\infty da\;a^{4-p-2q}\psi^*\hat{\mathcal{O}}\chi\\
		=&\bigg[a^{-p-2q}\bigg(\psi^*\frac{\partial\chi}{\partial a}-\frac{\partial\psi^*}{\partial a}\chi\bigg)\bigg]^\infty_0+\braket{\hat{\mathcal{O}}\psi|\chi},
	\end{align}
	the boundary term in the square bracket should vanish. For the set of wave packets considered in Eq. \eqref{wpS}, the upper limit vanishes due to the exponential factor, and for the lower limit, the boundary term goes as $a^{|1+p|}$ and vanishes as $a\rightarrow 0$, provided $|1+p|\neq0$.
	
	\section{Regularity of the expectation values of operators}\label{Regularity}
	In the subsection \ref{RicciBK}, we encountered the divergences in the expectation value of the Ricci scalar in Eq. \eqref{RBJE} and \eqref{RWE} that are for the parameter values outside of the domain of Hermiticity of Ricci scalar. A similar trend is observed in the case of the Riemann and Kretschmann scalar, as will be seen in the appendix \ref{HiCur}. In the last section, we have derived the Hermiticity condition for various operators, and here, we will derive the condition for the regularity of the expectation value of the various operators. As the states considered in this analysis are vanishing exponentially as  $a\rightarrow\infty$, the cause for divergences is the lower limit. For the discussion in this section, we will assume the states with asymptotic behavior $\psi(a)\rightarrow a^\alpha$ as $a\rightarrow 0$ and exponentially decaying as $a\rightarrow\infty$.
	First, we will discuss the case of the Hubble parameter for which the expectation value in state $\psi$ is
	\begin{align}
		\bar{\mathbb{H}}=\int_{0}^\infty da\psi^*\frac{\partial\psi}{\partial a}.
	\end{align} 
	Here, since the lower limit is of concern, we need to check the behavior of the integrand near $a=0$. Using the asymptotic expression for wave function, the integrand behaves as $a^{2\alpha-1}$ as $a\rightarrow 0$. Using the $p$-test\footnote{the integral of form $\displaystyle\int_0^adxx^{p}$ is convergent, if $p>-1$.} for the convergence, we get the condition $\alpha>0$. For the wave packet in Eq. \eqref{mwp}, this condition translates to $q\neq 0$. 
	
	In the case of the Ricci scalar operator for both orderings in Eq. \eqref{RBJDO} and \eqref{RWDO}, the expectation value in the state $\psi$ takes the form
	\begin{align}
		\overline{\mathcal{R}}=3\int_{0}^\infty da\;a^{-4}\psi^*\left(\beta(q)\psi+2a\partial_a\psi-a^2\partial^2_a\psi\right).
	\end{align}
	The behavior of the integrand for the given state is $a^{2\alpha-4}$ as $a\rightarrow 0$ (as the terms are of the form $a\partial_a$ and $a^2\partial_a^2$). Again using the $p$-test, the integral converges for the case when $\alpha>3/2$ translates to $|q|>3/2$ for the states under consideration in Eq. \eqref{mwp}. Similarly, for a cosmological constant driven universe as well, the expectation value of the Ricci scalar diverges for parameter values outside the domain of Hermiticity, with its origin also being the same as above.
	
	Here, we have derived the conditions for the Hermiticity of operators and the conditions for the regularity of the expectation value of these operators. Although the origin of these conditions is different, the domain of Hermiticity overlaps with the domain of regularity, thereby saving the quantum model from problematic divergences in a natural manner.
	
	\section{Classically anticipated expressions}\label{AppSchCl}
	\noindent In the case of a dust dominated universe, the classical expression for the scale factor is
	\begin{align}
		a(\tau)=\left(\frac{9P_T}{2}\right)^{\frac{1}{3}} \tau^{\frac{2}{3}},
	\end{align}
	and the Hubble parameter and curvature invariants are independent of the constant of motion $P_T$. The system is placed in a state described by the wave packet given in Eq. \eqref{mwp}, which is constructed using the energy distribution in Eq. \eqref{EnergyDis}. For this state, the classically anticipated expression for the scale factor is given by,
	\begin{align}
		a(\tau)=\left(\frac{9}{2}\right)^{\frac{1}{3}} \tau^{\frac{2}{3}}\braket{P_T^{\frac{1}{3}}}=\left(\frac{9(\lambda^2+4\tau^2)}{8\lambda}\right)^\frac{1}{3}\frac{\Gamma\left(\frac{2|q|}{3}+\frac{4}{3}\right)}{\Gamma\left(\frac{2|q|}{3}+1\right)}\bigg|_{\tau^2\gg\lambda^2},
	\end{align}
	where $\braket{P_T^{1/3}}$ is the ensemble average\footnote{Since $P_T=E$, the ensemble average  $\displaystyle\braket{P_T^n}=\braket{E^n}=\braket{\psi|\hat{H}^n|\psi}=\int_{0}^\infty d\sqrt{E} E^{n}A(\sqrt{E})^2$ is the $2n$-th moment of the distribution $A(\sqrt{E})$.} in the distribution \eqref{EnergyDis}. The expectation value of the scale factor in Eq. \eqref{sf} matches with the classically anticipated scale factor at the leading order, i.e., when $\tau^2\gg\lambda^2$. The observables of interest are independent of the constant of motion, and therefore, there is no ambiguity in their classically anticipated expression. 
	
	In the case of a dark energy dominated universe at the classical level, the scale factor, Hubble parameter, and Ricci scalar are related to the energy density via $a\propto\rho^{1/6}$, $\mathbb{H}=\sqrt{2\rho}$ and $\mathcal{R}=24\rho$, with energy density being equal to the momentum conjugate to the fluid variable, i.e., the cosmological constant $\rho=\Lambda$. First, we will address what should be the classically expected behavior of various objects for the universe that the wave packet in Eq. \eqref{wpS} represents. The wave packet is constructed using the Poisson-like energy distribution, and the ensemble averages with the distribution in Eq. \eqref{EnergyDis} for various objects are
	\begin{align}
		a_{cl}(\tau)&=\braket{(18\Lambda)^{1/6}}\tau^{1/3}=\left(\frac{18}{\lambda}\right)\frac{\Gamma \left(\frac{| p+1| }{6}+\frac{7}{6}\right)}{\Gamma \left(\frac{| p+1| }{6}+1\right)},\\
		\mathbb{H}_{cl}(\tau)&=\braket{\sqrt{2\Lambda}}=\sqrt{\frac{2}{\lambda}}\frac{\Gamma \left(\frac{| p+1| }{6}+\frac{3}{2}\right)}{\Gamma \left(\frac{| p+1| }{6}+1\right)},\\
		\mathcal{R}_{cl}(\tau)&=24\braket{\Lambda}=\frac{24 }{\lambda }\left(\frac{| p+1| }{6}+1\right).
	\end{align}
	These expressions correlate exactly with the asymptotic expression of the expectation values of these observables in Eqs. \eqref{sfSex}, \eqref{HPex}, \eqref{RSex1} and \eqref{RSex2} in the classical regime $\tau^2\gg\lambda^2$, but the semiclassical expressions do not correlate with the classically expected expressions. Interestingly, this exact quantum to classical correspondence is possible only in the context of the semiclassical state (e.g., coherent state or squeezed state), and it seems that the wave packets constructed in this analysis mimic the behavior as that of a semiclassical state \cite{Ashtekar:2005}.
	
	The difference between the semiclassical expressions and the quantum expectations of these observables comes from the fact that for the distribution in Eq. \eqref{EnergyDis}, the ensemble average follows $\braket{\Lambda^n}\neq\braket{\Lambda}^n$. The semiclassical expressions capture the distribution properties via the expectation value of the scale factor through $\braket{\Lambda}^{1/6}$, and the semiclassical expressions are the powers of this factor. The physical imprint of ambiguity of this kind is discussed in \cite{Malkiewicz:2020fvy}, where the expectation value $\braket{a^n}^{1/n}$ is used to describe the quantum corrected spacetime and the imprint of parameter $n$ is investigated on the primordial gravitational wave spectrum.
	
	\section{Higher curvature invariants}\label{HiCur}
	The method of constructing phase space expressions outlined in Sec. \ref{Sec3} can be used to find phase space expressions of the higher curvature invariants, and one can study these observables at the quantum level. Non-vanishing components of the Riemann and Ricci tensors are
	\begin{align}
		R_{0101}&=R_{0202}=R_{0303}=\frac{a}{\mathcal{N}}\left(\dot{a}\dot{\mathcal{N}}-\mathcal{N}\ddot{a}\right),\\
		R_{1212}&=R_{1313}=R_{2323}=\frac{a^2\dot{a}^2}{\mathcal{N}^2},\\
		R_{00}&=3\frac{\dot{a}\dot{\mathcal{N}}-\mathcal{N}\ddot{a}}{a\mathcal{N}},\qquad R_{ii}=\frac{a\mathcal{N}\ddot{a}-a\dot{a}\dot{\mathcal{N}}+2\mathcal{N}\dot{a}^2}{\mathcal{N}^3}.
	\end{align}
	We will term the scalar constructed out of the contraction of the Ricci tensor with itself as the Riemann scalar, 
	\begin{align}
		\text{Ri}=R_{\mu\nu}R^{\mu\nu}=9\frac{\left(\dot{a}\dot{\mathcal{N}}-\mathcal{N}\ddot{a}\right)^2}{a^2\mathcal{N}^6}+3\frac{\left(a\mathcal{N}\ddot{a}-a\dot{a}\dot{\mathcal{N}}+2\mathcal{N}\dot{a}^2\right)^2}{a^4\mathcal{N}^6}.
	\end{align}
	Using \eqref{da} and \eqref{dda} and gauge $\mathcal{N}=1$, we arrive at 
	\begin{align}
		\text{Ri}=\frac{12}{a^4}\left(\{p_a,\mathsf{H}\}^2+\frac{p_a^4}{a^4}+\frac{p_a^2}{a^2}\{p_a,\mathsf{H}\}\right).\label{RieCl}
	\end{align}
	Similarly, the canonical expression for the Kretschmann Scalar for this model is given by
	\begin{align}
		K&=R_{\mu\nu\alpha\beta}R^{\mu\nu\alpha\beta}=12\left(R_{0101}R^{0101}+R_{1212}R_{1212}\right)\nonumber\\
		&=12\left(\frac{\left(\dot{a}\dot{\mathcal{N}}-\mathcal{N}\ddot{a}\right)^2}{\mathcal{N}^6a^2}+\frac{\dot{a}^4}{a^4\mathcal{N}^4}\right)\nonumber\\
		&=\frac{12}{a^4}\left(\{p_a,\mathsf{H}\}^2+\frac{2p_a^4}{a^4}+\frac{2p_a^2}{a^2}\{p_a,\mathsf{H}\}\right).\label{KCl}
	\end{align}
	Classically for dust as matter, the Riemann and Kretschmann scalars are given by $\text{Ri}=16/9\tau^4,\;\text{and}\; K=80/27\tau^4$, again diverging at $\tau=0$.
	
	Here, we will do the analysis for the case of dust dominated universe only, and the case of a cosmological constant driven universe can be done in the same spirit. The expression for the Riemann and Kretschmann scalar computed from the expectation value of scale factor \eqref{sf} is 
	\begin{align}
		\text{Ri}(\bar{a})=&9\frac{\ddot{\bar{a}}^2}{\bar{a}^2}+3\frac{(\bar{a}\ddot{\bar{a}}+2\dot{\bar{a}}^2)^2}{\bar{a}^4}=\frac{256 \left(3 \lambda ^4+16 \tau ^4\right)}{9 \left(\lambda ^2+4 \tau ^2\right)^4},\\
		\text{K}(\bar{a})=&12\left(\frac{\ddot{\bar{a}}^2}{\bar{a}^2}+\frac{\dot{\bar{a}}^4}{\bar{a}^4}\right)=\frac{256 \left(9 \lambda ^4-24 \lambda ^2 \tau ^2+80 \tau ^4\right)}{27 \left(\lambda ^2+4 \tau ^2\right)^4}.
	\end{align}
	The semiclassical expressions for both curvature invariants are regularized functions with a maximum at the singularity and follow the classical behavior $	\text{Ri}\rightarrow16/9\tau^4$ and $\text{K}\rightarrow80/27\tau^4$ in the large $|\tau|$ regime. The phase space expressions for the Riemann and Kretschmann Scalar are given in equations \eqref{RieCl} and \eqref{KCl}.  There are three distinct terms appearing in both expressions, involving the powers of scale factor, momentum, and the Poisson bracket of momentum and Hamiltonian. In principle, there exist infinitely many options for symmetric ordering of these terms, but in this analysis, we will do a comparative analysis and write two sets of symmetrized orderings following the first ordering prescription.
	
	\begin{table}[H]
		\centering
		\bgroup
		\def\arraystretch{2.5}
		\scalebox{0.82}{\begin{tabular}{ |c|c|c|}
				\hline
				Classical expression&Ordering 1&Ordering 2\\
				\hline
				$\displaystyle \quad\{p_a,\mathsf{H}\}^2a^{-4}\quad$&$\displaystyle \quad-\hat{a}^{-2}[\hat{p}_a,\hat{\mathsf{H}}][\hat{p}_a,\hat{\mathsf{H}}]\hat{a}^{-2}\quad$&$\displaystyle \quad-[\hat{p}_a,\hat{\mathsf{H}}]\hat{a}^{-4}[\hat{p}_a,\hat{\mathsf{H}}]\quad$\\
				\hline
				$\displaystyle a^{-8}p_a^4$&$\displaystyle \hat{a}^{-4}\hat{p}^4_a\hat{a}^{-4}$&$\displaystyle \hat{p}_a^{2}\hat{a}^{-8}\hat{p}_a^{2}$\\
				\hline
				$\displaystyle a^{-6}p_a^2\{p_a,\mathsf{H}\}$&$\displaystyle -i\hat{a}^{-3}\hat{p}_a[\hat{p}_a,\hat{\mathsf{H}}]\hat{p}_aa^{-3}$&$\displaystyle -i\hat{p}_a\hat{a}^{-3}[\hat{p}_a,\hat{\mathsf{H}}]a^{-3}\hat{p}_a$\\
				\hline
		\end{tabular}}
		\egroup
		\caption{Three terms that the Riemann and Kretschmann scalar are comprised of and the symmetric orderings corresponding to each term.}
		\label{Table}
	\end{table}
\begin{figure*}
	\centering
	\includegraphics[scale=0.7]{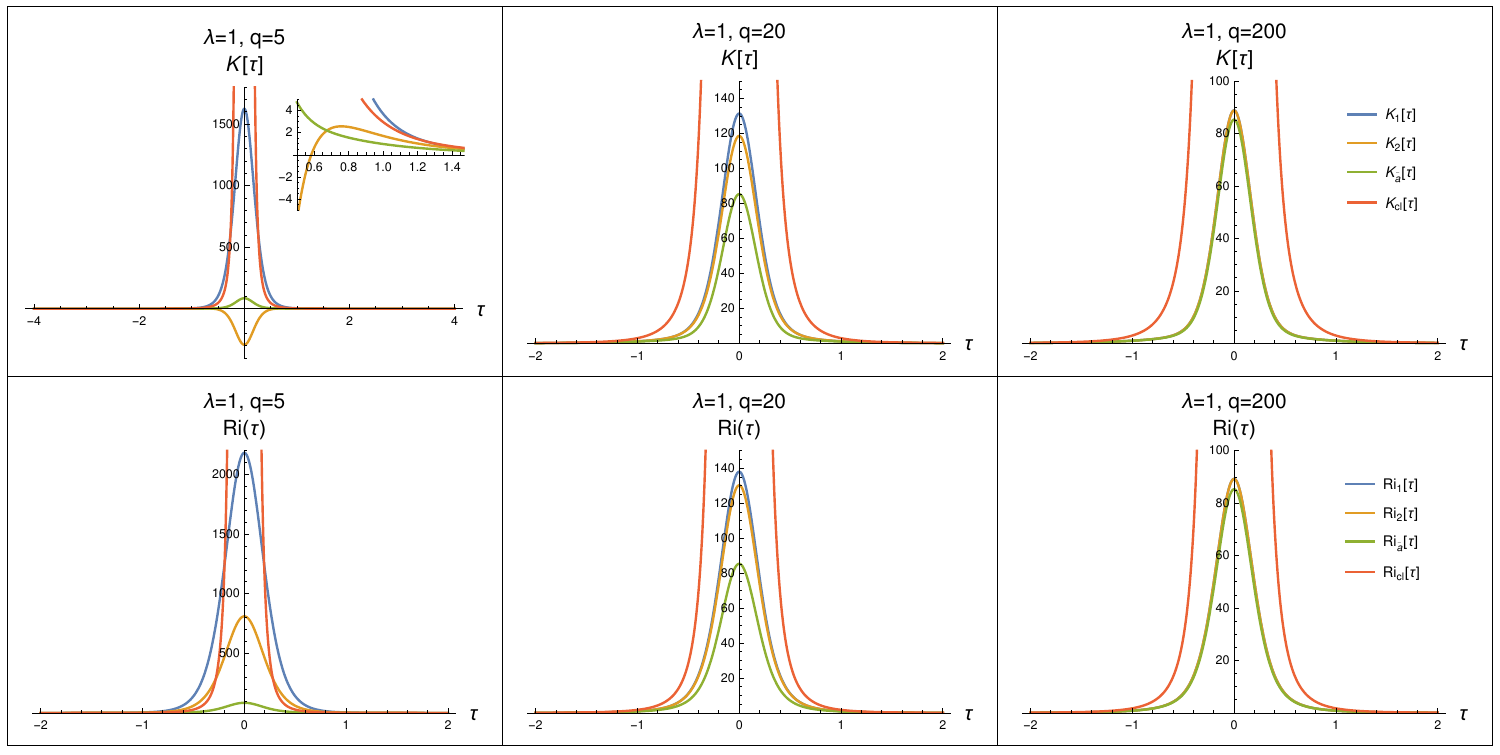}
	\caption{ Expectation value of Kretschmann scalar and Riemann scalar operator for various operator orderings.}\label{KF}
\end{figure*}
	In one set, the powers of the scale factor operator will be the leftmost and rightmost while other operators are sandwiched in between. Whereas in the other set, the powers of the momentum operator are placed on the leftmost and rightmost with other operators sandwiched, and in one term that is devoid of the bare momentum operator, the commutator operators are placed on the leftmost and rightmost, as shown in Table \ref{Table}. With these choices of ordering, the operators corresponding to the Riemann and Kretschmann scalars are
	
	\begin{strip}
	\begin{align}
		\begin{aligned}
			\hat{\text{Ri}}_1&=12\left(-\hat{a}^{-2}[\hat{p}_a,\hat{\mathsf{H}}][\hat{p}_a,\hat{\mathsf{H}}]\hat{a}^{-2}+\hat{a}^{-4}\hat{p}^4_a\hat{a}^{-4}-i\hat{a}^{-3}\hat{p}_a[\hat{p}_a,\hat{\mathsf{H}}]\hat{p}_aa^{-3}\right),\\
			\hat{\text{K}}_1&=12\left(-\hat{a}^{-2}[\hat{p}_a,\hat{\mathsf{H}}][\hat{p}_a,\hat{\mathsf{H}}]\hat{a}^{-2}+2\hat{a}^{-4}\hat{p}^4_a\hat{a}^{-4}-2i\hat{a}^{-3}\hat{p}_a[\hat{p}_a,\hat{\mathsf{H}}]\hat{p}_aa^{-3}\right),\\
			\hat{\text{Ri}}_2&=12\left(-[\hat{p}_a,\hat{\mathsf{H}}]\hat{a}^{-4}[\hat{p}_a,\hat{\mathsf{H}}]+\hat{p}_a^{2}\hat{a}^{-8}\hat{p}_a^{2}-i\hat{p}_a\hat{a}^{-3}[\hat{p}_a,\hat{\mathsf{H}}]a^{-3}\hat{p}_a\right),\\
			\hat{\text{K}}_2&=12\left(-[\hat{p}_a,\hat{\mathsf{H}}]\hat{a}^{-4}[\hat{p}_a,\hat{\mathsf{H}}]+2\hat{p}_a^{2}\hat{a}^{-8}\hat{p}_a^{2}-2i\hat{p}_a\hat{a}^{-3}[\hat{p}_a,\hat{\mathsf{H}}]a^{-3}\hat{p}_a\right).
		\end{aligned}
	\end{align} 
	Using the expression of the commutator of the momentum operator and Hamiltonian operator in Eq. \eqref{Comm}, the differential operators corresponding to these orderings take the form
	\begin{align}
		\begin{aligned}
			\hat{\text{Ri}}_1=&3a^{-12}\biggr((9|q|^4-66|q|^2+1153)-736a\partial_a+218a^2\partial_a^2-36a^3\partial_a^3+3a^4\partial_a^4\biggr),\\
			\hat{\text{Ri}}_2=&3a^{-12}\biggr((9|q|^4-282|q|^2+493)+608a\partial_a+50a^2\partial_a^2-36a^3\partial_a^3+3a^4\partial_a^4\biggr),\\
			\hat{K}_1=&3a^{-12}\biggr((9|q|^4+18|q|^2+2005)-16(3|q|^2+79)a\partial_a+(6|q|^2+368)a^2\partial_a^2-60a^3\partial_a^3+5a^4\partial_a^4\biggr),\\
			\hat{K}_2=&3a^{-12}\biggr((9|q|^4-342|q|^2+913)-16(3|q|^2-77)a\partial_a+(6|q|^2+56)a^2\partial_a^2-60a^3\partial_a^3+5a^4\partial_a^4\biggr).\label{K2}
		\end{aligned}
	\end{align}
	For different orderings, the third and fourth-order terms in the derivative match, while the other terms are distinct. The boundary terms arising from the requirement of Hermiticity of the Riemann scalar and Kretschmann scalar operator are 
	\begin{align}
		\begin{aligned}
			&\biggr[3a^{-6}\bigg(\psi^*\frac{\partial^3\chi}{\partial a^3}-\frac{\partial\psi^*}{\partial a}\frac{\partial^2\chi}{\partial a^2}+\frac{\partial^2\psi^*}{\partial a^2}\frac{\partial\chi}{\partial a}-\frac{\partial^3\psi^*}{\partial a^3}\chi\bigg)+18a^{-7}\bigg(\frac{\partial^2\psi^*}{\partial a^2}\chi-\psi^*\frac{\partial^2\chi}{\partial a^2}\bigg)+Aa^{-9}\bigg(\psi^*\frac{\partial\chi}{\partial a}-\frac{\partial\psi^*}{\partial a}\chi\bigg)\biggr]_0^\infty\\
			&\biggr[5a^{-6}\bigg(\psi^*\frac{\partial^3\chi}{\partial a^3}-\frac{\partial\psi^*}{\partial a}\frac{\partial^2\chi}{\partial a^2}+\frac{\partial^2\psi^*}{\partial a^2}\frac{\partial\chi}{\partial a}-\frac{\partial^3\psi^*}{\partial a^3}\chi\bigg)+30a^{-7}\bigg(\frac{\partial^2\psi^*}{\partial a^2}\chi-\psi^*\frac{\partial^2\chi}{\partial a^2}\bigg)+Ba^{-9}\bigg(\psi^*\frac{\partial\chi}{\partial a}-\frac{\partial\psi^*}{\partial a}\chi\bigg)\biggr]_0^\infty\label{KBC}
		\end{aligned}
	\end{align}
	respectively. The operators are Hermitian, provided that these boundary terms vanish. In the Hermiticity condition for different orderings, only the coefficients in the third term are different. The coefficient $A$ is 92 for $\hat{\text{Ri}}_1$ and 76 for $\hat{\text{Ri}}_2$ whereas $B$ is $2(3|q|^2+79)$ for $\hat{\text{K}}_1$  and $2(3|q|^2-77)$ for $\hat{\text{K}}_2$. Anyway, the coefficients do not matter for the vanishing of these boundary terms, and for the set of wave packets, the boundary terms vanish if $|q|>9/2$. We notice that the Hermiticity condition is indifferent to the operator orderings ambiguity. The expectation values of these operators for the wave packet in \eqref{mwp} are
	
	\begin{align}
		\overline{\text{Ri}}_1(\tau)=\frac{256 \left(\lambda ^2+4 \tau ^2\right)^{-4}}{27 (|q|-3) |q| (2 |q|-9) (2 |q|-3) }\bigg(&\lambda ^4 \left(3 |q| \left(12 |q|^3-61 |q|-205\right)+1264\right)+48 (|q|-3) |q| (2 |q|-9) (2 |q|\nonumber\\
		&-3) \tau ^4+8 \lambda ^2 (|q|-3) (2 |q|-9) (27 |q|-62) \tau ^2\bigg),\\
		\overline{\text{Ri}}_2(\tau)=\frac{256 \left(\lambda ^2+4 \tau ^2\right)^{-4}}{27(|q|-3) |q| (2 |q|-9) (2 |q|-3)}\bigg(&\lambda ^4 \left(3 |q| \left(12 |q|^3-349 |q|+131\right)+7696\right)+48 (|q|-3) |q| (2 |q|-9) (2 |q|\nonumber\\
		&-3) \tau ^4+8 \lambda ^2 (|q|-3) (2 |q|-9) (27 |q|+106) \tau ^2\bigg),\\
		\overline{K}_1(\tau)=\frac{256  \left(\lambda ^2+4 \tau ^2\right)^{-4}}{27 (|q|-3) |q| (2 |q|-9) (2 |q|-3) }\bigg(&-24 \lambda ^2 (|q|-3) (2 |q|-9) (|q| (2 |q|-15)+36) \tau ^2+\lambda ^4 (9 |q| (|q| (4 (|q|-1) |q|\nonumber\\
		&-13)-117)+2188)+80 (|q|-3) |q| (2 |q|-9) (2 |q|-3) \tau ^4\bigg),\\
		\overline{K}_2(\tau)=\frac{256  \left(\lambda ^2+4 \tau ^2\right)^{-4}}{27(|q|-3) |q| (2 |q|-9) (2 |q|-3)}\bigg(&-24 \lambda ^2 (|q|-3) (2 |q|-9) (|q| (2 |q|-15)-68) \tau ^2+(9 |q| (|q| (4 (|q|-1) |q|\nonumber\\
		&-173)+91)+14668)\lambda ^4 +80 (|q|-3) |q| (2 |q|-9) (2 |q|-3) \tau ^4\bigg).
	\end{align}
\end{strip}

	All expressions are regular functions for the case when $|q|>9/2$ and have a maximum at the classical singularity, except for $\overline{K}_2(\tau)$. Here as well, the regularity of the expectation value and the Hermiticity condition put the same constraint on the parameter $q$. In the large $q$ regime, different orderings give rise to the same expectation value, thereby reaffirming that there is no signature of operator ordering ambiguity in this regime. Far away from the singularity $\tau^2\gg\lambda^2$, we recover the classical expressions for both Riemann and Kretschmann scalar.
	
	We have plotted the expectation value of both curvature invariants in Fig. \ref{KF}. Here, as well, the expectation value of different orderings of the curvature invariants are in stark contrast for small $q$ values. On the other hand, different orderings merge onto the semiclassical expressions as we keep on increasing the $q$ parameter, following the trend observed in the case of Ricci scalar expectation value. The expectation value of the Kretschmann scalar in the second case acquires a negative value near singularity for small $q$, with a profile similar to the one observed for the Weyl-like ordering of the Ricci scalar.
	
	The analysis for higher curvature invariants yields the same trend as observed in the preceding subsections. Thus, for the ordering class of the Hamiltonian in Eq. \eqref{HO2} and wave packet in Eq. \eqref{mwp}, the main results can be summarized as the Hubble parameter matches the semiclassical expression, and for other observables, we can trust the semiclassical expressions only in the large $q$ parameter regime, i.e., for a sharply peaked distribution as argued in \cite{Agullo:2012sh}. All this was discussed with a fixed ordering scheme of the Hamiltonian. We can now relax the condition used to simplify the expression in Eq. \eqref{gwp} and consider the case when the model has the same energy distribution parameters but different ordering parameters of the Hamiltonian operator.


\begin{thebibliography}{92}%
		\makeatletter
		\providecommand \@ifxundefined [1]{%
			\@ifx{#1\undefined}
		}%
		\providecommand \@ifnum [1]{%
			\ifnum #1\expandafter \@firstoftwo
			\else \expandafter \@secondoftwo
			\fi
		}%
		\providecommand \@ifx [1]{%
			\ifx #1\expandafter \@firstoftwo
			\else \expandafter \@secondoftwo
			\fi
		}%
		\providecommand \natexlab [1]{#1}%
		\providecommand \enquote  [1]{``#1''}%
		\providecommand \bibnamefont  [1]{#1}%
		\providecommand \bibfnamefont [1]{#1}%
		\providecommand \citenamefont [1]{#1}%
		\providecommand \href@noop [0]{\@secondoftwo}%
		\providecommand \href [0]{\begingroup \@sanitize@url \@href}%
		\providecommand \@href[1]{\@@startlink{#1}\@@href}%
		\providecommand \@@href[1]{\endgroup#1\@@endlink}%
		\providecommand \@sanitize@url [0]{\catcode `\\12\catcode `\$12\catcode
			`\&12\catcode `\#12\catcode `\^12\catcode `\_12\catcode `\%12\relax}%
		\providecommand \@@startlink[1]{}%
		\providecommand \@@endlink[0]{}%
		\providecommand \url  [0]{\begingroup\@sanitize@url \@url }%
		\providecommand \@url [1]{\endgroup\@href {#1}{\urlprefix }}%
		\providecommand \urlprefix  [0]{URL }%
		\providecommand \Eprint [0]{\href }%
		\providecommand \doibase [0]{http://dx.doi.org/}%
		\providecommand \selectlanguage [0]{\@gobble}%
		\providecommand \bibinfo  [0]{\@secondoftwo}%
		\providecommand \bibfield  [0]{\@secondoftwo}%
		\providecommand \translation [1]{[#1]}%
		\providecommand \BibitemOpen [0]{}%
		\providecommand \bibitemStop [0]{}%
		\providecommand \bibitemNoStop [0]{.\EOS\space}%
		\providecommand \EOS [0]{\spacefactor3000\relax}%
		\providecommand \BibitemShut  [1]{\csname bibitem#1\endcsname}%
		\let\auto@bib@innerbib\@empty
		\bibitem [{\citenamefont {Tambornino}(2012)}]{Tambornino:2011vg}%
		\BibitemOpen
		\bibfield  {author} {\bibinfo {author} {\bibfnamefont {J.}~\bibnamefont
				{Tambornino}},\ }\href {\doibase 10.3842/SIGMA.2012.017} {\bibfield
			{journal} {\bibinfo  {journal} {SIGMA}\ }\textbf {\bibinfo {volume} {8}},\
			\bibinfo {pages} {017} (\bibinfo {year} {2012})}\BibitemShut
		{NoStop}%
		\bibitem [{\citenamefont {Anderson}(2012)}]{anderson_2012}%
		\BibitemOpen
		\bibfield  {author} {\bibinfo {author} {\bibfnamefont {E.}~\bibnamefont
				{Anderson}},\ }\href {\doibase https://doi.org/10.1002/andp.201200147}
		{\bibfield  {journal} {\bibinfo  {journal} {Annalen der Physik}\ }\textbf
			{\bibinfo {volume} {524}},\ \bibinfo {pages} {757} (\bibinfo {year}
			{2012})}\BibitemShut
		{NoStop}%
		\bibitem [{\citenamefont {Henneaux}\ and\ \citenamefont
			{Bunster}(1992)}]{henneaux_quantization_1992}%
		\BibitemOpen
		\bibfield  {author} {\bibinfo {author} {\bibfnamefont {M.}~\bibnamefont
				{Henneaux}}\ and\ \bibinfo {author} {\bibfnamefont {C.}~\bibnamefont
				{Bunster}},\ }\href {\doibase https://doi.org/10.1515/9780691213866} {\emph
			{\bibinfo {title} {Quantization of gauge systems}}},\ Physics\ (\bibinfo
		{publisher} {Princeton University Press},\ \bibinfo {address} {Princeton, New
			Jersey},\ \bibinfo {year} {1992})\BibitemShut {NoStop}%
		\bibitem [{\citenamefont {Gitman}\ and\ \citenamefont
			{Tyutin}(1990)}]{Gitman:1990qh}%
		\BibitemOpen
		\bibfield  {author} {\bibinfo {author} {\bibfnamefont {D.~M.}\ \bibnamefont
				{Gitman}}\ and\ \bibinfo {author} {\bibfnamefont {I.~V.}\ \bibnamefont
				{Tyutin}},\ }\href {\doibase 10.1007/978-3-642-83938-2} {\emph {\bibinfo
				{title} {{Quantization of Fields with Constraints}}}},\ Springer Series in
		Nuclear and Particle Physics\ (\bibinfo  {publisher} {Springer},\ \bibinfo
		{address} {Berlin, Germany},\ \bibinfo {year} {1990})\BibitemShut {NoStop}%
		\bibitem [{\citenamefont {Date}(2010)}]{Date:2010xr}%
		\BibitemOpen
		\bibfield  {author} {\bibinfo {author} {\bibfnamefont {G.}~\bibnamefont
				{Date}},\ }in\ \href@noop {} {\emph {\bibinfo {booktitle} {{Refresher Course
						for College Teachers}}}}\ (\bibinfo {year} {2010})\ \Eprint
		{http://arxiv.org/abs/1010.2062} {arXiv:1010.2062 [gr-qc]} \BibitemShut
		{NoStop}%
		\bibitem [{\citenamefont {Prokhorov}\ and\ \citenamefont
			{Shabanov}(2011)}]{prokhorov_hamiltonian_2011}%
		\BibitemOpen
		\bibfield  {author} {\bibinfo {author} {\bibfnamefont {L.~V.}\ \bibnamefont
				{Prokhorov}}\ and\ \bibinfo {author} {\bibfnamefont {S.~V.}\ \bibnamefont
				{Shabanov}},\ }\href {\doibase 10.1017/CBO9780511976209} {\emph {\bibinfo
				{title} {Hamiltonian {Mechanics} of {Gauge} {Systems}}}}\ (\bibinfo
		{publisher} {Cambridge University Press},\ \bibinfo {address} {Cambridge},\
		\bibinfo {year} {2011})\BibitemShut {NoStop}%
		\bibitem [{\citenamefont {Dirac}(1958)}]{Dirac:1958sc}%
		\BibitemOpen
		\bibfield  {author} {\bibinfo {author} {\bibfnamefont {P.~A.~M.}\
				\bibnamefont {Dirac}},\ }\href {\doibase 10.1098/rspa.1958.0142} {\bibfield
			{journal} {\bibinfo  {journal} {Proc. Roy. Soc. Lond. A}\ }\textbf {\bibinfo
				{volume} {246}},\ \bibinfo {pages} {333} (\bibinfo {year}
			{1958})}\BibitemShut {NoStop}%
		\bibitem [{\citenamefont {Arnowitt}\ \emph {et~al.}(1959)\citenamefont
			{Arnowitt}, \citenamefont {Deser},\ and\ \citenamefont
			{Misner}}]{arnowitt_dynamical_1959}%
		\BibitemOpen
		\bibfield  {author} {\bibinfo {author} {\bibfnamefont {R.}~\bibnamefont
				{Arnowitt}}, \bibinfo {author} {\bibfnamefont {S.}~\bibnamefont {Deser}}, \
			and\ \bibinfo {author} {\bibfnamefont {C.~W.}\ \bibnamefont {Misner}},\
		}\href {\doibase 10.1103/PhysRev.116.1322} {\bibfield  {journal} {\bibinfo
				{journal} {Physical Review}\ }\textbf {\bibinfo {volume} {116}},\ \bibinfo
			{pages} {1322} (\bibinfo {year} {1959})}\BibitemShut {NoStop}%
		\bibitem [{\citenamefont {Dirac}(1959)}]{PhysRev.114.924}%
		\BibitemOpen
		\bibfield  {author} {\bibinfo {author} {\bibfnamefont {P.~A.~M.}\
				\bibnamefont {Dirac}},\ }\href {\doibase 10.1103/PhysRev.114.924} {\bibfield
			{journal} {\bibinfo  {journal} {Phys. Rev.}\ }\textbf {\bibinfo {volume}
				{114}},\ \bibinfo {pages} {924} (\bibinfo {year} {1959})}\BibitemShut
		{NoStop}%
		\bibitem [{\citenamefont {Bergmann}\ and\ \citenamefont
			{Komar}(1972)}]{Bergmann:1972ud}%
		\BibitemOpen
		\bibfield  {author} {\bibinfo {author} {\bibfnamefont {P.~G.}\ \bibnamefont
				{Bergmann}}\ and\ \bibinfo {author} {\bibfnamefont {A.}~\bibnamefont
				{Komar}},\ }\href {\doibase 10.1007/BF00671650} {\bibfield  {journal}
			{\bibinfo  {journal} {Int. J. Theor. Phys.}\ }\textbf {\bibinfo {volume}
				{5}},\ \bibinfo {pages} {15} (\bibinfo {year} {1972})}\BibitemShut {NoStop}%
		\bibitem [{\citenamefont {Lee}\ and\ \citenamefont {Wald}(1990)}]{Lee:1990nz}%
		\BibitemOpen
		\bibfield  {author} {\bibinfo {author} {\bibfnamefont {J.}~\bibnamefont
				{Lee}}\ and\ \bibinfo {author} {\bibfnamefont {R.~M.}\ \bibnamefont {Wald}},\
		}\href {\doibase 10.1063/1.528801} {\bibfield  {journal} {\bibinfo  {journal}
				{J. Math. Phys.}\ }\textbf {\bibinfo {volume} {31}},\ \bibinfo {pages} {725}
			(\bibinfo {year} {1990})}\BibitemShut {NoStop}%
		\bibitem [{\citenamefont {Pons}\ \emph {et~al.}(1997)\citenamefont {Pons},
			\citenamefont {Salisbury},\ and\ \citenamefont {Shepley}}]{Pons:1996av}%
		\BibitemOpen
		\bibfield  {author} {\bibinfo {author} {\bibfnamefont {J.~M.}\ \bibnamefont
				{Pons}}, \bibinfo {author} {\bibfnamefont {D.~C.}\ \bibnamefont {Salisbury}},
			\ and\ \bibinfo {author} {\bibfnamefont {L.~C.}\ \bibnamefont {Shepley}},\
		}\href {\doibase 10.1103/PhysRevD.55.658} {\bibfield  {journal} {\bibinfo
				{journal} {Phys. Rev. D}\ }\textbf {\bibinfo {volume} {55}},\ \bibinfo
			{pages} {658} (\bibinfo {year} {1997})},\ \Eprint
		{http://arxiv.org/abs/gr-qc/9612037} {arXiv:gr-qc/9612037} \BibitemShut
		{NoStop}%
		\bibitem [{\citenamefont {Kucha\v{r}}(2011)}]{kuchar_time_2011}%
		\BibitemOpen
		\bibfield  {author} {\bibinfo {author} {\bibfnamefont {K.~V.}\ \bibnamefont
				{Kucha\v{r}}},\ }\href {\doibase 10.1142/S0218271811019347} {\bibfield
			{journal} {\bibinfo  {journal} {International Journal of Modern Physics D}\
			}\textbf {\bibinfo {volume} {20}},\ \bibinfo {pages} {3} (\bibinfo {year}
			{2011})}\BibitemShut {NoStop}%
		\bibitem [{\citenamefont {Isham}(1993)}]{Isham:1992ms}%
		\BibitemOpen
		\bibfield  {author} {\bibinfo {author} {\bibfnamefont {C.~J.}\ \bibnamefont
				{Isham}},\ }\href@noop {} {\bibfield  {journal} {\bibinfo  {journal} {NATO
					Sci. Ser. C}\ }\textbf {\bibinfo {volume} {409}},\ \bibinfo {pages} {157}
			(\bibinfo {year} {1993})},\ \Eprint {http://arxiv.org/abs/gr-qc/9210011}
		{arXiv:gr-qc/9210011} \BibitemShut {NoStop}%
		\bibitem [{\citenamefont {Chataignier}(2022)}]{Chataignier:2021ncn}%
		\BibitemOpen
		\bibfield  {author} {\bibinfo {author} {\bibfnamefont {L.}~\bibnamefont
				{Chataignier}},\ }\href {\doibase 10.1007/978-3-030-94448-3} {\emph {\bibinfo
				{title} {{Timeless Quantum Mechanics and the Early Universe}}}},\ Springer
		Theses\ (\bibinfo  {publisher} {Springer},\ \bibinfo {address} {Berlin,
			Germany},\ \bibinfo {year} {2022})\BibitemShut {NoStop}%
		\bibitem [{\citenamefont {Hoehn}\ \emph {et~al.}(2021)\citenamefont {Hoehn},
			\citenamefont {Smith},\ and\ \citenamefont {Lock}}]{Hoehn:2019fsy}%
		\BibitemOpen
		\bibfield  {author} {\bibinfo {author} {\bibfnamefont {P.~A.}\ \bibnamefont
				{H\"ohn}}, \bibinfo {author} {\bibfnamefont {A.~R.~H.}\ \bibnamefont {Smith}},
			\ and\ \bibinfo {author} {\bibfnamefont {M.~P.~E.}\ \bibnamefont {Lock}},\
		}\href {\doibase 10.1103/PhysRevD.104.066001} {\bibfield  {journal} {\bibinfo
				{journal} {Phys. Rev. D}\ }\textbf {\bibinfo {volume} {104}},\ \bibinfo
			{pages} {066001} (\bibinfo {year} {2021})},\ \Eprint
		{http://arxiv.org/abs/1912.00033} {arXiv:1912.00033 [quant-ph]} \BibitemShut
		{NoStop}%
		\bibitem [{\citenamefont {Hoehn}\ \emph
			{et~al.}(2021{\natexlab{a}})\citenamefont {Hoehn}, \citenamefont {Smith},\
			and\ \citenamefont {Lock}}]{Hoehn:2020epv}%
		\BibitemOpen
		\bibfield  {author} {\bibinfo {author} {\bibfnamefont {P.~A.}\ \bibnamefont
				{Hoehn}}, \bibinfo {author} {\bibfnamefont {A.~R.~H.}\ \bibnamefont {Smith}},
			\ and\ \bibinfo {author} {\bibfnamefont {M.~P.~E.}\ \bibnamefont {Lock}},\
		}\href {\doibase 10.3389/fphy.2021.587083} {\bibfield  {journal} {\bibinfo
				{journal} {Front. in Phys.}\ }\textbf {\bibinfo {volume} {9}},\ \bibinfo
			{pages} {181} (\bibinfo {year} {2021}{\natexlab{a}})},\ \Eprint
		{http://arxiv.org/abs/2007.00580} {arXiv:2007.00580 [gr-qc]} \BibitemShut
		{NoStop}%
		\bibitem [{\citenamefont {Page}\ and\ \citenamefont
			{Wootters}(1983)}]{PhysRevD.27.2885}%
		\BibitemOpen
		\bibfield  {author} {\bibinfo {author} {\bibfnamefont {D.~N.}\ \bibnamefont
				{Page}}\ and\ \bibinfo {author} {\bibfnamefont {W.~K.}\ \bibnamefont
				{Wootters}},\ }\href {\doibase 10.1103/PhysRevD.27.2885} {\bibfield
			{journal} {\bibinfo  {journal} {Phys. Rev. D}\ }\textbf {\bibinfo {volume}
				{27}},\ \bibinfo {pages} {2885} (\bibinfo {year} {1983})}\BibitemShut
		{NoStop}%
		\bibitem [{\citenamefont {Wootters}(1984)}]{Wootters1984}%
		\BibitemOpen
		\bibfield  {author} {\bibinfo {author} {\bibfnamefont {W.~K.}\ \bibnamefont
				{Wootters}},\ }\href {\doibase 10.1007/BF02214098} {\bibfield  {journal}
			{\bibinfo  {journal} {International Journal of Theoretical Physics}\ }\textbf
			{\bibinfo {volume} {23}},\ \bibinfo {pages} {701} (\bibinfo {year}
			{1984})}\BibitemShut {NoStop}%
		\bibitem [{\citenamefont {Höhn}(2019)}]{universe5050116}%
		\BibitemOpen
		\bibfield  {author} {\bibinfo {author} {\bibfnamefont {P.~A.}\ \bibnamefont
				{Höhn}},\ }\href {\doibase 10.3390/universe5050116} {\bibfield  {journal}
			{\bibinfo  {journal} {Universe}\ }\textbf {\bibinfo {volume} {5}} (\bibinfo
			{year} {2019}),\ 10.3390/universe5050116}\BibitemShut {NoStop}%
		\bibitem [{\citenamefont {H\"ohn}\ and\ \citenamefont
			{Vanrietvelde}(2020)}]{Hohn:2018toe}%
		\BibitemOpen
		\bibfield  {author} {\bibinfo {author} {\bibfnamefont {P.~A.}\ \bibnamefont
				{H\"ohn}}\ and\ \bibinfo {author} {\bibfnamefont {A.}~\bibnamefont
				{Vanrietvelde}},\ }\href {\doibase 10.1088/1367-2630/abd1ac} {\bibfield
			{journal} {\bibinfo  {journal} {New J. Phys.}\ }\textbf {\bibinfo {volume}
				{22}},\ \bibinfo {pages} {123048} (\bibinfo {year} {2020})},\ \Eprint
		{http://arxiv.org/abs/1810.04153} {arXiv:1810.04153 [gr-qc]} \BibitemShut
		{NoStop}%
		\bibitem [{\citenamefont {Ashtekar}\ \emph {et~al.}(2009)\citenamefont
			{Ashtekar}, \citenamefont {Kaminski},\ and\ \citenamefont
			{Lewandowski}}]{Ashtekar:2009mb}%
		\BibitemOpen
		\bibfield  {author} {\bibinfo {author} {\bibfnamefont {A.}~\bibnamefont
				{Ashtekar}}, \bibinfo {author} {\bibfnamefont {W.}~\bibnamefont {Kaminski}},
			\ and\ \bibinfo {author} {\bibfnamefont {J.}~\bibnamefont {Lewandowski}},\
		}\href {\doibase 10.1103/PhysRevD.79.064030} {\bibfield  {journal} {\bibinfo
				{journal} {Phys. Rev. D}\ }\textbf {\bibinfo {volume} {79}},\ \bibinfo
			{pages} {064030} (\bibinfo {year} {2009})},\ \Eprint
		{http://arxiv.org/abs/0901.0933} {arXiv:0901.0933 [gr-qc]} \BibitemShut
		{NoStop}%
		\bibitem [{\citenamefont {Agullo}\ \emph {et~al.}(2012)\citenamefont {Agullo},
			\citenamefont {Ashtekar},\ and\ \citenamefont {Nelson}}]{Agullo:2012sh}%
		\BibitemOpen
		\bibfield  {author} {\bibinfo {author} {\bibfnamefont {I.}~\bibnamefont
				{Agullo}}, \bibinfo {author} {\bibfnamefont {A.}~\bibnamefont {Ashtekar}}, \
			and\ \bibinfo {author} {\bibfnamefont {W.}~\bibnamefont {Nelson}},\ }\href
		{\doibase 10.1103/PhysRevLett.109.251301} {\bibfield  {journal} {\bibinfo
				{journal} {Phys. Rev. Lett.}\ }\textbf {\bibinfo {volume} {109}},\ \bibinfo
			{pages} {251301} (\bibinfo {year} {2012})},\ \Eprint
		{http://arxiv.org/abs/1209.1609} {arXiv:1209.1609 [gr-qc]} \BibitemShut
		{NoStop}%
		\bibitem [{\citenamefont {Agullo}\ \emph
			{et~al.}(2013{\natexlab{a}})\citenamefont {Agullo}, \citenamefont
			{Ashtekar},\ and\ \citenamefont {Nelson}}]{Agullo:2012fc}%
		\BibitemOpen
		\bibfield  {author} {\bibinfo {author} {\bibfnamefont {I.}~\bibnamefont
				{Agullo}}, \bibinfo {author} {\bibfnamefont {A.}~\bibnamefont {Ashtekar}}, \
			and\ \bibinfo {author} {\bibfnamefont {W.}~\bibnamefont {Nelson}},\ }\href
		{\doibase 10.1103/PhysRevD.87.043507} {\bibfield  {journal} {\bibinfo
				{journal} {Phys. Rev. D}\ }\textbf {\bibinfo {volume} {87}},\ \bibinfo
			{pages} {043507} (\bibinfo {year} {2013}{\natexlab{a}})},\ \Eprint
		{http://arxiv.org/abs/1211.1354} {arXiv:1211.1354 [gr-qc]} \BibitemShut
		{NoStop}%
		\bibitem [{\citenamefont {Agullo}\ \emph
			{et~al.}(2013{\natexlab{b}})\citenamefont {Agullo}, \citenamefont
			{Ashtekar},\ and\ \citenamefont {Nelson}}]{Agullo:2013ai}%
		\BibitemOpen
		\bibfield  {author} {\bibinfo {author} {\bibfnamefont {I.}~\bibnamefont
				{Agullo}}, \bibinfo {author} {\bibfnamefont {A.}~\bibnamefont {Ashtekar}}, \
			and\ \bibinfo {author} {\bibfnamefont {W.}~\bibnamefont {Nelson}},\ }\href
		{\doibase 10.1088/0264-9381/30/8/085014} {\bibfield  {journal} {\bibinfo
				{journal} {Class. Quant. Grav.}\ }\textbf {\bibinfo {volume} {30}},\ \bibinfo
			{pages} {085014} (\bibinfo {year} {2013}{\natexlab{b}})},\ \Eprint
		{http://arxiv.org/abs/1302.0254} {arXiv:1302.0254 [gr-qc]} \BibitemShut
		{NoStop}%
		\bibitem [{\citenamefont {DeWitt}(1967)}]{dewitt_quantum_1967}%
		\BibitemOpen
		\bibfield  {author} {\bibinfo {author} {\bibfnamefont {B.~S.}\ \bibnamefont
				{DeWitt}},\ }\href {\doibase 10.1103/PhysRev.160.1113} {\bibfield  {journal}
			{\bibinfo  {journal} {Physical Review}\ }\textbf {\bibinfo {volume} {160}},\
			\bibinfo {pages} {1113} (\bibinfo {year} {1967})}\BibitemShut {NoStop}%
		\bibitem [{\citenamefont {Kiefer}(2012)}]{kiefer_quantum_2012}%
		\BibitemOpen
		\bibfield  {author} {\bibinfo {author} {\bibfnamefont {C.}~\bibnamefont
				{Kiefer}},\ }\href@noop {} {\emph {\bibinfo {title} {Quantum gravity}}},\
		\bibinfo {edition} {third edition}\ ed.,\ \bibinfo {series} {International
			series of monographs on physics}\ No.\ \bibinfo {number} {155}\ (\bibinfo
		{publisher} {Oxford University Press},\ \bibinfo {address} {Oxford},\
		\bibinfo {year} {2012})\BibitemShut {NoStop}%
		\bibitem [{\citenamefont {Kiefer}\ \emph
			{et~al.}(2019{\natexlab{a}})\citenamefont {Kiefer}, \citenamefont
			{Kwidzinski},\ and\ \citenamefont {Piontek}}]{kiefer_singularity_2019-1}%
		\BibitemOpen
		\bibfield  {author} {\bibinfo {author} {\bibfnamefont {C.}~\bibnamefont
				{Kiefer}}, \bibinfo {author} {\bibfnamefont {N.}~\bibnamefont {Kwidzinski}},
			\ and\ \bibinfo {author} {\bibfnamefont {D.}~\bibnamefont {Piontek}},\ }\href
		{\doibase 10.1140/epjc/s10052-019-7193-6} {\bibfield  {journal} {\bibinfo
				{journal} {The European Physical Journal C}\ }\textbf {\bibinfo {volume}
				{79}},\ \bibinfo {pages} {686} (\bibinfo {year}
			{2019}{\natexlab{a}})}\BibitemShut {NoStop}%
		\bibitem [{\citenamefont {Ashtekar}\ \emph
			{et~al.}(2006{\natexlab{a}})\citenamefont {Ashtekar}, \citenamefont
			{Pawlowski},\ and\ \citenamefont {Singh}}]{Ashtekar:2006uz}%
		\BibitemOpen
		\bibfield  {author} {\bibinfo {author} {\bibfnamefont {A.}~\bibnamefont
				{Ashtekar}}, \bibinfo {author} {\bibfnamefont {T.}~\bibnamefont {Pawlowski}},
			\ and\ \bibinfo {author} {\bibfnamefont {P.}~\bibnamefont {Singh}},\ }\href
		{\doibase 10.1103/PhysRevD.73.124038} {\bibfield  {journal} {\bibinfo
				{journal} {Phys. Rev. D}\ }\textbf {\bibinfo {volume} {73}},\ \bibinfo
			{pages} {124038} (\bibinfo {year} {2006}{\natexlab{a}})},\ \Eprint
		{http://arxiv.org/abs/gr-qc/0604013} {arXiv:gr-qc/0604013} \BibitemShut
		{NoStop}%
		\bibitem [{\citenamefont {Ashtekar}\ \emph
			{et~al.}(2006{\natexlab{b}})\citenamefont {Ashtekar}, \citenamefont
			{Pawlowski},\ and\ \citenamefont {Singh}}]{Ashtekar:2006wn}%
		\BibitemOpen
		\bibfield  {author} {\bibinfo {author} {\bibfnamefont {A.}~\bibnamefont
				{Ashtekar}}, \bibinfo {author} {\bibfnamefont {T.}~\bibnamefont {Pawlowski}},
			\ and\ \bibinfo {author} {\bibfnamefont {P.}~\bibnamefont {Singh}},\ }\href
		{\doibase 10.1103/PhysRevD.74.084003} {\bibfield  {journal} {\bibinfo
				{journal} {Phys. Rev. D}\ }\textbf {\bibinfo {volume} {74}},\ \bibinfo
			{pages} {084003} (\bibinfo {year} {2006}{\natexlab{b}})},\ \Eprint
		{http://arxiv.org/abs/gr-qc/0607039} {arXiv:gr-qc/0607039} \BibitemShut
		{NoStop}%
		\bibitem [{\citenamefont {Peter}\ \emph {et~al.}(2005)\citenamefont {Peter},
			\citenamefont {Pinho},\ and\ \citenamefont {Pinto-Neto}}]{Peter:2005hm}%
		\BibitemOpen
		\bibfield  {author} {\bibinfo {author} {\bibfnamefont {P.}~\bibnamefont
				{Peter}}, \bibinfo {author} {\bibfnamefont {E.}~\bibnamefont {Pinho}}, \ and\
			\bibinfo {author} {\bibfnamefont {N.}~\bibnamefont {Pinto-Neto}},\ }\href
		{\doibase 10.1088/1475-7516/2005/07/014} {\bibfield  {journal} {\bibinfo
				{journal} {JCAP}\ }\textbf {\bibinfo {volume} {07}},\ \bibinfo {pages} {014}
			(\bibinfo {year} {2005})},\ \Eprint {http://arxiv.org/abs/hep-th/0509232}
		{arXiv:hep-th/0509232} \BibitemShut {NoStop}%
		\bibitem [{\citenamefont {Peter}\ \emph {et~al.}(2006)\citenamefont {Peter},
			\citenamefont {Pinho},\ and\ \citenamefont {Pinto-Neto}}]{Peter:2006id}%
		\BibitemOpen
		\bibfield  {author} {\bibinfo {author} {\bibfnamefont {P.}~\bibnamefont
				{Peter}}, \bibinfo {author} {\bibfnamefont {E.~J.~C.}\ \bibnamefont {Pinho}},
			\ and\ \bibinfo {author} {\bibfnamefont {N.}~\bibnamefont {Pinto-Neto}},\
		}\href {\doibase 10.1103/PhysRevD.73.104017} {\bibfield  {journal} {\bibinfo
				{journal} {Phys. Rev. D}\ }\textbf {\bibinfo {volume} {73}},\ \bibinfo
			{pages} {104017} (\bibinfo {year} {2006})},\ \Eprint
		{http://arxiv.org/abs/gr-qc/0605060} {arXiv:gr-qc/0605060} \BibitemShut
		{NoStop}%
		\bibitem [{\citenamefont {Peter}\ \emph {et~al.}(2007)\citenamefont {Peter},
			\citenamefont {Pinho},\ and\ \citenamefont {Pinto-Neto}}]{Peter:2006hx}%
		\BibitemOpen
		\bibfield  {author} {\bibinfo {author} {\bibfnamefont {P.}~\bibnamefont
				{Peter}}, \bibinfo {author} {\bibfnamefont {E.~J.~C.}\ \bibnamefont {Pinho}},
			\ and\ \bibinfo {author} {\bibfnamefont {N.}~\bibnamefont {Pinto-Neto}},\
		}\href {\doibase 10.1103/PhysRevD.75.023516} {\bibfield  {journal} {\bibinfo
				{journal} {Phys. Rev. D}\ }\textbf {\bibinfo {volume} {75}},\ \bibinfo
			{pages} {023516} (\bibinfo {year} {2007})},\ \Eprint
		{http://arxiv.org/abs/hep-th/0610205} {arXiv:hep-th/0610205} \BibitemShut
		{NoStop}%
		\bibitem [{\citenamefont {Peter}\ and\ \citenamefont
			{Pinto-Neto}(2008)}]{Peter:2008qz}%
		\BibitemOpen
		\bibfield  {author} {\bibinfo {author} {\bibfnamefont {P.}~\bibnamefont
				{Peter}}\ and\ \bibinfo {author} {\bibfnamefont {N.}~\bibnamefont
				{Pinto-Neto}},\ }\href {\doibase 10.1103/PhysRevD.78.063506} {\bibfield
			{journal} {\bibinfo  {journal} {Phys. Rev. D}\ }\textbf {\bibinfo {volume}
				{78}},\ \bibinfo {pages} {063506} (\bibinfo {year} {2008})},\ \Eprint
		{http://arxiv.org/abs/0809.2022} {arXiv:0809.2022 [gr-qc]} \BibitemShut
		{NoStop}%
		\bibitem [{\citenamefont {Bergeron}\ \emph {et~al.}(2018)\citenamefont
			{Bergeron}, \citenamefont {Gazeau},\ and\ \citenamefont
			{Ma\l{}kiewicz}}]{Bergeron:2017ddo}%
		\BibitemOpen
		\bibfield  {author} {\bibinfo {author} {\bibfnamefont {H.}~\bibnamefont
				{Bergeron}}, \bibinfo {author} {\bibfnamefont {J.~P.}\ \bibnamefont
				{Gazeau}}, \ and\ \bibinfo {author} {\bibfnamefont {P.}~\bibnamefont
				{Ma\l{}kiewicz}},\ }\href {\doibase 10.1088/1475-7516/2018/05/057} {\bibfield
			{journal} {\bibinfo  {journal} {JCAP}\ }\textbf {\bibinfo {volume} {05}},\
			\bibinfo {pages} {057} (\bibinfo {year} {2018})},\ \Eprint
		{http://arxiv.org/abs/1709.05851} {arXiv:1709.05851 [gr-qc]} \BibitemShut
		{NoStop}%
		\bibitem [{\citenamefont {Ma\l{}kiewicz}\ and\ \citenamefont
			{Miroszewski}(2021)}]{Malkiewicz:2020fvy}%
		\BibitemOpen
		\bibfield  {author} {\bibinfo {author} {\bibfnamefont {P.}~\bibnamefont
				{Ma\l{}kiewicz}}\ and\ \bibinfo {author} {\bibfnamefont {A.}~\bibnamefont
				{Miroszewski}},\ }\href {\doibase 10.1103/PhysRevD.103.083529} {\bibfield
			{journal} {\bibinfo  {journal} {Phys. Rev. D}\ }\textbf {\bibinfo {volume}
				{103}},\ \bibinfo {pages} {083529} (\bibinfo {year} {2021})},\ \Eprint
		{http://arxiv.org/abs/2011.03487} {arXiv:2011.03487 [gr-qc]} \BibitemShut
		{NoStop}%
		\bibitem [{\citenamefont {Martin}\ \emph
			{et~al.}(2022{\natexlab{a}})\citenamefont {Martin}, \citenamefont
			{Ma\l{}kiewicz},\ and\ \citenamefont {Peter}}]{Martin:2021dbz}%
		\BibitemOpen
		\bibfield  {author} {\bibinfo {author} {\bibfnamefont {J.~d.~C.}\
				\bibnamefont {Martin}}, \bibinfo {author} {\bibfnamefont {P.}~\bibnamefont
				{Ma\l{}kiewicz}}, \ and\ \bibinfo {author} {\bibfnamefont {P.}~\bibnamefont
				{Peter}},\ }\href {\doibase 10.1103/PhysRevD.105.023522} {\bibfield
			{journal} {\bibinfo  {journal} {Phys. Rev. D}\ }\textbf {\bibinfo {volume}
				{105}},\ \bibinfo {pages} {023522} (\bibinfo {year} {2022}{\natexlab{a}})},\
		\Eprint {http://arxiv.org/abs/2111.02963} {arXiv:2111.02963 [gr-qc]}
		\BibitemShut {NoStop}%
		\bibitem [{\citenamefont {Martin}\ \emph
			{et~al.}(2022{\natexlab{b}})\citenamefont {Martin}, \citenamefont
			{Ma\l{}kiewicz},\ and\ \citenamefont {Peter}}]{Martin:2022ptk}%
		\BibitemOpen
		\bibfield  {author} {\bibinfo {author} {\bibfnamefont {J.~d.~C.}\
				\bibnamefont {Martin}}, \bibinfo {author} {\bibfnamefont {P.}~\bibnamefont
				{Ma\l{}kiewicz}}, \ and\ \bibinfo {author} {\bibfnamefont {P.}~\bibnamefont
				{Peter}},\ }\href@noop {} {\  (\bibinfo {year} {2022}{\natexlab{b}})},\
		\Eprint {http://arxiv.org/abs/2212.12484} {arXiv:2212.12484 [gr-qc]}
		\BibitemShut {NoStop}%
		\bibitem [{\citenamefont {Blyth}\ and\ \citenamefont
			{Isham}(1975)}]{PhysRevD.11.768}%
		\BibitemOpen
		\bibfield  {author} {\bibinfo {author} {\bibfnamefont {W.~F.}\ \bibnamefont
				{Blyth}}\ and\ \bibinfo {author} {\bibfnamefont {C.~J.}\ \bibnamefont
				{Isham}},\ }\href {\doibase 10.1103/PhysRevD.11.768} {\bibfield  {journal}
			{\bibinfo  {journal} {Phys. Rev. D}\ }\textbf {\bibinfo {volume} {11}},\
			\bibinfo {pages} {768} (\bibinfo {year} {1975})}\BibitemShut {NoStop}%
		\bibitem [{\citenamefont {Gotay}\ and\ \citenamefont
			{Demaret}(1983)}]{PhysRevD.28.2402}%
		\BibitemOpen
		\bibfield  {author} {\bibinfo {author} {\bibfnamefont {M.~J.}\ \bibnamefont
				{Gotay}}\ and\ \bibinfo {author} {\bibfnamefont {J.}~\bibnamefont
				{Demaret}},\ }\href {\doibase 10.1103/PhysRevD.28.2402} {\bibfield  {journal}
			{\bibinfo  {journal} {Phys. Rev. D}\ }\textbf {\bibinfo {volume} {28}},\
			\bibinfo {pages} {2402} (\bibinfo {year} {1983})}\BibitemShut {NoStop}%
		\bibitem [{\citenamefont {H{\'a}j{\'i}{\v c}ek}\ and\ \citenamefont
			{Kiefer}(2001)}]{hajicek_singularity_2001}%
		\BibitemOpen
		\bibfield  {author} {\bibinfo {author} {\bibfnamefont {P.}~\bibnamefont
				{H{\'a}j{\'i}{\v c}ek}}\ and\ \bibinfo {author} {\bibfnamefont
				{C.}~\bibnamefont {Kiefer}},\ }\href {\doibase 10.1142/S0218271801001578}
		{\bibfield  {journal} {\bibinfo  {journal} {International Journal of Modern
					Physics D}\ }\textbf {\bibinfo {volume} {10}},\ \bibinfo {pages} {775}
			(\bibinfo {year} {2001})}\BibitemShut {NoStop}%
		\bibitem [{\citenamefont {Dabrowski}\ \emph {et~al.}(2006)\citenamefont
			{Dabrowski}, \citenamefont {Kiefer},\ and\ \citenamefont
			{Sandh{\"o}fer}}]{Dabrowski:2006dd}%
		\BibitemOpen
		\bibfield  {author} {\bibinfo {author} {\bibfnamefont {M.~P.}\ \bibnamefont
				{Dabrowski}}, \bibinfo {author} {\bibfnamefont {C.}~\bibnamefont {Kiefer}}, \
			and\ \bibinfo {author} {\bibfnamefont {B.}~\bibnamefont {Sandh{\"o}fer}},\
		}\href {\doibase 10.1103/PhysRevD.74.044022} {\bibfield  {journal} {\bibinfo
				{journal} {Phys. Rev. D}\ }\textbf {\bibinfo {volume} {74}},\ \bibinfo
			{pages} {044022} (\bibinfo {year} {2006})},\ \Eprint
		{http://arxiv.org/abs/hep-th/0605229} {arXiv:hep-th/0605229} \BibitemShut
		{NoStop}%
		\bibitem [{\citenamefont {Bergeron}\ \emph {et~al.}(2015)\citenamefont
			{Bergeron}, \citenamefont {Czuchry}, \citenamefont {Gazeau}, \citenamefont
			{Ma{\l }kiewicz},\ and\ \citenamefont
			{Piechocki}}]{bergeron_singularity_2015}%
		\BibitemOpen
		\bibfield  {author} {\bibinfo {author} {\bibfnamefont {H.}~\bibnamefont
				{Bergeron}}, \bibinfo {author} {\bibfnamefont {E.}~\bibnamefont {Czuchry}},
			\bibinfo {author} {\bibfnamefont {J.-P.}\ \bibnamefont {Gazeau}}, \bibinfo
			{author} {\bibfnamefont {P.}~\bibnamefont {Ma{\l }kiewicz}}, \ and\ \bibinfo
			{author} {\bibfnamefont {W.}~\bibnamefont {Piechocki}},\ }\href {\doibase
			10.1103/PhysRevD.92.124018} {\bibfield  {journal} {\bibinfo  {journal}
				{Physical Review D}\ }\textbf {\bibinfo {volume} {92}},\ \bibinfo {pages}
			{124018} (\bibinfo {year} {2015})}\BibitemShut {NoStop}%
		\bibitem [{\citenamefont {Liu}\ \emph {et~al.}(2014)\citenamefont {Liu},
			\citenamefont {Malafarina}, \citenamefont {Modesto},\ and\ \citenamefont
			{Bambi}}]{liu_singularity_2014}%
		\BibitemOpen
		\bibfield  {author} {\bibinfo {author} {\bibfnamefont {Y.}~\bibnamefont
				{Liu}}, \bibinfo {author} {\bibfnamefont {D.}~\bibnamefont {Malafarina}},
			\bibinfo {author} {\bibfnamefont {L.}~\bibnamefont {Modesto}}, \ and\
			\bibinfo {author} {\bibfnamefont {C.}~\bibnamefont {Bambi}},\ }\href
		{\doibase 10.1103/PhysRevD.90.044040} {\bibfield  {journal} {\bibinfo
				{journal} {Physical Review D}\ }\textbf {\bibinfo {volume} {90}},\ \bibinfo
			{pages} {044040} (\bibinfo {year} {2014})}\BibitemShut {NoStop}%
		\bibitem [{\citenamefont {Kamenshchik}(2013)}]{kamenshchik_quantum_2013}%
		\BibitemOpen
		\bibfield  {author} {\bibinfo {author} {\bibfnamefont {A.~Y.}\ \bibnamefont
				{Kamenshchik}},\ }\href {\doibase 10.1088/0264-9381/30/17/173001} {\bibfield
			{journal} {\bibinfo  {journal} {Classical and Quantum Gravity}\ }\textbf
			{\bibinfo {volume} {30}},\ \bibinfo {pages} {173001} (\bibinfo {year}
			{2013})}\BibitemShut {NoStop}%
		\bibitem [{\citenamefont {Sami}\ \emph {et~al.}(2006)\citenamefont {Sami},
			\citenamefont {Singh},\ and\ \citenamefont
			{Tsujikawa}}]{sami_avoidance_2006}%
		\BibitemOpen
		\bibfield  {author} {\bibinfo {author} {\bibfnamefont {M.}~\bibnamefont
				{Sami}}, \bibinfo {author} {\bibfnamefont {P.}~\bibnamefont {Singh}}, \ and\
			\bibinfo {author} {\bibfnamefont {S.}~\bibnamefont {Tsujikawa}},\ }\href
		{\doibase 10.1103/PhysRevD.74.043514} {\bibfield  {journal} {\bibinfo
				{journal} {Physical Review D}\ }\textbf {\bibinfo {volume} {74}},\ \bibinfo
			{pages} {043514} (\bibinfo {year} {2006})}\BibitemShut {NoStop}%
		\bibitem [{\citenamefont {Kiefer}(2015)}]{kiefer_quantum_2015}%
		\BibitemOpen
		\bibfield  {author} {\bibinfo {author} {\bibfnamefont {C.}~\bibnamefont
				{Kiefer}},\ }\href {http://arxiv.org/abs/1512.08346} {\bibfield  {journal}
			{\bibinfo  {journal} {arXiv:1512.08346 [gr-qc]}\ } (\bibinfo {year}
			{2015})},\ \bibinfo {note} {arXiv: 1512.08346}\BibitemShut {NoStop}%
		\bibitem [{\citenamefont {Kamenshchik}\ \emph {et~al.}(2016)\citenamefont
			{Kamenshchik}, \citenamefont {Kiefer},\ and\ \citenamefont
			{Kwidzinski}}]{Kamenshchik:2016rtr}%
		\BibitemOpen
		\bibfield  {author} {\bibinfo {author} {\bibfnamefont {A.}~\bibnamefont
				{Kamenshchik}}, \bibinfo {author} {\bibfnamefont {C.}~\bibnamefont {Kiefer}},
			\ and\ \bibinfo {author} {\bibfnamefont {N.}~\bibnamefont {Kwidzinski}},\
		}\href {\doibase 10.1103/PhysRevD.93.083519} {\bibfield  {journal} {\bibinfo
				{journal} {Phys. Rev. D}\ }\textbf {\bibinfo {volume} {93}},\ \bibinfo
			{pages} {083519} (\bibinfo {year} {2016})},\ \Eprint
		{http://arxiv.org/abs/1602.01319} {arXiv:1602.01319 [gr-qc]} \BibitemShut
		{NoStop}%
		\bibitem [{\citenamefont {Alonso-Serrano}\ \emph {et~al.}(2018)\citenamefont
			{Alonso-Serrano}, \citenamefont {Bouhmadi-L\'opez},\ and\ \citenamefont
			{Mart\'\i{}n-Moruno}}]{Alonso-Serrano:2018zpi}%
		\BibitemOpen
		\bibfield  {author} {\bibinfo {author} {\bibfnamefont {A.}~\bibnamefont
				{Alonso-Serrano}}, \bibinfo {author} {\bibfnamefont {M.}~\bibnamefont
				{Bouhmadi-L\'opez}}, \ and\ \bibinfo {author} {\bibfnamefont
				{P.}~\bibnamefont {Mart\'\i{}n-Moruno}},\ }\href {\doibase
			10.1103/PhysRevD.98.104004} {\bibfield  {journal} {\bibinfo  {journal} {Phys.
					Rev. D}\ }\textbf {\bibinfo {volume} {98}},\ \bibinfo {pages} {104004}
			(\bibinfo {year} {2018})},\ \Eprint {http://arxiv.org/abs/1802.03290}
		{arXiv:1802.03290 [gr-qc]} \BibitemShut {NoStop}%
		\bibitem [{\citenamefont {Bouhmadi-L\'opez}\ \emph {et~al.}(2019)\citenamefont
			{Bouhmadi-L\'opez}, \citenamefont {Kiefer},\ and\ \citenamefont
			{Mart\'\i{}n-Moruno}}]{Bouhmadi-Lopez:2019zvz}%
		\BibitemOpen
		\bibfield  {author} {\bibinfo {author} {\bibfnamefont {M.}~\bibnamefont
				{Bouhmadi-L\'opez}}, \bibinfo {author} {\bibfnamefont {C.}~\bibnamefont
				{Kiefer}}, \ and\ \bibinfo {author} {\bibfnamefont {P.}~\bibnamefont
				{Mart\'\i{}n-Moruno}},\ }\href {\doibase 10.1007/s10714-019-2618-y}
		{\bibfield  {journal} {\bibinfo  {journal} {Gen. Rel. Grav.}\ }\textbf
			{\bibinfo {volume} {51}},\ \bibinfo {pages} {135} (\bibinfo {year} {2019})}
		\BibitemShut {NoStop}%
		\bibitem [{\citenamefont {Kiefer}\ and\ \citenamefont
			{Schmitz}(2019)}]{kiefer_singularity_2019}%
		\BibitemOpen
		\bibfield  {author} {\bibinfo {author} {\bibfnamefont {C.}~\bibnamefont
				{Kiefer}}\ and\ \bibinfo {author} {\bibfnamefont {T.}~\bibnamefont
				{Schmitz}},\ }\href {\doibase 10.1103/PhysRevD.99.126010} {\bibfield
			{journal} {\bibinfo  {journal} {Physical Review D}\ }\textbf {\bibinfo
				{volume} {99}},\ \bibinfo {pages} {126010} (\bibinfo {year}
			{2019})}\BibitemShut {NoStop}%
		\bibitem [{\citenamefont {Jalalzadeh}(2022)}]{Jalalzadeh:2022bgz}%
		\BibitemOpen
		\bibfield  {author} {\bibinfo {author} {\bibfnamefont {S.}~\bibnamefont
				{Jalalzadeh}},\ }\href {\doibase 10.1016/j.physletb.2022.137285} {\bibfield
			{journal} {\bibinfo  {journal} {Phys. Lett. B}\ }\textbf {\bibinfo {volume}
				{833}},\ \bibinfo {pages} {137285} (\bibinfo {year} {2022})},\ \Eprint
		{http://arxiv.org/abs/2207.00727} {arXiv:2207.00727 [gr-qc]} \BibitemShut
		{NoStop}%
		\bibitem [{\citenamefont {Kiefer}\ \emph
			{et~al.}(2019{\natexlab{b}})\citenamefont {Kiefer}, \citenamefont
			{Kwidzinski},\ and\ \citenamefont {Piontek}}]{Kiefer:2019bxk}%
		\BibitemOpen
		\bibfield  {author} {\bibinfo {author} {\bibfnamefont {C.}~\bibnamefont
				{Kiefer}}, \bibinfo {author} {\bibfnamefont {N.}~\bibnamefont {Kwidzinski}},
			\ and\ \bibinfo {author} {\bibfnamefont {D.}~\bibnamefont {Piontek}},\ }\href
		{\doibase 10.1140/epjc/s10052-019-7193-6} {\bibfield  {journal} {\bibinfo
				{journal} {Eur. Phys. J. C}\ }\textbf {\bibinfo {volume} {79}},\ \bibinfo
			{pages} {686} (\bibinfo {year} {2019}{\natexlab{b}})},\ \Eprint
		{http://arxiv.org/abs/1903.04391} {arXiv:1903.04391 [gr-qc]} \BibitemShut
		{NoStop}%
		\bibitem [{\citenamefont {Bojowald}(2001{\natexlab{a}})}]{Bojowald:2001xe}%
		\BibitemOpen
		\bibfield  {author} {\bibinfo {author} {\bibfnamefont {M.}~\bibnamefont
				{Bojowald}},\ }\href {\doibase 10.1103/PhysRevLett.86.5227} {\bibfield
			{journal} {\bibinfo  {journal} {Phys. Rev. Lett.}\ }\textbf {\bibinfo
				{volume} {86}},\ \bibinfo {pages} {5227} (\bibinfo {year}
			{2001}{\natexlab{a}})},\ \Eprint {http://arxiv.org/abs/gr-qc/0102069}
		{arXiv:gr-qc/0102069} \BibitemShut {NoStop}%
		\bibitem [{\citenamefont {Alvarenga}\ \emph {et~al.}(2002)\citenamefont
			{Alvarenga}, \citenamefont {Fabris}, \citenamefont {Lemos},\ and\
			\citenamefont {Monerat}}]{Alvarenga:2001nm}%
		\BibitemOpen
		\bibfield  {author} {\bibinfo {author} {\bibfnamefont {F.~G.}\ \bibnamefont
				{Alvarenga}}, \bibinfo {author} {\bibfnamefont {J.~C.}\ \bibnamefont
				{Fabris}}, \bibinfo {author} {\bibfnamefont {N.~A.}\ \bibnamefont {Lemos}}, \
			and\ \bibinfo {author} {\bibfnamefont {G.~A.}\ \bibnamefont {Monerat}},\
		}\href {\doibase 10.1023/A:1015986011295} {\bibfield  {journal} {\bibinfo
				{journal} {Gen. Rel. Grav.}\ }\textbf {\bibinfo {volume} {34}},\ \bibinfo
			{pages} {651} (\bibinfo {year} {2002})},\ \Eprint
		{http://arxiv.org/abs/gr-qc/0106051} {arXiv:gr-qc/0106051} \BibitemShut
		{NoStop}%
		\bibitem [{\citenamefont {Bojowald}(2001{\natexlab{b}})}]{Bojowald:2001vw}%
		\BibitemOpen
		\bibfield  {author} {\bibinfo {author} {\bibfnamefont {M.}~\bibnamefont
				{Bojowald}},\ }\href {\doibase 10.1103/PhysRevD.64.084018} {\bibfield
			{journal} {\bibinfo  {journal} {Phys. Rev. D}\ }\textbf {\bibinfo {volume}
				{64}},\ \bibinfo {pages} {084018} (\bibinfo {year} {2001}{\natexlab{b}})},\
		\Eprint {http://arxiv.org/abs/gr-qc/0105067} {arXiv:gr-qc/0105067}
		\BibitemShut {NoStop}%
		\bibitem [{\citenamefont {Bojowald}(2002)}]{Bojowald:2002gz}%
		\BibitemOpen
		\bibfield  {author} {\bibinfo {author} {\bibfnamefont {M.}~\bibnamefont
				{Bojowald}},\ }\href {\doibase 10.1088/0264-9381/19/10/313} {\bibfield
			{journal} {\bibinfo  {journal} {Class. Quant. Grav.}\ }\textbf {\bibinfo
				{volume} {19}},\ \bibinfo {pages} {2717} (\bibinfo {year} {2002})},\ \Eprint
		{http://arxiv.org/abs/gr-qc/0202077} {arXiv:gr-qc/0202077} \BibitemShut
		{NoStop}%
		\bibitem [{\citenamefont {Gryb}\ and\ \citenamefont
			{Th\'ebault}(2019)}]{Gryb:2018whn}%
		\BibitemOpen
		\bibfield  {author} {\bibinfo {author} {\bibfnamefont {S.}~\bibnamefont
				{Gryb}}\ and\ \bibinfo {author} {\bibfnamefont {K.~P.~Y.}\ \bibnamefont
				{Th\'ebault}},\ }\href {\doibase 10.1088/1361-6382/aaf823} {\bibfield
			{journal} {\bibinfo  {journal} {Class. Quant. Grav.}\ }\textbf {\bibinfo
				{volume} {36}},\ \bibinfo {pages} {035009} (\bibinfo {year} {2019})},\
		\Eprint {http://arxiv.org/abs/1801.05789} {arXiv:1801.05789 [gr-qc]}
		\BibitemShut {NoStop}%
		\bibitem [{\citenamefont {Gielen}\ and\ \citenamefont
			{Men\'endez-Pidal}(2020)}]{Gielen:2020abd}%
		\BibitemOpen
		\bibfield  {author} {\bibinfo {author} {\bibfnamefont {S.}~\bibnamefont
				{Gielen}}\ and\ \bibinfo {author} {\bibfnamefont {L.}~\bibnamefont
				{Men\'endez-Pidal}},\ }\href {\doibase 10.1088/1361-6382/abb14f} {\bibfield
			{journal} {\bibinfo  {journal} {Class. Quant. Grav.}\ }\textbf {\bibinfo
				{volume} {37}},\ \bibinfo {pages} {205018} (\bibinfo {year} {2020})},\
		\Eprint {http://arxiv.org/abs/2005.05357} {arXiv:2005.05357 [gr-qc]}
		\BibitemShut {NoStop}%
		\bibitem [{\citenamefont {Gielen}\ and\ \citenamefont
			{Men\'endez-Pidal}(2022{\natexlab{a}})}]{Gielen:2021igw}%
		\BibitemOpen
		\bibfield  {author} {\bibinfo {author} {\bibfnamefont {S.}~\bibnamefont
				{Gielen}}\ and\ \bibinfo {author} {\bibfnamefont {L.}~\bibnamefont
				{Men\'endez-Pidal}},\ }\href {\doibase 10.1088/1361-6382/ac504f} {\bibfield
			{journal} {\bibinfo  {journal} {Class. Quant. Grav.}\ }\textbf {\bibinfo
				{volume} {39}},\ \bibinfo {pages} {075011} (\bibinfo {year}
			{2022}{\natexlab{a}})},\ \Eprint {http://arxiv.org/abs/2109.02660}
		{arXiv:2109.02660 [gr-qc]} \BibitemShut {NoStop}%
		\bibitem [{\citenamefont {Gielen}\ and\ \citenamefont
			{Men\'endez-Pidal}(2022{\natexlab{b}})}]{Gielen:2022tzi}%
		\BibitemOpen
		\bibfield  {author} {\bibinfo {author} {\bibfnamefont {S.}~\bibnamefont
				{Gielen}}\ and\ \bibinfo {author} {\bibfnamefont {L.}~\bibnamefont
				{Men\'endez-Pidal}},\ }\href@noop {} {\  (\bibinfo {year}
			{2022}{\natexlab{b}})},\ \Eprint {http://arxiv.org/abs/2205.15387}
		{arXiv:2205.15387 [gr-qc]} \BibitemShut {NoStop}%
		\bibitem [{\citenamefont {Ma\l{}kiewicz}\ \emph {et~al.}(2020)\citenamefont
			{Ma\l{}kiewicz}, \citenamefont {Peter},\ and\ \citenamefont
			{Vitenti}}]{Malkiewicz:2019azw}%
		\BibitemOpen
		\bibfield  {author} {\bibinfo {author} {\bibfnamefont {P.}~\bibnamefont
				{Ma\l{}kiewicz}}, \bibinfo {author} {\bibfnamefont {P.}~\bibnamefont
				{Peter}}, \ and\ \bibinfo {author} {\bibfnamefont {S.~D.~P.}\ \bibnamefont
				{Vitenti}},\ }\href {\doibase 10.1103/PhysRevD.101.046012} {\bibfield
			{journal} {\bibinfo  {journal} {Phys. Rev. D}\ }\textbf {\bibinfo {volume}
				{101}},\ \bibinfo {pages} {046012} (\bibinfo {year} {2020})},\ \Eprint
		{http://arxiv.org/abs/1911.09892} {arXiv:1911.09892 [gr-qc]} \BibitemShut
		{NoStop}%
		\bibitem [{\citenamefont {Thebault}(2022)}]{Thebault:2022dmv}%
		\BibitemOpen
		\bibfield  {author} {\bibinfo {author} {\bibfnamefont {K.~P.~Y.}\
				\bibnamefont {Thebault}},\ }\href@noop {} {\  (\bibinfo {year} {2022})},\
		\Eprint {http://arxiv.org/abs/2209.05905} {arXiv:2209.05905 [gr-qc]}
		\BibitemShut {NoStop}%
		\bibitem [{\citenamefont {Kuchar}(1993)}]{kuchar_canonical_1993}%
		\BibitemOpen
		\bibfield  {author} {\bibinfo {author} {\bibfnamefont {K.}~\bibnamefont
				{Kuchar}},\ }\href {http://arxiv.org/abs/gr-qc/9304012} {\enquote {\bibinfo
				{title} {Canonical quantum gravity},}\ } (\bibinfo {year} {1993})\BibitemShut
		{NoStop}%
		\bibitem [{\citenamefont {Barbour}\ and\ \citenamefont
			{Foster}(2008)}]{barbour2008constraints}%
		\BibitemOpen
		\bibfield  {author} {\bibinfo {author} {\bibfnamefont {J.}~\bibnamefont
				{Barbour}}\ and\ \bibinfo {author} {\bibfnamefont {B.~Z.}\ \bibnamefont
				{Foster}},\ }\href@noop {} {\enquote {\bibinfo {title} {Constraints and gauge
					transformations: Dirac's theorem is not always valid},}\ } (\bibinfo {year}
		{2008}),\ \Eprint {http://arxiv.org/abs/0808.1223} {arXiv:0808.1223 [gr-qc]}
		\BibitemShut {NoStop}%
		\bibitem [{\citenamefont {Bojowald}\ \emph {et~al.}(2011)\citenamefont
			{Bojowald}, \citenamefont {Höhn},\ and\ \citenamefont
			{Tsobanjan}}]{Bojowald_2011}%
		\BibitemOpen
		\bibfield  {author} {\bibinfo {author} {\bibfnamefont {M.}~\bibnamefont
				{Bojowald}}, \bibinfo {author} {\bibfnamefont {P.~A.}\ \bibnamefont {Höhn}},
			\ and\ \bibinfo {author} {\bibfnamefont {A.}~\bibnamefont {Tsobanjan}},\
		}\href {\doibase 10.1103/physrevd.83.125023} {\bibfield  {journal} {\bibinfo
				{journal} {Physical Review D}\ }\textbf {\bibinfo {volume} {83}} (\bibinfo
			{year} {2011}),\ 10.1103/physrevd.83.125023}\BibitemShut {NoStop}%
		\bibitem [{\citenamefont {Brown}\ and\ \citenamefont
			{Kuchar}(1995)}]{brown_dust_1995}%
		\BibitemOpen
		\bibfield  {author} {\bibinfo {author} {\bibfnamefont {J.~D.}\ \bibnamefont
				{Brown}}\ and\ \bibinfo {author} {\bibfnamefont {K.~V.}\ \bibnamefont
				{Kuchar}},\ }\href {\doibase 10.1103/PhysRevD.51.5600} {\bibfield  {journal}
			{\bibinfo  {journal} {Physical Review D}\ }\textbf {\bibinfo {volume} {51}},\
			\bibinfo {pages} {5600} (\bibinfo {year} {1995})}\BibitemShut {NoStop}%
		\bibitem [{\citenamefont {Henneaux}\ and\ \citenamefont
			{Teitelboim}(1989)}]{Henneaux:1989zc}%
		\BibitemOpen
		\bibfield  {author} {\bibinfo {author} {\bibfnamefont {M.}~\bibnamefont
				{Henneaux}}\ and\ \bibinfo {author} {\bibfnamefont {C.}~\bibnamefont
				{Teitelboim}},\ }\href {\doibase 10.1016/0370-2693(89)91251-3} {\bibfield
			{journal} {\bibinfo  {journal} {Phys. Lett. B}\ }\textbf {\bibinfo {volume}
				{222}},\ \bibinfo {pages} {195} (\bibinfo {year} {1989})}\BibitemShut
		{NoStop}%
		\bibitem [{\citenamefont {Unruh}(1989)}]{Unruh:1988in}%
		\BibitemOpen
		\bibfield  {author} {\bibinfo {author} {\bibfnamefont {W.~G.}\ \bibnamefont
				{Unruh}},\ }\href {\doibase 10.1103/PhysRevD.40.1048} {\bibfield  {journal}
			{\bibinfo  {journal} {Phys. Rev. D}\ }\textbf {\bibinfo {volume} {40}},\
			\bibinfo {pages} {1048} (\bibinfo {year} {1989})}\BibitemShut {NoStop}%
		\bibitem [{\citenamefont {Kuchar}(1991)}]{Kuchar:1991xd}%
		\BibitemOpen
		\bibfield  {author} {\bibinfo {author} {\bibfnamefont {K.~V.}\ \bibnamefont
				{Kuchar}},\ }\href {\doibase 10.1103/PhysRevD.43.3332} {\bibfield  {journal}
			{\bibinfo  {journal} {Phys. Rev. D}\ }\textbf {\bibinfo {volume} {43}},\
			\bibinfo {pages} {3332} (\bibinfo {year} {1991})}\BibitemShut {NoStop}%
		\bibitem [{\citenamefont {Amemiya}\ and\ \citenamefont
			{Koike}(2009)}]{Amemiya_2009}%
		\BibitemOpen
		\bibfield  {author} {\bibinfo {author} {\bibfnamefont {F.}~\bibnamefont
				{Amemiya}}\ and\ \bibinfo {author} {\bibfnamefont {T.}~\bibnamefont
				{Koike}},\ }\href {\doibase 10.1103/physrevd.80.103507} {\bibfield  {journal}
			{\bibinfo  {journal} {Physical Review D}\ }\textbf {\bibinfo {volume} {80}}
			(\bibinfo {year} {2009}),\ 10.1103/physrevd.80.103507}\BibitemShut {NoStop}%
		\bibitem [{\citenamefont {Gieres}(2000)}]{Gieres_2000}%
		\BibitemOpen
		\bibfield  {author} {\bibinfo {author} {\bibfnamefont {F.}~\bibnamefont
				{Gieres}},\ }\href {\doibase 10.1088/0034-4885/63/12/201} {\bibfield
			{journal} {\bibinfo  {journal} {Reports on Progress in Physics}\ }\textbf
			{\bibinfo {volume} {63}},\ \bibinfo {pages} {1893–1931} (\bibinfo {year}
			{2000})}\BibitemShut {NoStop}%
		\bibitem [{\citenamefont {Bonneau}\ \emph {et~al.}(2001)\citenamefont
			{Bonneau}, \citenamefont {Faraut},\ and\ \citenamefont
			{Valent}}]{Bonneau_2001}%
		\BibitemOpen
		\bibfield  {author} {\bibinfo {author} {\bibfnamefont {G.}~\bibnamefont
				{Bonneau}}, \bibinfo {author} {\bibfnamefont {J.}~\bibnamefont {Faraut}}, \
			and\ \bibinfo {author} {\bibfnamefont {G.}~\bibnamefont {Valent}},\ }\href
		{\doibase 10.1119/1.1328351} {\bibfield  {journal} {\bibinfo  {journal}
				{American Journal of Physics}\ }\textbf {\bibinfo {volume} {69}},\ \bibinfo
			{pages} {322–331} (\bibinfo {year} {2001})}\BibitemShut {NoStop}%
		\bibitem [{\citenamefont {Gitman}\ \emph {et~al.}(2012)\citenamefont {Gitman},
			\citenamefont {Tyutin},\ and\ \citenamefont {Voronov}}]{Gitman2012}%
		\BibitemOpen
		\bibfield  {author} {\bibinfo {author} {\bibfnamefont {D.~M.}\ \bibnamefont
				{Gitman}}, \bibinfo {author} {\bibfnamefont {I.~V.}\ \bibnamefont {Tyutin}},
			\ and\ \bibinfo {author} {\bibfnamefont {B.~L.}\ \bibnamefont {Voronov}},\
		}\enquote {\bibinfo {title} {Free one-dimensional particle on an interval},}\
		in\ \href {\doibase 10.1007/978-0-8176-4662-2_6} {\emph {\bibinfo {booktitle}
				{Self-adjoint Extensions in Quantum Mechanics: General Theory and
					Applications to Schr{\"o}dinger and Dirac Equations with Singular
					Potentials}}}\ (\bibinfo  {publisher} {Birkh{\"a}user Boston},\ \bibinfo
		{address} {Boston},\ \bibinfo {year} {2012})\ pp.\ \bibinfo {pages}
		{207--236}\BibitemShut {NoStop}%
		\bibitem [{\citenamefont {Kucha\ifmmode~\check{r}\else \v{r}\fi{}}\ and\
			\citenamefont {Ryan}(1989)}]{PhysRevD.40.3982}%
		\BibitemOpen
		\bibfield  {author} {\bibinfo {author} {\bibfnamefont {K.~V.}\ \bibnamefont
				{Kucha\ifmmode~\check{r}\else \v{r}\fi{}}}\ and\ \bibinfo {author}
			{\bibfnamefont {M.~P.}\ \bibnamefont {Ryan}},\ }\href {\doibase
			10.1103/PhysRevD.40.3982} {\bibfield  {journal} {\bibinfo  {journal} {Phys.
					Rev. D}\ }\textbf {\bibinfo {volume} {40}},\ \bibinfo {pages} {3982}
			(\bibinfo {year} {1989})}\BibitemShut {NoStop}%
		\bibitem [{\citenamefont {Rovelli}(1990)}]{Rovelli_1990}%
		\BibitemOpen
		\bibfield  {author} {\bibinfo {author} {\bibfnamefont {C.}~\bibnamefont
				{Rovelli}},\ }\href {\doibase 10.1103/PhysRevD.42.2638} {\bibfield  {journal}
			{\bibinfo  {journal} {Phys. Rev. D}\ }\textbf {\bibinfo {volume} {42}},\
			\bibinfo {pages} {2638} (\bibinfo {year} {1990})}\BibitemShut {NoStop}%
		\bibitem [{\citenamefont {Rovelli}(1991)}]{Rovelli_1991}%
		\BibitemOpen
		\bibfield  {author} {\bibinfo {author} {\bibfnamefont {C.}~\bibnamefont
				{Rovelli}},\ }\href {\doibase 10.1103/PhysRevD.43.442} {\bibfield  {journal}
			{\bibinfo  {journal} {Phys. Rev. D}\ }\textbf {\bibinfo {volume} {43}},\
			\bibinfo {pages} {442} (\bibinfo {year} {1991})}\BibitemShut {NoStop}%
		\bibitem [{\citenamefont {Dittrich}(2007)}]{Dittrich:2004cb}%
		\BibitemOpen
		\bibfield  {author} {\bibinfo {author} {\bibfnamefont {B.}~\bibnamefont
				{Dittrich}},\ }\href {\doibase 10.1007/s10714-007-0495-2} {\bibfield
			{journal} {\bibinfo  {journal} {Gen. Rel. Grav.}\ }\textbf {\bibinfo {volume}
				{39}},\ \bibinfo {pages} {1891} (\bibinfo {year} {2007})},\ \Eprint
		{http://arxiv.org/abs/gr-qc/0411013} {arXiv:gr-qc/0411013} \BibitemShut
		{NoStop}%
		\bibitem [{\citenamefont {Dittrich}(2006)}]{Dittrich:2005kc}%
		\BibitemOpen
		\bibfield  {author} {\bibinfo {author} {\bibfnamefont {B.}~\bibnamefont
				{Dittrich}},\ }\href {\doibase 10.1088/0264-9381/23/22/006} {\bibfield
			{journal} {\bibinfo  {journal} {Class. Quant. Grav.}\ }\textbf {\bibinfo
				{volume} {23}},\ \bibinfo {pages} {6155} (\bibinfo {year} {2006})},\ \Eprint
		{http://arxiv.org/abs/gr-qc/0507106} {arXiv:gr-qc/0507106} \BibitemShut
		{NoStop}%
		\bibitem [{\citenamefont {Barvinsky}(1993)}]{BARVINSKY1993237}%
		\BibitemOpen
		\bibfield  {author} {\bibinfo {author} {\bibfnamefont {A.}~\bibnamefont
				{Barvinsky}},\ }\href {\doibase https://doi.org/10.1016/0370-1573(93)90032-9}
		{\bibfield  {journal} {\bibinfo  {journal} {Physics Reports}\ }\textbf
			{\bibinfo {volume} {230}},\ \bibinfo {pages} {237} (\bibinfo {year}
			{1993})}\BibitemShut {NoStop}%
		\bibitem [{\citenamefont {Barvinsky}\ and\ \citenamefont
			{Kamenshchik}(2014)}]{Barvinsky:2013aya}%
		\BibitemOpen
		\bibfield  {author} {\bibinfo {author} {\bibfnamefont {A.~O.}\ \bibnamefont
				{Barvinsky}}\ and\ \bibinfo {author} {\bibfnamefont {A.~Y.}\ \bibnamefont
				{Kamenshchik}},\ }\href {\doibase 10.1103/PhysRevD.89.043526} {\bibfield
			{journal} {\bibinfo  {journal} {Phys. Rev. D}\ }\textbf {\bibinfo {volume}
				{89}},\ \bibinfo {pages} {043526} (\bibinfo {year} {2014})},\ \Eprint
		{http://arxiv.org/abs/1312.3147} {arXiv:1312.3147 [gr-qc]} \BibitemShut
		{NoStop}%
            \bibitem [{\citenamefont {Torre}(1993)}]{Torre:1993fq}%
  \BibitemOpen
  \bibfield  {author} {\bibinfo {author} {\bibfnamefont {C.~G.}\ \bibnamefont
  {Torre}},\ }\href {\doibase 10.1103/PhysRevD.48.R2373} {\bibfield  {journal}
  {\bibinfo  {journal} {Phys. Rev. D}\ }\textbf {\bibinfo {volume} {48}},\
  \bibinfo {pages} {R2373} (\bibinfo {year} {1993})},\ \Eprint
  {http://arxiv.org/abs/gr-qc/9306030} {arXiv:gr-qc/9306030} \BibitemShut
  {NoStop}%
		\bibitem [{\citenamefont {Maeda}(2015)}]{maeda_unitary_2015}%
		\BibitemOpen
		\bibfield  {author} {\bibinfo {author} {\bibfnamefont {H.}~\bibnamefont
				{Maeda}},\ }\href {\doibase 10.1088/0264-9381/32/23/235023} {\bibfield
			{journal} {\bibinfo  {journal} {Classical and Quantum Gravity}\ }\textbf
			{\bibinfo {volume} {32}},\ \bibinfo {pages} {235023} (\bibinfo {year}
			{2015})}\BibitemShut {NoStop}%
		\bibitem [{\citenamefont {Vilenkin}(1988)}]{PhysRevD.37.888}%
		\BibitemOpen
		\bibfield  {author} {\bibinfo {author} {\bibfnamefont {A.}~\bibnamefont
				{Vilenkin}},\ }\href {\doibase 10.1103/PhysRevD.37.888} {\bibfield  {journal}
			{\bibinfo  {journal} {Phys. Rev. D}\ }\textbf {\bibinfo {volume} {37}},\
			\bibinfo {pages} {888} (\bibinfo {year} {1988})}\BibitemShut {NoStop}%
		\bibitem [{\citenamefont {Sahota}\ and\ \citenamefont
			{Lochan}(2021)}]{PhysRevD.104.126027}%
		\BibitemOpen
		\bibfield  {author} {\bibinfo {author} {\bibfnamefont {H.~S.}\ \bibnamefont
				{Sahota}}\ and\ \bibinfo {author} {\bibfnamefont {K.}~\bibnamefont
				{Lochan}},\ }\href {\doibase 10.1103/PhysRevD.104.126027} {\bibfield
			{journal} {\bibinfo  {journal} {Phys. Rev. D}\ }\textbf {\bibinfo {volume}
				{104}},\ \bibinfo {pages} {126027} (\bibinfo {year} {2021})}\BibitemShut
		{NoStop}%
		\bibitem [{\citenamefont {Mukhanov}\ \emph {et~al.}(1992)\citenamefont
			{Mukhanov}, \citenamefont {Feldman},\ and\ \citenamefont
			{Brandenberger}}]{Mukhanov:1990me}%
		\BibitemOpen
		\bibfield  {author} {\bibinfo {author} {\bibfnamefont {V.~F.}\ \bibnamefont
				{Mukhanov}}, \bibinfo {author} {\bibfnamefont {H.~A.}\ \bibnamefont
				{Feldman}}, \ and\ \bibinfo {author} {\bibfnamefont {R.~H.}\ \bibnamefont
				{Brandenberger}},\ }\href {\doibase 10.1016/0370-1573(92)90044-Z} {\bibfield
			{journal} {\bibinfo  {journal} {Phys. Rept.}\ }\textbf {\bibinfo {volume}
				{215}},\ \bibinfo {pages} {203} (\bibinfo {year} {1992})}\BibitemShut
		{NoStop}%
		\bibitem [{\citenamefont {Kodama}\ and\ \citenamefont
			{Sasaki}(1984)}]{10.1143/PTPS.78.1}%
		\BibitemOpen
		\bibfield  {author} {\bibinfo {author} {\bibfnamefont {H.}~\bibnamefont
				{Kodama}}\ and\ \bibinfo {author} {\bibfnamefont {M.}~\bibnamefont
				{Sasaki}},\ }\href {\doibase 10.1143/PTPS.78.1} {\bibfield  {journal}
			{\bibinfo  {journal} {Progress of Theoretical Physics Supplement}\ }\textbf
			{\bibinfo {volume} {78}},\ \bibinfo {pages} {1} (\bibinfo {year} {1984})},\
		\BibitemShut {NoStop}%
		\bibitem [{\citenamefont {Kerner}\ and\ \citenamefont
			{Sutcliffe}(1970)}]{Kerner_1970}%
		\BibitemOpen
		\bibfield  {author} {\bibinfo {author} {\bibfnamefont {E.~H.}\ \bibnamefont
				{Kerner}}\ and\ \bibinfo {author} {\bibfnamefont {W.~G.}\ \bibnamefont
				{Sutcliffe}},\ }\href {\doibase 10.1063/1.1665150} {\bibfield  {journal}
			{\bibinfo  {journal} {Journal of Mathematical Physics}\ }\textbf {\bibinfo
				{volume} {11}},\ \bibinfo {pages} {391} (\bibinfo {year} {1970})},\ \Eprint
		{http://arxiv.org/abs/https://doi.org/10.1063/1.1665150}
		{https://doi.org/10.1063/1.1665150} \BibitemShut {NoStop}%
		\bibitem [{\citenamefont {De~Gosson}(2018)}]{e20110869}%
		\BibitemOpen
		\bibfield  {author} {\bibinfo {author} {\bibfnamefont {M.~A.}\ \bibnamefont
				{De~Gosson}},\ }\href {\doibase 10.3390/e20110869} {\bibfield  {journal}
			{\bibinfo  {journal} {Entropy}\ }\textbf {\bibinfo {volume} {20}} (\bibinfo
			{year} {2018}),\ 10.3390/e20110869}\BibitemShut {NoStop}%
		\bibitem [{\citenamefont {Smolin}(2009)}]{Smolin:2009ti}%
		\BibitemOpen
		\bibfield  {author} {\bibinfo {author} {\bibfnamefont {L.}~\bibnamefont
				{Smolin}},\ }\href {\doibase 10.1103/PhysRevD.80.084003} {\bibfield
			{journal} {\bibinfo  {journal} {Phys. Rev. D}\ }\textbf {\bibinfo {volume}
				{80}},\ \bibinfo {pages} {084003} (\bibinfo {year} {2009})},\ \Eprint
		{http://arxiv.org/abs/0904.4841} {arXiv:0904.4841 [hep-th]} \BibitemShut
		{NoStop}%
		\bibitem [{\citenamefont {Birrell}\ and\ \citenamefont
			{Davies}(1984)}]{Birrell:1982ix}%
		\BibitemOpen
		\bibfield  {author} {\bibinfo {author} {\bibfnamefont {N.~D.}\ \bibnamefont
				{Birrell}}\ and\ \bibinfo {author} {\bibfnamefont {P.~C.~W.}\ \bibnamefont
				{Davies}},\ }\href {\doibase 10.1017/CBO9780511622632} {\emph {\bibinfo
				{title} {{Quantum Fields in Curved Space}}}},\ Cambridge Monographs on
		Mathematical Physics\ (\bibinfo  {publisher} {Cambridge Univ. Press},\
		\bibinfo {address} {Cambridge, UK},\ \bibinfo {year} {1984})\BibitemShut
		{NoStop}%
		\bibitem [{\citenamefont {Alexandre}\ and\ \citenamefont
			{Magueijo}(2022)}]{Alexandre:2022ijm}%
		\BibitemOpen
		\bibfield  {author} {\bibinfo {author} {\bibfnamefont {B.}~\bibnamefont
				{Alexandre}}\ and\ \bibinfo {author} {\bibfnamefont {J.}~\bibnamefont
				{Magueijo}},\ }\href {\doibase 10.1103/PhysRevD.106.063520} {\bibfield
			{journal} {\bibinfo  {journal} {Phys. Rev. D}\ }\textbf {\bibinfo {volume}
				{106}},\ \bibinfo {pages} {063520} (\bibinfo {year} {2022})},\ \Eprint
		{http://arxiv.org/abs/2207.03854} {arXiv:2207.03854 [gr-qc]} \BibitemShut
		{NoStop}%
		\bibitem [{\citenamefont {Dhanuka}\ and\ \citenamefont
			{Lochan}(2020)}]{Dhanuka:2020yxp}%
		\BibitemOpen
		\bibfield  {author} {\bibinfo {author} {\bibfnamefont {A.}~\bibnamefont
				{Dhanuka}}\ and\ \bibinfo {author} {\bibfnamefont {K.}~\bibnamefont
				{Lochan}},\ }\href {\doibase 10.1103/PhysRevD.102.085009} {\bibfield
			{journal} {\bibinfo  {journal} {Phys. Rev. D}\ }\textbf {\bibinfo {volume}
				{102}},\ \bibinfo {pages} {085009} (\bibinfo {year} {2020})},\ \Eprint
		{http://arxiv.org/abs/2003.07380} {arXiv:2003.07380 [gr-qc]} \BibitemShut
		{NoStop}%
		\bibitem [{\citenamefont {Dhanuka}\ and\ \citenamefont
			{Lochan}(2022)}]{Dhanuka:2022ggi}%
		\BibitemOpen
		\bibfield  {author} {\bibinfo {author} {\bibfnamefont {A.}~\bibnamefont
				{Dhanuka}}\ and\ \bibinfo {author} {\bibfnamefont {K.}~\bibnamefont
				{Lochan}},\ }\href {\doibase 10.1103/PhysRevD.106.125006} {\bibfield
			{journal} {\bibinfo  {journal} {Phys. Rev. D}\ }\textbf {\bibinfo {volume}
				{106}},\ \bibinfo {pages} {125006} (\bibinfo {year} {2022})},\ \Eprint
		{http://arxiv.org/abs/2210.11186} {arXiv:2210.11186 [gr-qc]} \BibitemShut
		{NoStop}%
		\bibitem [{\citenamefont {Ashtekar}\ \emph {et~al.}(2005)\citenamefont
			{Ashtekar}, \citenamefont {Bombelli},\ and\ \citenamefont
			{Corichi}}]{Ashtekar:2005}%
		\BibitemOpen
		\bibfield  {author} {\bibinfo {author} {\bibfnamefont {A.}~\bibnamefont
				{Ashtekar}}, \bibinfo {author} {\bibfnamefont {L.}~\bibnamefont {Bombelli}},
			\ and\ \bibinfo {author} {\bibfnamefont {A.}~\bibnamefont {Corichi}},\ }\href
		{\doibase 10.1103/PhysRevD.72.025008} {\bibfield  {journal} {\bibinfo
				{journal} {Phys. Rev. D}\ }\textbf {\bibinfo {volume} {72}},\ \bibinfo
			{pages} {025008} (\bibinfo {year} {2005})},\ \Eprint
		{http://arxiv.org/abs/gr-qc/0504052} {arXiv:gr-qc/0504052} \BibitemShut
		{NoStop}%
	\end{thebibliography}
\end{document}